\documentclass{aa}
\usepackage{graphics}
\usepackage{epsfig}
\newcommand{\JminK}{\mbox{$(J-K_\mathrm{s})$}}

\newcommand{\jk}{\mbox{$J\!-\!K$}}

\newcommand{\mh}{\mbox{\rm [{\rm M}/{\rm H}]}}

\newcommand{\Teff}{\mbox{$T_{\rm eff}$}}
\newcommand{\Tbdred}{\mbox{$T_{\rm b}^{\rm dred}$}}

\newcommand{\comment}[1]{}

\begin{document}

   \title{AGB stars in the Magellanic Clouds. II. The rate of star
   formation across the LMC.}

   \author{M.-R.L. Cioni\inst{1}
          \and
           L. Girardi\inst{2}
          \and
           P. Marigo\inst{3}
           \and
           H.J. Habing\inst{4}
          }

   \offprints{mrc@roe.ac.uk}

   \institute{SUPA, School of Physics, University of Edinburgh,
              IfA, Blackford Hill,
              Edinburgh EH9 3HJ, UK
              \and
              Osservatorio Astronomico di Trieste, INAF,
              Via G. B. Tiepolo 11,
              34131 Trieste, Italy
              \and
              Dipartimento di Astronomia, Universit\`{a} di Padova,
              Vicolo dell'Osservatorio 2,
              35122 Padova, Italy
              \and
              Sterrewacht Leiden,
              Niels Bohrweg 2,
              2333 RA Leiden, The Netherlands
             }

   \date{Received 27 Jyly 2005/ Accepted ...}

   \titlerunning{SFR across the LMC}

   \authorrunning{Cioni et al.}

   \abstract{This article compares the distribution of $K_\mathrm{s}$
magnitudes of Large Magellanic Cloud (LMC) asymptotic giant branch
(AGB) stars obtained from the DENIS and 2MASS data with theoretical
distributions. These have been constructed using up-to-date stellar
evolution calculations for low and  intermediate-mass stars, and
in particular for thermally pulsing AGB stars. A fit of the
magnitude distribution of both carbon- and oxygen-rich AGB stars allowed us
to constrain the metallicity distribution across the LMC 
and its star formation rate (SFR). The LMC stellar 
population is found to be
on average $5-6$ Gyr old and is consistent with a mean metallicity
corresponding to $Z=0.006$. These values may however be affected by 
systematic errors in the underlying stellar models, and by the limited
exploration of the possible SFR histories. Instead our method should 
be particularly useful for detecting variations in 
the mean metallicity and SFR across the LMC disk. 
There are well defined regions
where both the metallicity and the mean-age of the underlying stellar
population span the whole range of grid parameters. The C/M ratio
discussed in paper I is a tracer of the metallicity distribution if
the underlying stellar population is older than about a few Gyr. 
A similar study across the Small Magellanic Cloud is given
in paper III of this series.}

\maketitle

\section{Introduction}
Asymptotic Giant  Branch (AGB) stars are among  the brightest
infrared members of  a galaxy. They  are useful indicators of
galactic structure because they are widely distributed and easily
noticed, and the ratio between carbon-rich (C-rich or C-type) and
oxygen-rich (O-rich or M-type) AGB stars is an indicator of
metallicity. Cioni \& Habing (\cite{crat}) -- Paper I --
showed that this ratio varies over the face of the Large Magellanic 
Cloud (LMC) and concluded that a metallicity gradient is present in 
this galaxy; within the Small Magellanic Cloud (SMC)
there are variations in this ratio but there is no clear pattern.
Because AGB stars trace the stellar populations from $\sim0.1$ to 
several Gyr, we want to constrain the global star
formation rate (SFR) of the Magellanic Clouds, interpreting the
large-scale magnitude distribution of AGB stars in the  $K_s$ band
with the help of up-to-date stellar models. This work focuses on
the stellar population in the field; we do not discuss clusters
and associations. Although almost all previous work, a summary of which 
is given below, has been based on information from small
regions within each galaxy, they more or less agree on the
time-scale of major star formation events but it is not yet clear 
what these events were and what caused them.

Section \ref{obs} describes our selection of the sample of AGB
stars from the DENIS catalogue towards the Magellanic Clouds
(DCMC) and from the 2MASS catalogue. Section \ref{theo} describes
the theoretical models used  to construct a magnitude distribution
while Sect. \ref{comp} compares the observed and theoretical
distributions. Section \ref{dis} contains a comparison of our
results with the information known from the literature while
section \ref{fin} concludes this paper. A similar study of the star
formation rate in the SMC is presented in paper III of the series.

\subsection{SFR in the LMC - Review}
\label{intro}
% Magellanic Clouds general properties
The LMC is a disk galaxy within which a bar is embedded. The disk,
elongated toward the Galactic center, perhaps because of the tidal
force induced by the  Milky Way, is elliptical with ellipticity
$\epsilon=0.199\pm0.008$;  it  is viewed  at an inclination  angle
$i=34^{\circ}.7\pm6^{\circ}.2$ and a position angle of the major
axis ${\mathrm{PA}}_\mathrm{maj}=189^{\circ}.3\pm1^{\circ}.4$
(van der Marel \cite{vdm}, van der Marel \& Cioni \cite{maci}).
The off-center bar is about $3^{\circ}$ long, it is inclined at 
$15^{\circ}.1\pm2^{\circ}.7$ such  that the eastern side  is
closer to us than the western side (Subramaniam
\cite{subra}).  The LMC thickness (about
$2.4$ kpc, Lah et al. \cite{lah}) is almost negligible compared 
to its distance. 

% LMC globular clusters
The majority of the LMC globular cluster population is about 
$3-4$ Gyr old and perhaps results from a  tidal interaction
with  the SMC (van den  Bergh  \cite{vdbe}, Bekki et al.
\cite{bek}). A few globular clusters formed before $11.5$ Gyr ago and
during recent distinct episodes of cluster formation (Hodge
\cite{hod73}, Subramaniam \cite{subra}, and references therein).
Most star clusters
dissolve in the LMC after about $1$ Gyr (Hodge \cite{hod88}) and then
contribute to a stellar population in the field that is of varying
age and metallicity. A decoupling between the rate of cluster
formation and the SFR suggests that clusters may not be good tracers
of the star formation in general (van den Bergh \cite{vdb98}).
%Note that the difference in age between pairs
%of clusters increases with their separation. This suggests that
%star formation is hierarchial in space and in time (Efremov \&
%Elmegreen \cite{efrelm}).

% The LMC
Pioneering studies of different types of stars (Tifft  \& Snell
\cite{tifsne}) and of their distribution as a function of age
(Isserstedt \cite{iss}) lead us to discover that the bulk of the LMC
disk stellar population formed $3-5$ Gyr  ago (Butcher \cite{but}).  
Some M giants (mostly AGB stars) as well C-rich AGB stars in the
centre of the bar formed during this burst and others during another
episode about $0.1$ Gyr ago (Frogel \& Blanco \cite{frobla}, Wood et
al. \cite{woo} -- Fig. \ref{lmcagb}). 
This latter episode, possibly
associated with a close  encounter with the SMC, has produced a
burst  of star formation  in both galaxies.  It is supported  by
the distribution of early-type stars  (Kontizas et al. \cite{kon})
and has been confirmed  by the MACHO  project comparing stellar
evolution and pulsation  models of Cepheids with their observed
period-frequency distribution (Alcock  et al. \cite{alc}). The
same study also suggests that during the last $0.1$ Gyr star
formation has been propagating from the South-East (SE) to the 
North-West (NW) along the bar where it is still ongoing (Hardy et 
al. \cite{har}, Fig. \ref{lmcagb}).
The formation of supergiant (massive) stars occured more recently,
about $7.5$ Myr ago (Ardeberg \cite{ard76}, Feitzinger et
al. \cite{fei}). 

Olson \& Pefia (\cite{olspef}) postulated a constant
birthrate interrupted by periods of higher than average star
formation. However, Rocca-Volmerange  et al. (\cite{roc}), by
comparing integrated colours of  the Magellanic Clouds with  an
evolutionary model, concluded that  the evolution has been  smooth,
with a steady  or slightly  decreasing SFR in the past. 
Significant variations in the bolometric luminosity of AGB stars
point to differences in the star formation history across the field
(Reid \& Mould \cite{reimou}, Fig. \ref{lmcagb}). In particular carbon
stars at the periphery might be younger (Costa \& Frogel \cite{cost},
Fig. \ref{lmcagb}).  

Deep $B$ and $V$ band observations were compared by Bertelli et
al. (\cite{ber}, Fig. \ref{lmcagb}) and by Vallenari et al. (\cite{vala},
\cite{valb}, Fig. \ref{lmcagb}) with synthetic
colour-magnitude-diagrams and luminosity functions (LFs).
The authors concluded that from
region to region star formation took place at different epochs. 
In particular, the SFR in some fields was low until $4\pm0.5$ Gyr
ago, but this  burst would occur about $2$ Gyr later if the models
include overshooting; the SFR was enhanced $6-8$ Gyr ago in the region East
(E) of the bar while in the region West (W) of the center there was star
formation $2-3$ Gyr ago.
About $8^{\circ}$ from the  LMC center Gallart et al. (\cite{gal04},
Fig. \ref{lmcagb}) using high quality $V$ and $I$ photometry derived 
an intermediate age ($> 2.5$ Gyr old) stellar 
population as well as a younger ($\approx 1.5$ Gyr old) component which is
perhaps the result of an LMC-SMC encounter or from a merger of the LMC with a
smaller galaxy. Although there is
an indication of an age gradient, in the sense: ``older stars
towards the periphery'' ($> 8$ Gyr old), there is only a small
population as old and metal poor as that of the Milky Way halo
globular clusters and dwarf spheroidal galaxies (see also Stryker
\cite{str}).

%HST & the LMC
Observations with the Hubble Space Telescope (HST)
 in the LMC outer disk  revealed for  the
first  time the  oldest  main-sequence turnoff point  (Gallagher
et  al.  \cite{gal}, Fig. \ref{lmcagb}). 
A best fit model of the main-sequence suggests a roughly constant SFR
over $10$ Gyr with an increase around $2$ Gyr (Holtzman et
al. \cite{hol}, Geha et al. \cite{geha} -- Fig. \ref{lmcagb}) although
in the vicinity of the globular cluster NGC 1866 an  increase in  the
star formation   rate   probably occurred a   Gyr earlier (Stappers
et al. \cite{sta}, Fig.~\ref{lmcagb}). A relatively larger component
 of older stars around globular clusters was also found by Olsen
 (\cite{ols}, Fig. \ref{lmcagb}). 
In the inner disk a burst at about $1$  Gyr may correspond  to the
 formation  of the bar (Elson et al. \cite{els},
 Fig. \ref{lmcagb}) but an older bar ($4-6$ Gyr) 
was claimed by Smecker-Hane et  al. (\cite{sme}, Fig. \ref{lmcagb}).
Differences
in either age or metallicity were detected between this inner disk
field and the outer disk field studied by  Gallagher et al.
(\cite{gal}). 
Near the bar Javiel et al. (\cite{jav},
 Fig. \ref{lmcagb}) confirm a stars forming event at $1$ Gyr and one
 before $10$ Gyr (see also Holtzman et al. \cite{holt}, Fig.
 \ref{lmcagb}) with a clear gap from $3-6$ Gyr.  
In the  centre of  the  bar Ardeberg  et al. (\cite{ard97},
Fig. \ref{lmcagb}) identify two strong populations: a  young component that
originated  about $0.5$ Gyr ago and  an older component that
 originated  between $2$  and $9$  Gyr ago.

\vspace{1.5cm}
Thus, information about the star formation history of the LMC has been
obtained from the study of many relatively small regions located
in the outer and inner disk as well as along the bar. 
There are regions of the galaxy that can be
described by a relatively uniform SFR across several Gyr. The dominant
stellar population is of intermediate-age ($>2.5$ Gyr) and extends
to the remote periphery. A burst of star formation has occurred
between $1$ and $3$ Gyr; although some authors attribute to this
event the formation of the bar, there is evidence that stars as
old as $4-8$ Gyr exist in the bar as well as in the disk. Only a
detailed kinematic study of the stellar population will reveal its
distinct components. This as well as more recent bursts of star
formation are probably due to a close passage with the Mikly Way
and the SMC. In the outer disk, searches for stars similar to those
in the halo of the Milky Way have shown that metal poor old giants
are lacking. AGB stars formed essentially during two major epochs:
around $10^8$ yr ago for the most massive and a few Gyr ago for
lower masses. Globular clusters are older than about $11$ Gyr or
younger than about $4$ Gyr.
 
\section{Observational sample of AGB stars}
\label{obs}

\subsection{Selection of AGB stars from DCMC}

The DENIS catalogue towards the Magellanic Clouds (DCMC - Cioni et
al. \cite{ccat}) provides simultaneous observations in $IJK_s$ of
over one million point sources in  the direction of the LMC.
Sources detected in all three wavebands have been used in Cioni \&
Habing (\cite{crat}, paper I)  to study the metallicity
distribution across  the surface of the galaxy.

Following the same criteria as in paper I we selected an almost
complete sample of AGB  stars. In the
colour-magnitude  diagram ($I-J$, $I$) the AGB  stars,
irrespective of their chemical nature,  lie in the region of 
objects  brighter than the tip of  the red giant  branch (TRGB --
Cioni et al. \cite{ctip}) and redder than a diagonal line that
distinguishes them  from younger stars or  foreground stars (see
Fig. 1 of Cioni, Habing  \&  Israel \cite{cmor}). Our sample
includes O-rich AGB  stars of  early M spectral sub-type  (i.e.
M0-M1), close  to the TRGB, but  excludes AGB stars  with thick
circumstellar envelopes that  are fainter  than the TRGB location
\footnote{AGB stars  with thick
circumstellar envelopes are expected to be just a small fraction
($<10$\%) of the total number of AGB stars. This upper limit has been
estimated from the number of sources with $(J-K_{\mathrm s})>2.5$.}.
The spectral  sub-type  of  O-rich  AGB stars is  a function of
their ($I-J$) colour  (Fluks et al. \cite{fluk}, Blanco, McCarthy
\& Blanco \cite{bmb}).

The  TRGB is at $I=14.54$ and to account for the extinction along the 
line of sight we  adopt an average extinction \footnote{Differential
reddening has a negligible
effect in the $K_{\mathrm s}$ band.} corresponding to
$E(B-V)=0.15$ 
as given by  Westerlund (\cite{west}). Using the extinction law of 
Glass (\cite{glas}) we obtain the following absorptions: $A_I=0.27$,
$A_J=0.11$ and $A_{K_s}=0.04$.
Before selection of the AGB population the data have
been  dereddened using the extinction values listed above. Because
we use both DCMC and 2MASS data (see below) the DCMC $JK_s$ data
have been transformed into the 2MASS photometry by applying the
systematic shifts derived by Delmotte et al. (\cite{delm}):
$J^\mathrm{2MASS}=J^\mathrm{DCMC}+0.11$ and $K_s^{\mathrm
2MASS}=K_s^\mathrm{  DCMC}+0.14$. A diagonal line
separates AGB stars from younger and foreground objects:
\begin{equation}
I_0=-4.64\times(I-J^\mathrm{2MASS})_0+19.78 \,\,\,.
\end{equation}
where the subscript $0$ indicates that magnitudes and colours are
corrected for extinction. 
These selection criteria  result in $32801$
candidate AGB stars (Fig.~\ref{gen}).

\subsection{Selection of AGB stars from 2MASS}

\begin{figure}
\resizebox{\hsize}{!}{\includegraphics{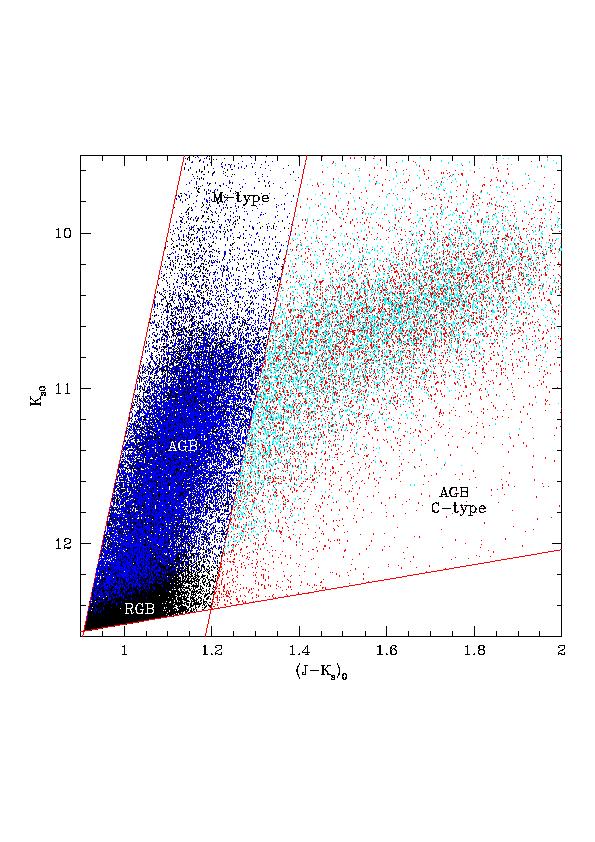}}
\caption{Distribution  of  candidate  LMC  AGB  stars  in  the  colour
magnitude diagram  ($J-K_s$, $K_s$). Because  of the large amount of
data this figure can only  be appreciated in colour.  Stars selected
from the 2MASS data are shown in  black (M-type) and red (C-type) while
stars selected  from the DCMC data  are shown in blue  (M-type) and
cyan (C-type). Note that 2MASS selected sources most probably include
genuine RGB  stars at the faintest  magnitudes of M-type candidates as
well  as   more  stars  than  selected  from the  DCMC  data  distributed
overall. This is probably a result of the different selection criteria
as well as the different photometric accuracy.}
\label{gen}
\end{figure}

The 2MASS all sky  survey (Skrutskie \cite{skru}) provides
simultanous $JHK_s$ observations of  sources distributed in the
whole  sky as well as in the direction of  the Magellanic Clouds.
In the ($J-K_s$, $K_s$) colour-magnitude-diagram most AGB stars
are brighter than the TRGB and are redder  than younger  and
foreground objects.  There are  some AGB stars as faint as the
TRGB itself  or slightly fainter  and they cannot be distinguished
from evolved red giant branch (RGB) stars by near-infrared
photometric criteria. These stars are usually O-rich and of early
spectral subtype (i.e. M0-M1).

After dereddening the data using  the absorption values derived in the
previous sub-section, candidate AGB stars occupy the region
of  the colour  magnitude  diagram ($J-K_s$,  $K_s$)  enclosed by  two
lines:
\begin{eqnarray}
K_s{_0} & = & -0.48\times(J-K_s)_0+13\\
K_s{_0} & = & -13.333\times(J-K_s)_0+24.666.
\end{eqnarray}
A discussion  about the  precise location of  these lines is  given in
Sect.~\ref{features}. Figure \ref{gen}
shows the location of these lines  as well a third line to distinguish
between O-rich and C-rich AGB stars presented in the next sub-section.

In summary $61078$ candidate AGB stars were selected in
the LMC. We included only sources
detected  in all three  photometric wavebands and with good
photometry and accurate positions (this is equivalent to selecting
stars with 2MASS rd\_flg$=1$,  $2$ or $3$ in  each wave band),  to
strengthen the reliability of  the sample (Fig. \ref{gen}).  These
stars  occupy the same surface area as DCMC stars and in
particular were extracted from the  all  sky 2MASS data  release
as  those  included  in a  polygon:
$61.61^{\circ}<\alpha<101.79^{\circ}$ and $-61^{\circ}
<\delta<-77^{\circ}$. 

\subsection{Photometric selection of M and C stars}
\label{sec_photosel}

AGB  stars can be either  O-rich or  C-rich. The
distinction between spectral types can be made because
different molecules dominate the stellar spectra. O-rich stars are
essentially dominated by VO, TiO and H$_2$O molecules while CN and C$_2$
molecules dominate the spectra of C-rich AGB stars. O-rich AGB
stars have an approximately constant ($J-K_{\mathrm s}$) colour and span a
large range of $K_s$ magnitudes. C-rich AGB stars have redder
colours and are constrained to a smaller magnitude interval.
However there is a range of colours populated by both O-rich and
C-rich AGB stars (Groenewegen \cite{groe}).

We discriminate the two spectral classes of AGB
stars using a line at
\begin{equation}
K_s=-13.333\times(J-K_s)+29.333,
\end{equation}
which is also shown in Fig.~\ref{gen}. In Cioni \& Habing
(\cite{crat}) we use spectroscopically confirmed AGB stars and
theoretical models to discriminte between O-rich and C-rich AGB stars
using a vertical line at $J-K_{\mathrm s}=1.4$. 
Here we improve the discrimination criteria
using a diagonal line with the same slope as the line that
discriminates AGB stars from younger and foreground stars that also
agrees with the slope of the O-rich AGB branch. C-rich AGB stars
depart from this branch towards redder colours under the effect of
molecular blanketing. 

The use of  sharp boundaries to select a group of stars may be
affected by ``migration''  of objects across the boundaries because of 
uncertainties due to photometric errors or to variability. The
errors are small ($< 0.02$ mag), but the ``migration'' due to
variability may be significant. Most AGB stars are classified as
long-period variables and vary over several magnitudes in the
visual with a periodicity of a few hundred days and a more or less
regular light-curve shape (e.g. Cioni et al. \cite{cvar},
\cite{ciso}). At near-infrared wavelengths the amplitude of the
variation is below $1$ magnitude and the variation in at 
$J-K_{\mathrm s}$ colour is least a factor of ten smaller.
However observations at a single epoch,
like those available for this paper, caught stars at a specific
position in the light curve, some close to minimum brightness,
others close to maximum brightness and some in the middle. Because of
these differences both magnitude and colour will be affected by random
and not systematic shifts considerably reducing the effect of
variability. 

Age and metallicity also contribute to  
the location of individual AGB stars  in the near-infrared
colour-magnitude diagram. 
Cioni \& Habing (\cite{crat}) have shown that the ratio
between C-rich  and O-rich AGB stars  (the C/M ratio) indicates a
variation $\Delta\mathrm{{[Fe/H]}}=0.75$ dex within the LMC.
This covers a range of metallicities from the SMC to the Milky Way and
corresponds to a variation in the position of the TRGB of
$\pm0.2$ mag for a stellar population $3-10$ Gyr old. 
We assume that AGB stars above the TRGB will
be affected by a similar shift in $K_{\mathrm s}$ magnitude (Glass et
al. \cite{comp}). Similarly $J-K_{\mathrm s}$ varies by $\pm 0.15$
mag. 

In this paper we assume that within a small
region ($0.16$ deg$^2$;  Sect. 4) age and metallicity are constant (i.e. stars
formed at the same time in a homogeneous chemical environment) and we
study differences between these regions. While variability affects all
AGB stars regardless of metallicity, an error in the choice of the
diagonal line that divides O-rich from C-rich AGB stars may
systematically affect the interpretation of the number of stars in
each group. This effect can be quantified as follows: within 
$\Delta(J-K_{\mathrm s})=0.15$ mag of the line there is a number of
stars equivalent to  $5\%$ of the total number of AGB 
candidates; these cover about $10$ bins of $\Delta K_{\mathrm s}=0.2$ mag each.
In a sample region ($0.16$ deg$^2$) this corresponds to
mis-placing, on average, not more than $1$ star per bin
(Fig. \ref{fart}). We conclude that only where too few stars occupy a
single bin our conclusions may be uncertain. 

\subsection{Subdivision of the LMC area}
\label{sec_surface}

\begin{figure}
\resizebox{\hsize}{!}{\includegraphics{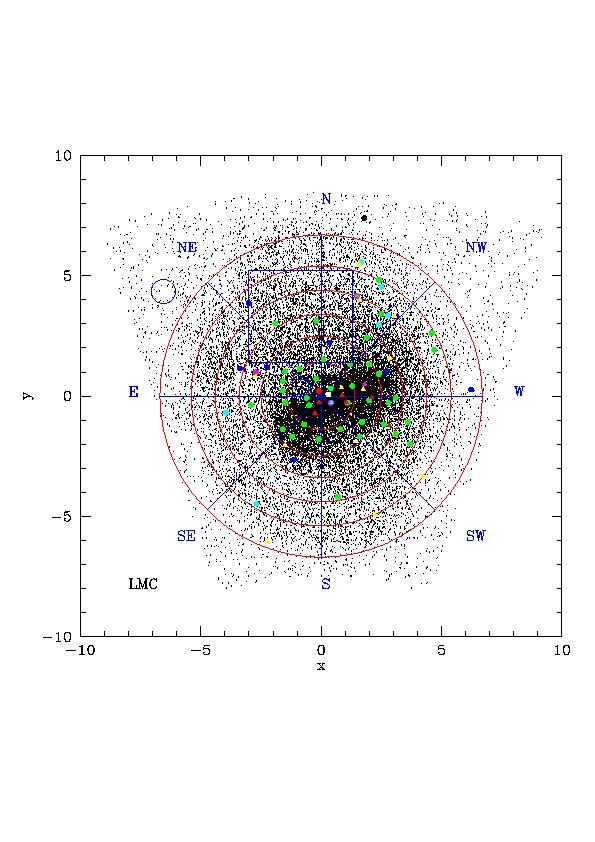}}
\caption{Distribution of the AGB stars  in the LMC.  Data are from
the DCMC and include only sources detected in three DENIS wave
bands.  The division into rings and sectors used in this article
is indicated. The center   is    at   $\alpha    = 82.25^{\circ}$
and    $\delta   = -69.5^{\circ}$.  Coloured symbols indicate  the
location of  regions analysed by other authors, in particular:
thick dots for Costa \& Frogel (\cite{cost}, green),  Javiel et al.
(\cite{jav};  cyan), Vallenari et al. (\cite{valb},  blue),
Holtzman  et al.  (\cite{holt},  magenta), Gallart  et  al.
(\cite{gal04},  black), Elson  et  al. (\cite{els}, white),
Ardeberg  et  al.   (\cite{ard97},  red) and  triangles for Hodge
(\cite{hod88},  yellow), Olsen  et al. (\cite{ols}, red),
Gallagher et al.  (\cite{gal}), Holtzman et al. (\cite{hol}) and Geha et
al. (\cite{geha}) (green), Tifft \&  Snell
(\cite{tifsne},  magenta), Smecker-Hane  et   al. (\cite{sme},
cyan).  The   NGC  1866  cluster corresponds to  a concentration
of  four different symbols  (N-NW) while large empty regions refer
to studies by: Reid \& Mould (\cite{reimou}, big rectangle),
Stryker (\cite{str}, circle)  and Wood et al. (\cite{woo}, small
square).}
\label{lmcagb}
\end{figure}

In  order to  study  the shape  of  the magnitude  distributions as  a
function of spatial coordinates  we divided the LMC area into
rings and sectors: six concentric rings centered at $\alpha
=  82.25^{\circ}$  and  $\delta   =  -69.5^{\circ}$  at  a  radius  of
$1.4^{\circ}$,     $2.5^{\circ}$,     $3.4^{\circ}$,    $4.4^{\circ}$,
$5.4^{\circ}$  and  $6.7^{\circ}$,  respectively,  as  well  as  eight
sectors   at   an  angular   aperture   of   $45^{\circ}$. 
Rings  are numbered $0-5$ with increasing  radius and sectors
by compass   direction  (E,   NE,   N,  NW,   W,   SW,   S,  SE);   see
Fig. \ref{lmcagb}. The outer four  rings coincide with the division in
van der  Marel \& Cioni (\cite{maci});  see also paper  I, while their
innermost  region has  been  split in  two.  

\subsection{Comparison between DCMC and 2MASS samples}

The  important advantage  of having  simultaneous $IJK_s$
band data is that it is possible to include O-rich  AGB
stars of early spectral type in the selected AGB sample. 
$JK_s$ photometry does 
not contain enough information for this task because while all AGB stars
are located above the TRGB in the $I$ band, some will be below the TRGB in
the $K_s$ band, therefore they are mixed with RGB stars. This 
is clearly demonstrated in Fig. \ref{gen}. 

On the other hand the 2MASS photometry is somewhat more precise
than the DENIS photometry and 2MASS observations reach about $1$
mag fainter  objects. The important  consequence is  that the
sequences occupied  by  different  stars  in the colour-magnitude
diagram  are narrower and better defined.

A comparison of DCMC and 2MASS  selected AGB stars  allow us
to: (i)  test  the  contamination  of the sample of AGB stars by
RGB stars of early spectral subtype, and (ii) determine the
completeness of AGB stars above  the TRGB in  the $K_s$  band that
should be equally recovered using both photometric selection
criteria in both datasets.

The AGB stars are selected in the  ($I-J$, $I$)
diagram from DCMC-data as described in Sect. 2.1. The same
stars then are plotted in the ($J-K_s$, $K_s$) diagram and it is
this latter distribution that has lead us to define three lines to
characterize the region that all O-rich and C-rich AGB stars
occupy.

\begin{figure}
\resizebox{\hsize}{!}{\includegraphics{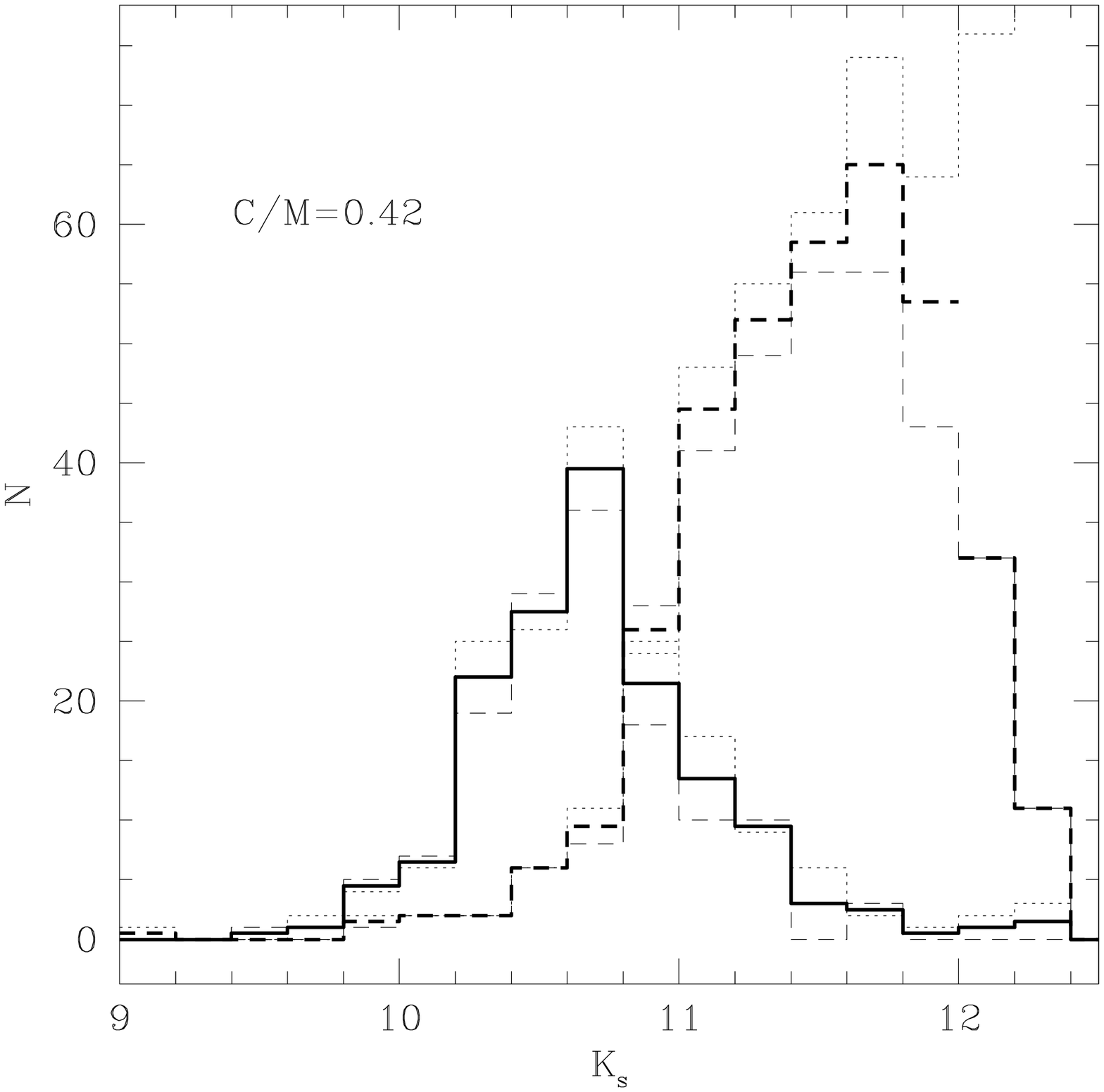}}
\caption{Distribution of C  and M stars as a function  of $K_s$ mag in
sector N of  ring number $3$ (Fig. \ref{lmcagb}). 
The histogram peaking at  about $K_s=10.7$ is due to C stars
while  M stars  populate the  peak  at fainter  mag ($K_s=11.7$).  The
dotted  line indicates  the distribution  of stars  selected  from the
2MASS data while  the dashed line indicates the  distribution of stars
selected from the DCMC data. The thick continuous and dashed lines are
the  average of both  data sets  for $K_s<12$  while for  $K_s>12$ the
distribution of  M stars is strictly  that of DCMC  data.  These thick
lines show  the final histogram  obtained from the  observational data
that  is later  compared with  a theoretical  distribution.  The ratio
between  C and  M stars  obtained from  this final  histogram  is also
indicated.}
\label{fart}
\end{figure}

Let  us focus on a  particular ring  (number  $3$)
and sector (N).  Fig. \ref{fart}  shows the distribution of
O-rich and C-rich  stars  selected from  each  data set  as  well
as a  combined histogram produced as follows.  At magnitudes above
the TRGB ($K_s<12$) a simple average was calculated of 
stars selected from 2MASS and DCMC. This  is justified because at
these magnitudes  both data sets should have detected the same AGB
star candidates irrespective of their $I$ or $H$ magnitude.
Therefore we attribute  differences to a different photometric
quality, sensitivity  and possibly different arrays 
and  flags of image or reduction  origin that may have lead  to
exclusion of  some  detections, even  at  these relatively bright
magnitudes.   Averaging  measurements  from  both catalogues has
the important advantage of reducing  migration effects between
adjacent bins and smoothning the effect of variability. 2MASS and
DCMC observations were obtained on average from a  few to several
months apart and their mean magnitudes should closely resemble the
average magnitude that one would obtain  if a near-infrared
light curve were available  for each source.  The distributions of
the C stars from DCMC and from 2MASS are very similar. This is
also the case in other sectors whether they contain more or fewer
stars. A Kolmogorov-Smirnov test indicates  a probability of about
$80-90$\% that  both C-star distributions originate  from the same
population.

DCMC and 2MASS samples also agree when one considers the O-rich
stars brighter than $K_s=12$ mag. At $K_s>12$ the distribution of
O-rich stars extracted from the 2MASS sample departs considerably
from that of the stars extracted from the DCMC sample and we
attribute this to the strong contamination of the AGB sample by
RGB stars in the 2MASS data. Simply averaging the DCMC and 2MASS
samples will produce erroneous results, namely a too large number
of candidate AGB stars of early spectral sub-type. Because of the
availability of  the $I$-DCMC measurement we are confident that we
avoided most of these RGB/pseudo-AGB stars and we conclude that
the best distribution at $K_s>12$ is given by DCMC stars only.
Note that at these magnitudes amplitudes of variability decrease 
while photometric errors increase; it is likely that the 
effect of migration between adjacent bins becomes more important 
than the effect of variability. 

\section{Theoretical models}
\label{theo}

\subsection{General description}

The $JK_s$-band photometry of the Large Magellanic Cloud has been
simulated using  a  population  synthesis   code  called  TRILEGAL
(Girardi  et al. \cite{gi05}) that randomly generates a population
of stars following a given SFR,
age-metallicity relation (AMR) and initial  mass   function
(IMF).  The   stellar  intrinsic  properties (luminosity $L$,
effective  temperature $T_{\rm eff}$, surface gravity $g$, etc.)
are interpolated over  a large grid of stellar evolutionary
tracks,  based on  Bertelli et  al. (\cite{bert})  for  massive
stars, Girardi et al. (\cite{gi00}) for low- and intermediate-mass
stars, and complemented with grids of thermal pulsing AGB (TP-AGB) 
tracks calculated by
means of Marigo et al.'s (\cite{ma99}) synthetic code.  This latter
code has been improved as follows.

First, the TP-AGB computations now adopt low-temperature opacities
consistent with the actual chemical composition of the stellar
envelope, rather than for the (usually assumed) solar-scaled
compositions. As first shown by Marigo
(\cite{ma02}) this choice has dramatic consequences for the
evolution of TP-AGB stars, that include significant changes in
their \Teff\ and lifetimes, providing a natural explanation
for the appearance of a red tail of C stars in colour-magnitude
diagrams (CMDs) involving the \JminK\ colour (see Marigo et al.
\cite{ma03}).

Second, the TP-AGB tracks adopt a new formalism to describe the 
efficiency of the third dredge-up process and the mass loss rates 
(Marigo et al., in preparation). The main novelties, with respect 
to the formalism described in Marigo et al. (\cite{ma99}), are:
the ratio between the mass of dredged-up material 
and the total core mass growth at each interpulse phase, $\lambda$,
is a function of mass and $Z$ according to the Izzard et al.
(2004) stellar models. As in Marigo et al. (\cite{ma99})
the criterion for activating the third dredge up is based on
the minimum temperature reached at the basis of the convective 
envelope after shell He-flash episodes, \Tbdred; this parameter,
however, now includes a small metallicity dependence as suggested by
the calibration of the carbon star luminosities in the Magellanic
Clouds. The mass loss 
rates are still described using the Vassiliadis \& Wood 
(1993) prescription, but now using the period values given by
either fundamental mode or first overtone. The transition between
these two pulsation modes is predicted according to a series of
theoretical indications from pulsation models (Ostlie \& Cox 1986). 
The details of these TP-AGB models will be given in a forthcoming paper. 
They provide a reasonable reproduction of
the C star LFs in both Magellanic Clouds, and of the lifetime of
the carbon phase as a function of mass as deduced from C-star 
counts in LMC star clusters (see Girardi \& Marigo \cite{ma02}). 

In this paper we use complete grids of tracks for $5$
different metallicities, namely $Z=0.0004$, $0.001$, $0.004$,
$0.008$ and $0.019$. The $Z=0.004$, $0.008$ and $0.019$ cases
are close to the metallicity values found in young populations of
the SMC, LMC and Solar Neighbourhood, respectively.
Interpolations among these grids allow us to generate stars for
any intermediate metallicity between $Z=0.0004$ and $0.019$.

The evolution along the TP-AGB phase is initially described in
terms of the properties in the quiescent stages between thermal
pulses. We then include in the simulations the effect of
luminosity variations driven by thermal pulses, in the same way as 
Marigo et al. (\cite{ma03}). This produces a
broadening -- especially to lower luminosities and high \Teff\ 
-- of the distribution of TP-AGB stars in the HR diagram.

\subsection{Construction of the theoretical $K_s$-band distribution}

Once the intrinsic stellar properties ($L$, \Teff, C/O) have been 
determined, we simulate the photometry by applying the extended 
tables of bolometric corrections (BCs) from Girardi et al. (\cite{gi02}) 
for O-rich stars (with C/O$<1$), and by empirical relations for C-rich
stars (with C/O$>1$).

The BCs for M-type stars are based on the
empirical spectra and \Teff--colour scale by Fluks et al.
(\cite{fluk}). For other O-rich stars with $\Teff\ga4000$~K,
instead, they are entirely based on ATLAS9 model atmospheres (see
Kurucz \cite{kuru}, Castelli et al. \cite{cast}). Since at LMC 
metallicities the M-type fraction of the AGB phase develops
mostly at \Teff\ lower than 4000~K, the Fluks et al. (\cite{fluk})
set of corrections is actually the most relevant for the present
work. The magnitudes and colours we simulate in this way
correspond to the DENIS $I,J,K_\mathrm{s}$ and to the 2MASS
$JHK_\mathrm{s}$ filters.

Regarding C-type stars, we rely on a few
empirical relations. The BC in the $K$ band is taken from the
relation by Costa \& Frogel (\cite{cost}). The $(J-K)$ colour is
derived from the \Teff--(\jk)--C/O relation from Marigo et al.
(\cite{ma03}), which itself is based on a fit to the empirical
data by Bergeat et al. (\cite{berg}).\footnote{The fitting relations
to BC in the $K$-band (Costa \& Frogel \cite{cost} show a dispersion
of just 0.05~mag, which is comparable to the
photometric errors in the observed samples). The
possible errors in the attribution of \jk\ will affect only the 
selection between C- and M-type stars in the fainter end of their
magnitude distribution, but not the bulk of the AGB stars.}

Finally, we simulate the photometric errors in the following way:
for each star of $J$ and $K_\mathrm{s}$ we compute $\sigma_{J}$ and 
$\sigma_{K_\mathrm{s}}$ using the fitting formulas by Bonatto et al. 
(\cite{bona}, and private communication). Then, the single errors are 
randomly generated from Gaussian distributions of widths $\sigma_{J}$ 
and $\sigma_{K_\mathrm{s}}$, and added to $J$ and $K_\mathrm{s}$. 
This helps in producing clean-looking CMDs,
but is not a critical step since at the magnitudes involved in
this study, the photometric errors are typically very low (less
than 0.03 mag).

\subsection{Main features and behaviour of the theoretical distributions}
\label{features}

Before the comparison with observational data, we 
discuss the basic behaviour of present models, as a function of
galactic properties. The main parameters describing a galaxy are
the SFR, $\psi(t)$, and the mean metallicity $Z(t)$, as a function of
stellar age\footnote{The IMF of stars
is assumed to be independent of age and equal to the log-normal
function of Chabrier (\cite{chab}). None of the results of this
paper depend on the detailed shape of the IMF.} $t$.

Of course, we can test only a limited number of all possible SFR
and AMR functions. We will use existing,
and hopefully realistic models for these functions for the
LMC. However, we found it
useful to begin with simple functions that may be somewhat
unrealistic but very instructive.

\subsubsection{Varying the metallicity at a constant SFR}
\label{constSFH}

In this case, $\psi(t)=\psi_0$, where $\psi_0$ is an arbitrary constant
chosen so as to give a reasonable number (several thousand) of AGB stars.
$Z$ is also kept at a constant value, $Z(t)=Z_0$, inside each simulation.
We then analyse the effect of varying $Z_0$ between our limiting values of
$0.0004$ and $0.019$.

\begin{figure*}
\begin{minipage}{0.325\textwidth}
\resizebox{\hsize}{!}{\includegraphics{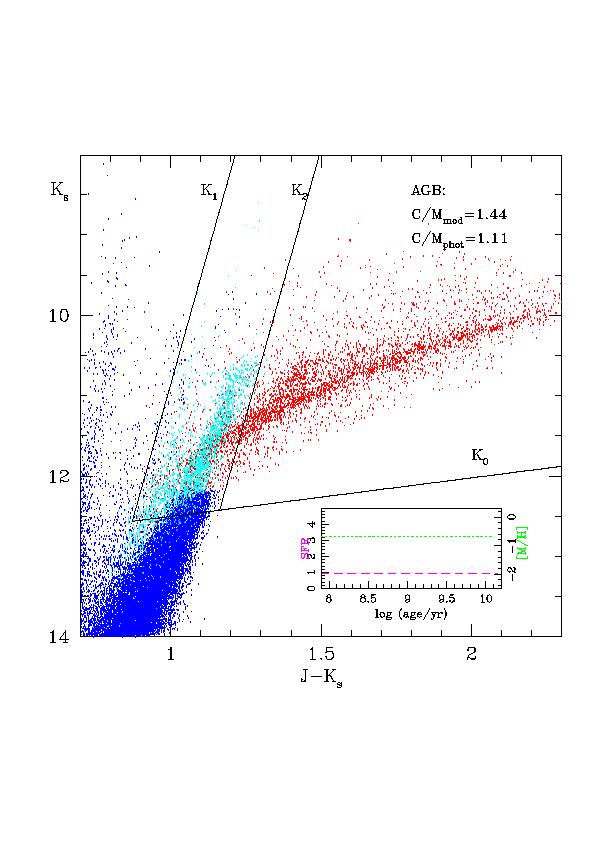}}
\end{minipage}
\hfill
\begin{minipage}{0.325\textwidth}
\resizebox{\hsize}{!}{\includegraphics{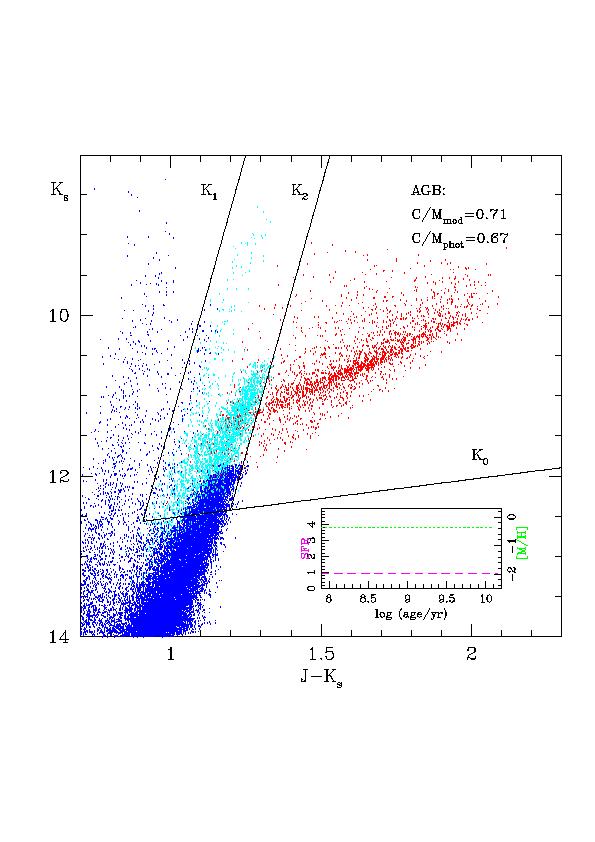}}
\end{minipage}
\hfill
\begin{minipage}{0.325\textwidth}
\resizebox{\hsize}{!}{\includegraphics{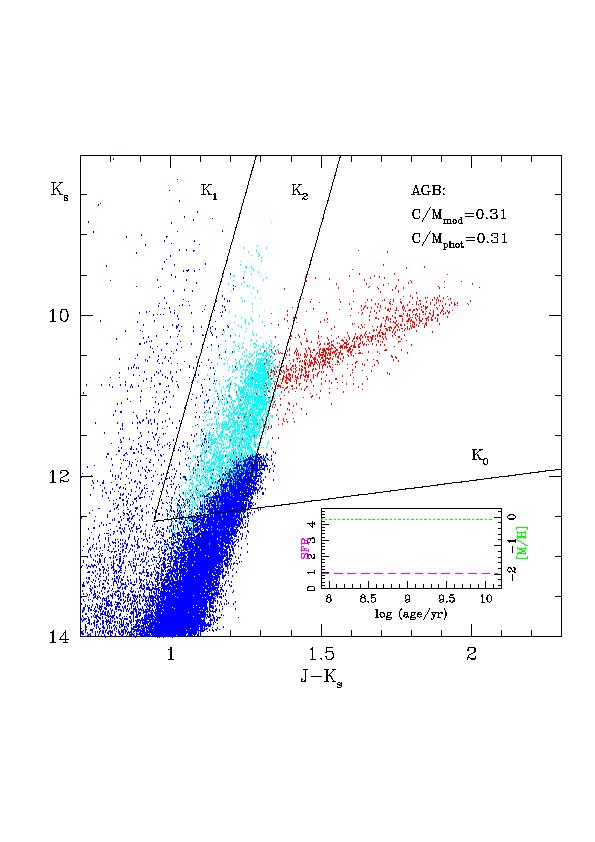}}
\end{minipage}
\caption{Simulated CMDs for models that assume constant SFR, and
three different values of metallicity for stars of all ages:
$Z=0.004$ (left panel), $Z=0.008$ (middle) and $Z=0.016$ (right).
For comparison, the inserts illustrate the shape of
SFR (dashed line, left axis) and AMR for each model 
(dotted line, right axis). A distance of $50$ kpc is assumed
(this is approximately the distance to the LMC). The straight
lines delimit regions of the CMD that were used in the photometric
selection of C- and M-type stars (see text). In the electronic
version of this paper only, different kinds of stars are
denoted with different colours, namely blue for all stars before
the TP-AGB phase, cyan for O-rich and red for C-rich TP-AGB stars.
Notice the marked differences in the location and relative number
of M and C stars. The TRGB is easily noticed at $K_s\sim12.3$,
$\sim12.0$ and $\sim11.7$ in the left, middle and right panels,
respectively. } 
\label{fig_compZ}
\end{figure*}

A sequence of such models is illustrated in Fig.~\ref{fig_compZ}.
They show some remarkable changes in the appearance and location of the
several CMD features, that will be commented on later in this work.

\begin{figure}
\resizebox{\hsize}{!}{\includegraphics{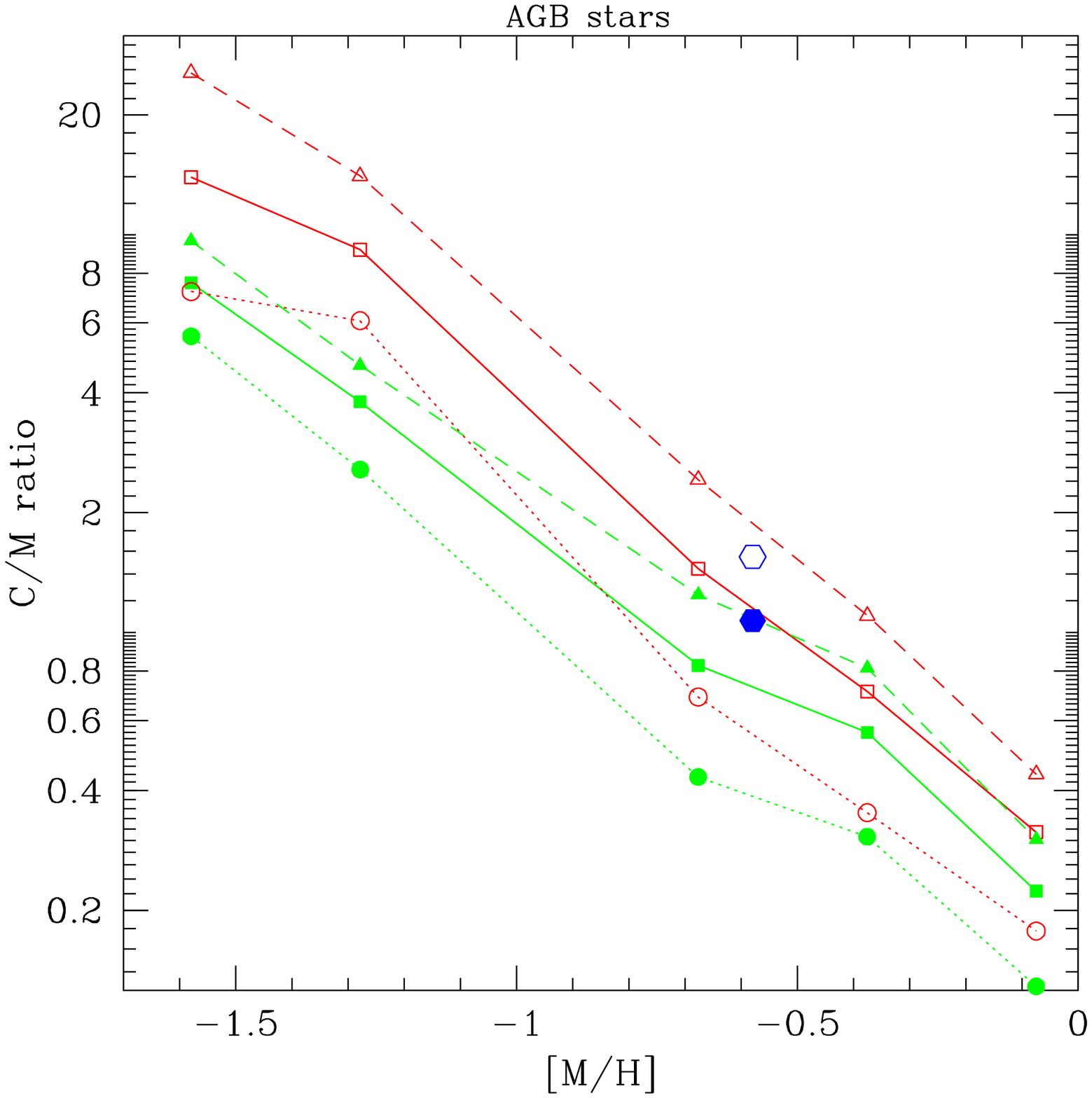}}
\caption{Variation of C/M with metallicity \mh\ for 
three different cases of SFR -- from top to bottom: 
$\propto\exp(-t/5{\mathrm Gyr})$ (triangles and dashed lines), 
constant (squares and continuous lines), and
$\propto\exp(t/5{\mathrm Gyr})$ (circles and dotted lines) -- 
and using both the model 
(empty symbols) and photometric (filled symbols) selection criteria 
for AGB stars above the TRGB. Notice that both criteria produce
the same behaviour of C/M as a function of [M/H]. The hexagons
represent the C/M for a ``realistic model'' of the LMC 
(see Sect.~\protect\ref{sec_realistic}).} 
\label{fig_CM_SFR}
\end{figure}

The continuous lines connecting the squares 
in Fig.~\ref{fig_CM_SFR} illustrate the variation of C/M with 
metallicity $\mh=\log(Z_0/Z_\odot)$, for two different selection
criteria of C and M stars. The most significant of these criteria
is the {\em model C/M ratio}, that we obtain by simply counting the
simulated TP-AGB stars of different C/O ratio (C/O$<1$ for M-type,
C/O$>1$ for C-type).
Then we have the {\em photometric C/M ratio}, defined in the following way:
first, we draw three different lines in the CMD, namely:
\begin{eqnarray}
  K_0 & = & -0.48\,\JminK + 13.022 + 0.056\,\mh\\
  K_1 & = & -13.333\,\JminK + 25.293 + 1.568\,\mh\\
  K_2 & = & -13.333\,\JminK + 29.960 + 1.568\,\mh
\end{eqnarray} 
C stars are defined by $K<K_0$ and $K>K_2$, M stars by $K<K_0$ and 
$K_1< K < K_2$. The significance of these three lines is clear when 
looking at Fig.~\ref{fig_compZ}. The photometric C/M ratio is obtained 
by the division of the number of stars counted in these two different
regions of the CMD. The lines $K_1$ and $K_2$ are parallel with a
separation $\Delta\JminK=0.26$~mag. This is wide enough to
completely encompass the sequences of O-rich TP-AGB stars
(especially at high metallicities where their colour distribution
is broader), avoiding as much as possible the contamination by
other kinds of stars. In these equations, the terms with
\mh\ reflect how the theoretical sequences shift with the metallicity
assumed in the simulations.

Fig.~\ref{fig_CM_SFR} shows that the photometric C/M ratio as
function of \mh\ is parallel to the the model C/M ratio, in spite 
of the incomplete elimination of RGB stars from the
AGB sample. Because the bluest C-stars will be misclassified as
M-type stars we will always have 
$\mathrm{C/M(phot)}<\mathrm{C/M(model)}$. Both C/M
values depend on \mh\ and, for the case of constant SFR, 
have values typically between $0.25$ and $14$.

\begin{figure*}
\begin{minipage}{0.325\textwidth}
\resizebox{\hsize}{!}{\includegraphics{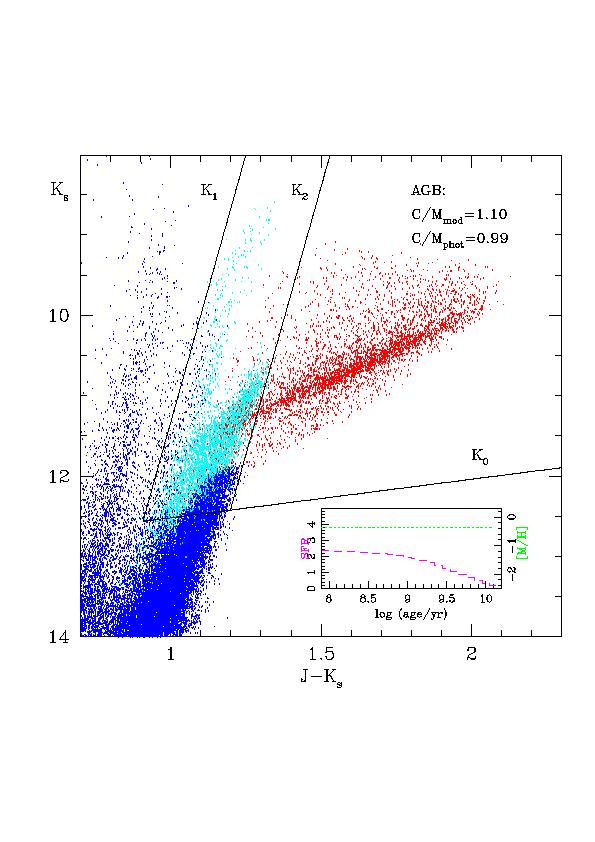}}
\end{minipage}
\hfill
\begin{minipage}{0.325\textwidth}
\resizebox{\hsize}{!}{\includegraphics{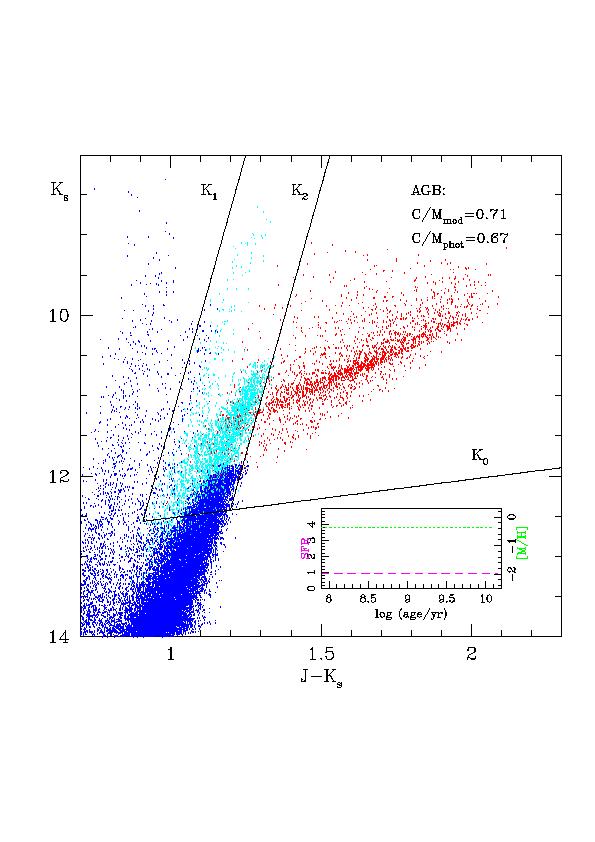}}
\end{minipage}
\hfill
\begin{minipage}{0.325\textwidth}
\resizebox{\hsize}{!}{\includegraphics{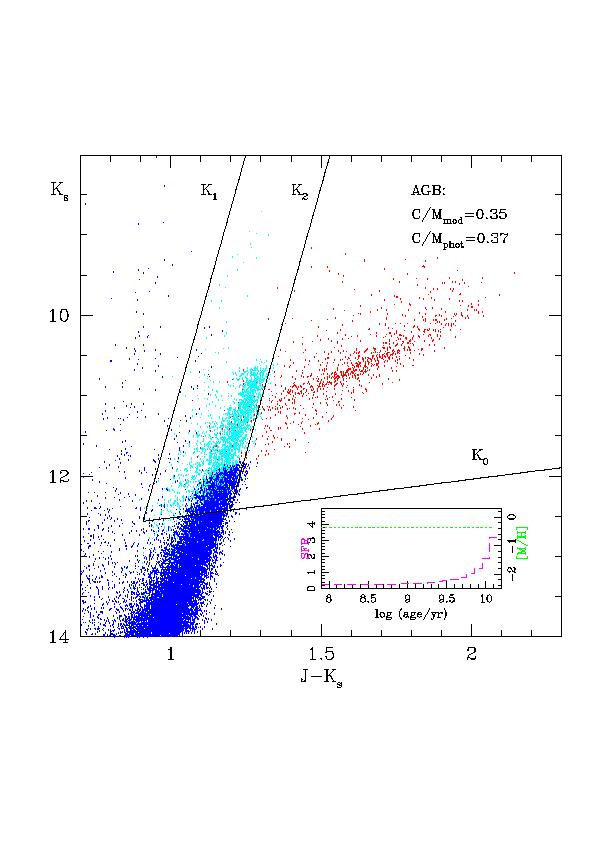}}
\end{minipage}
\caption{Simulated CMDs for models with a constant metallicity,
$Z_0=0.008$, and three different cases of varying SFR: decreasing
($\alpha=-5$~Gyr, left panel), constant (middle panel) and
increasing ($\alpha=+5$~Gyr, right panel) with stellar age. The
insert, straight lines and C/M definitions are described
in Fig.~\protect\ref{fig_compZ}. Notice the large changes in
relative numbers of different kinds of stars.}
\label{fig_compSFR}
\end{figure*}

\subsubsection{Varying the SFR at a constant metallicity}
\label{constZ}

As before we take $Z(t)=Z_0$, and keep it at the same value inside
each simulation. 
What changes from case to case is the SFR. We adopt the
following family of exponentially increasing/decreasing SFRs:
\begin{equation}
\psi(t)=\psi_0 \exp(t/\alpha)
\label{eq_sfr}
\end{equation}
where $t$ is the look-back time (stellar age) and $\alpha$ 
is positive/negative for increasing/decreasing look-back time. In
our calculations $\alpha$ will take values $2,5,\infty,-2,-5$.
Again, $\psi_0$ is an arbitrary constant chosen so as to produce a
reasonable number of stars.

A sequence of such models is illustrated in
Fig.~\ref{fig_compSFR}. The location of the important CMD features
is the same in the three panels but the relative numbers of stars
change as does the C/M ratio. Fig.~\ref{fig_CM_SFR}
illustrates the variation of C/M for a few SFR functions as a
function of metallicity.

\begin{figure}
\resizebox{\hsize}{!}{\includegraphics{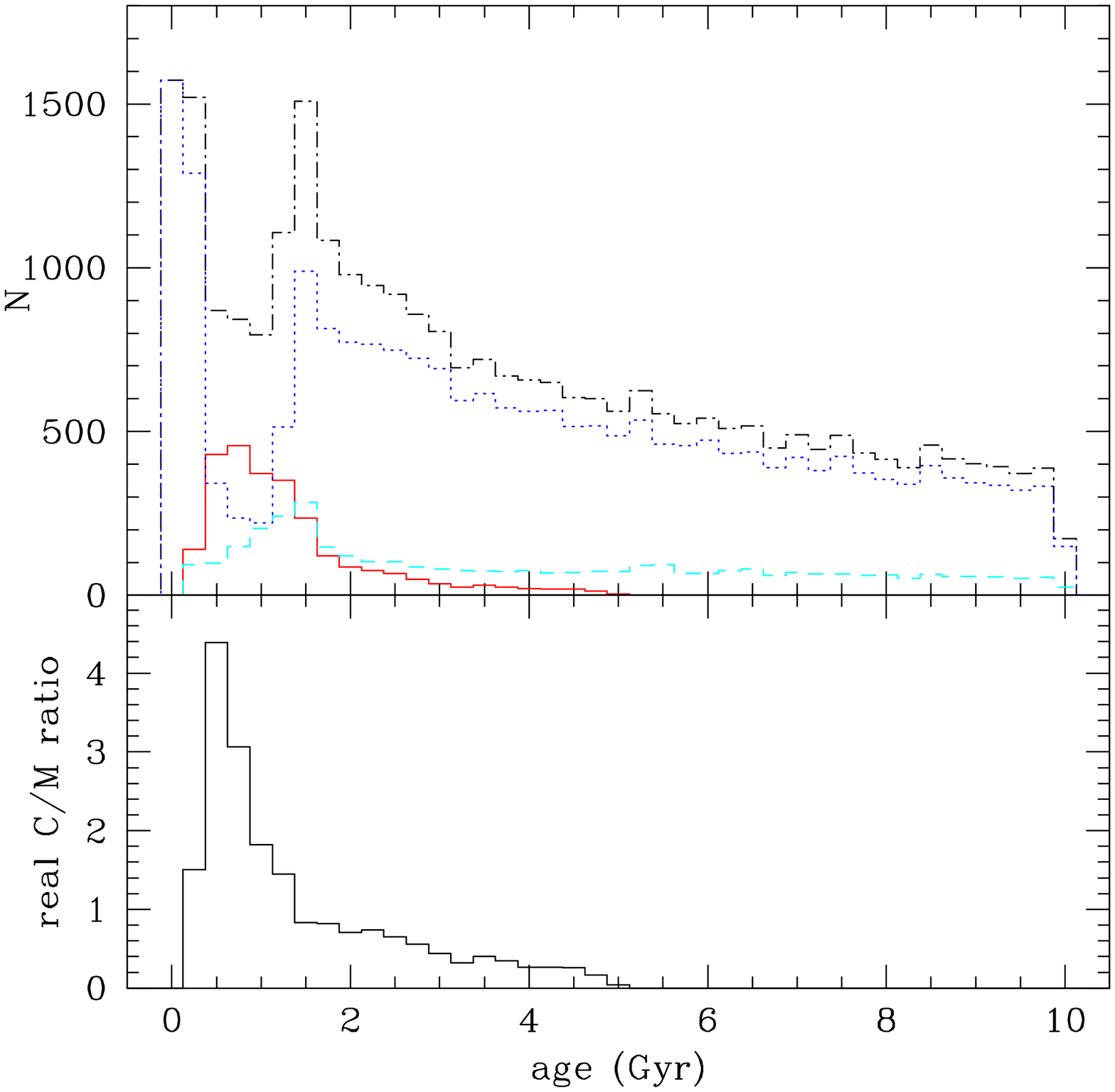}} \caption{ {\bf
Top panel:} Age distribution of different stars when the SFR and
$Z=0.008$ have been constant over the last 10 Gyr. Only stars with
$M_K<-4.5$ ($K<14$ at the LMC distance) have been considered. The
C-rich (red, continuous line) and O-rich (blue, dashed line)
TP-AGB stars are in general less numerous than stars from previous
evolutionary phases (blue, dotted lines; these are mostly core-He
burning and early-AGB stars for $t<1$~Gyr, and RGB stars for
$t>1$~Gyr). The blue, upper dot-dashed line indicates the total
number of $K<14$ stars. {\bf Bottom panel:} The C/M ratio of
TP-AGB stars, as a function of stellar age, for the same model. }
\label{fig_agedist}
\end{figure}

These simulations thus indicate that the C/M ratio of TP-AGB stars
depends strongly on the shape of SFR, a dependence that so far has
not been much discussed. The top panel of Fig.~\ref{fig_agedist}
illustrates this effect. 
It presents a histogram of the ages of the stars in a
sample generated under the assumption that the SFR and the
metallicity both have been constant ($Z=0.008$) over a period of
10 Gyr. The figure shows that the age distribution of C stars
(red, continuous curve) has a somewhat marked peak at 0.75 Gyr and
that there are no C stars younger than 0.25 Gyr and older than
5 Gyr. The M stars (blue, dashed) show a peak at an age of about
1.5 Gyr, but they have a much broader distribution to both younger
and older ages. Thus, the C/M ratio (bottom panel of
Fig.~\ref{fig_agedist}) changes significantly with the population age,
going from 0 for very young stars to a maximum at about 0.5 Gyr,
then falling to 0 again for stars older than $5$ Gyr.
This behaviour agrees with observational data for AGB stars in LMC
star clusters (see e.g. Frogel et al. \cite{frog}).

Changing now from a constant to a variable SFR, the age
distribution can be constructed from that in
Fig.~\ref{fig_agedist} by giving relative weights to age intervals
that have different C/M ratios. This makes it clear that the C/M
ratio must depend on the shape of the SFR function. It is also
evident, from the figure, that by giving more weight to stars
recently formed (decreasing SFR), we get higher C/M
ratios (and vice-versa for increasing SFR).

Thus, the C/M ratio of a stellar population depends both on
the metallicity and on the SFR. At lower metallicities the C/M
ratio is on average higher than at higher metallicities. Moreover
variations of the SFR cause variations of the C/M ratio that are
slightly higher at higher metallicities than at lower
metallicities. 

\subsubsection{A more realistic LMC model}
\label{sec_realistic}

\begin{figure}
\resizebox{\hsize}{!}{\includegraphics{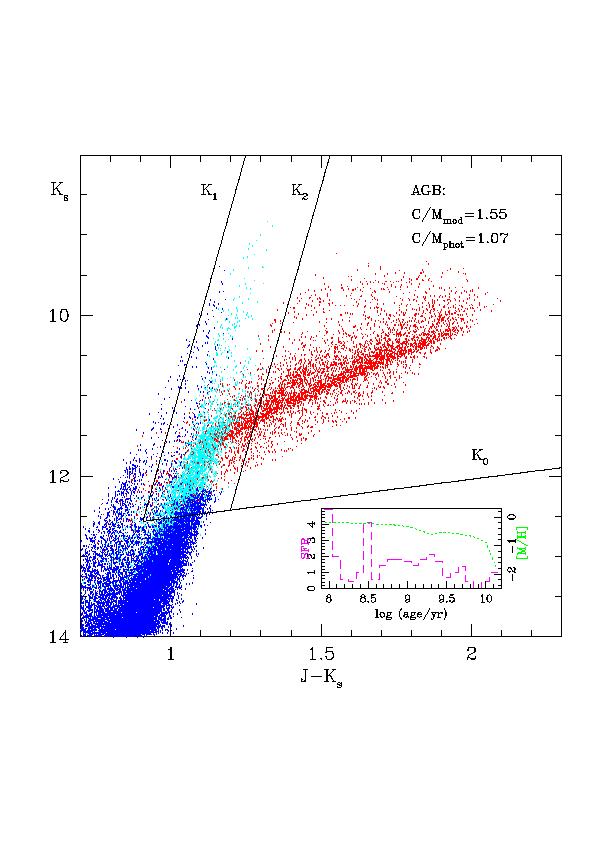}}
\caption{Simulated CMD for a model that assumes observed SFR and
AMR for a LMC bar 
field using Holtzman et al.'s (\cite{holt}) SFR. 
The AMR of Pagel \& Tautvaisiene (\cite{page}) was used. In
this simulation the O-rich TP-AGB stars make a single and
straight sequence in the CMD, being much better defined than in
previous constant-metallicity simulations of Fig.~\ref{fig_compZ}
and Fig. \ref{fig_compSFR}.} 
\label{fig_compMCs}
\end{figure}

So far, we have used the quite unrealistic cases of constant AMR,
i.e. no chemical evolution over the galaxy history, together with
very simplified prescriptions for the SFR. We will now describe a
simulation that adopts a SFR and an AMR derived from observations  
(Fig.~\ref{fig_compMCs}). The SFR is from the analysis, via an
objective method, of optical HST photometry reaching the oldest 
main sequence turn-off in a bar field (Holtzman et al. \cite{holt}). 
The AMR is from chemical evolution models (Pagel \& Tautvaisiene 
\cite{page}), but is shown to agree very well with
independent determinations of chemical abundances and ages of star
clusters in the LMC.

The resulting CMDs (Fig.~\ref{fig_compMCs}) represent the observations
rather well,
and should be compared to actual near-IR data, 
for instance the figures in van der Marel \& Cioni
(\cite{ctip}) and Marigo et al. (\cite{ma03}) in Fig.~\ref{fig_compZ}
and Fig. \ref{fig_compSFR}, the O-rich TP-AGB stars form two distinct
almost-vertical sequences in the CMD: the first to the blue (left)
and the second to the red (right). The first extends to bright
magnitudes; the second sequence is right above the RGB. It is
somewhat limited in magnitude (the exact extension depending
mostly on the assumed metallicity); both sequences merge at lower
luminosities, at about $K\sim 11.6$ for $Z=0.008$. In
Fig.~\ref{fig_compSFR}, it is clear that the bluest sequence
is caused by the younger populations (age less than a few Myr),
whereas the reddest is caused by a population older than a few
Gyr. The two features are separate in the CMD because the stars of
intermediate age expected in between the two sequences have
become C-type TP-AGB stars and they now form the red tail thus
leaving an empty "valley" in the RG- and the AG-Branch.

In the more realistic simulation of Fig.~\ref{fig_compMCs}, these
two sequences of O-rich TP-AGB stars actually merge into a single
sequence. This happens because the metallicity difference between
young and old populations has now been taken into account: the
young metal-rich and old metal-poor populations -- with age and
metallicity as defined by the underlying AMR -- now happen to have
about the same colours. And in fact, Magellanic Cloud observations
(DENIS, 2MASS) present a single sequence of M-type AGB stars on
top of the RGB, exactly as in our simulation of
Fig.~\ref{fig_compMCs}.

\section{Comparison between observed and theoretical distribution}
\label{comp}

\begin{figure*}
\begin{minipage}{0.5\textwidth}
\resizebox{\hsize}{!}{\includegraphics{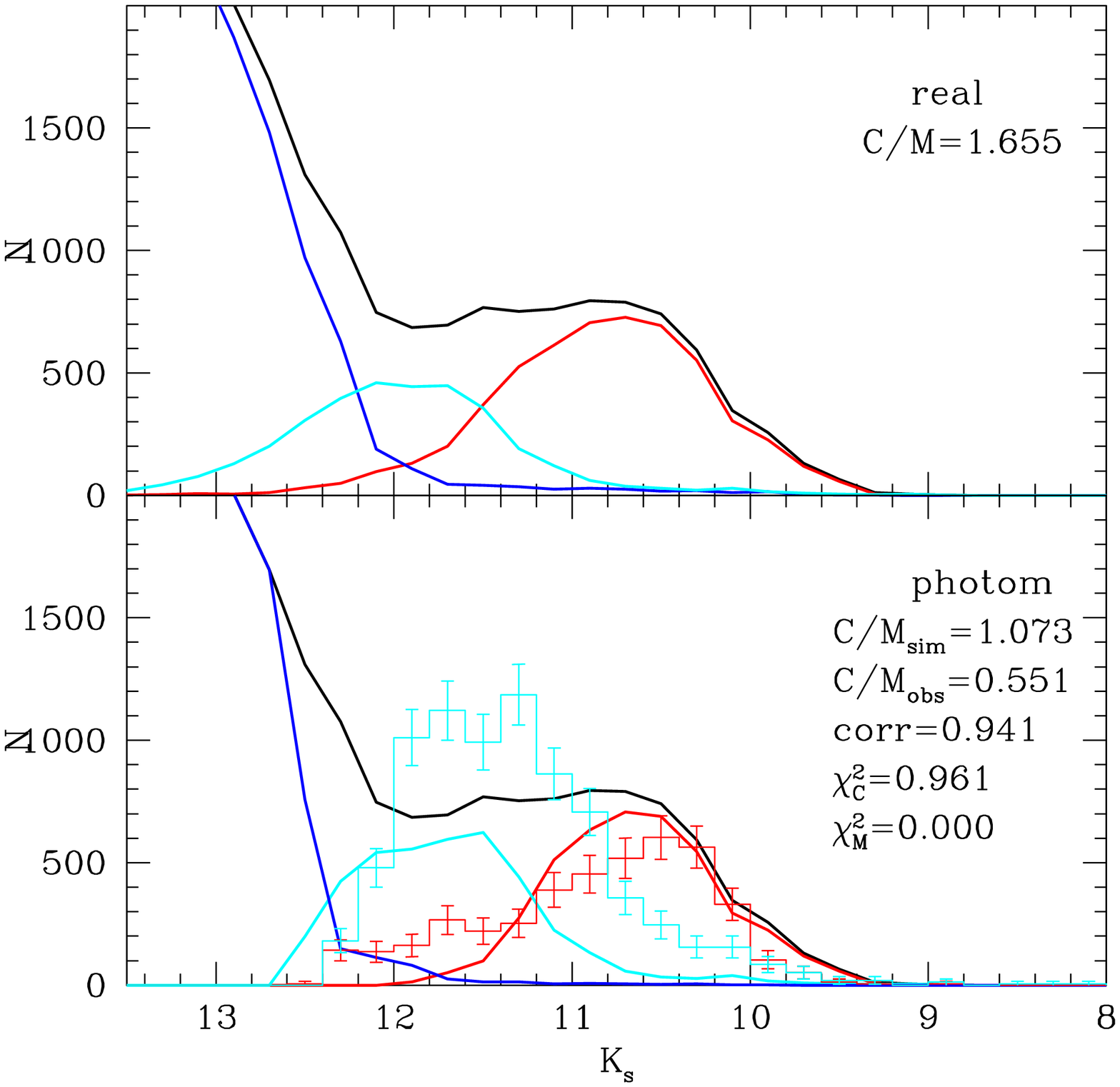}}
\end{minipage}
\begin{minipage}{0.5\textwidth}
\resizebox{\hsize}{!}{\includegraphics{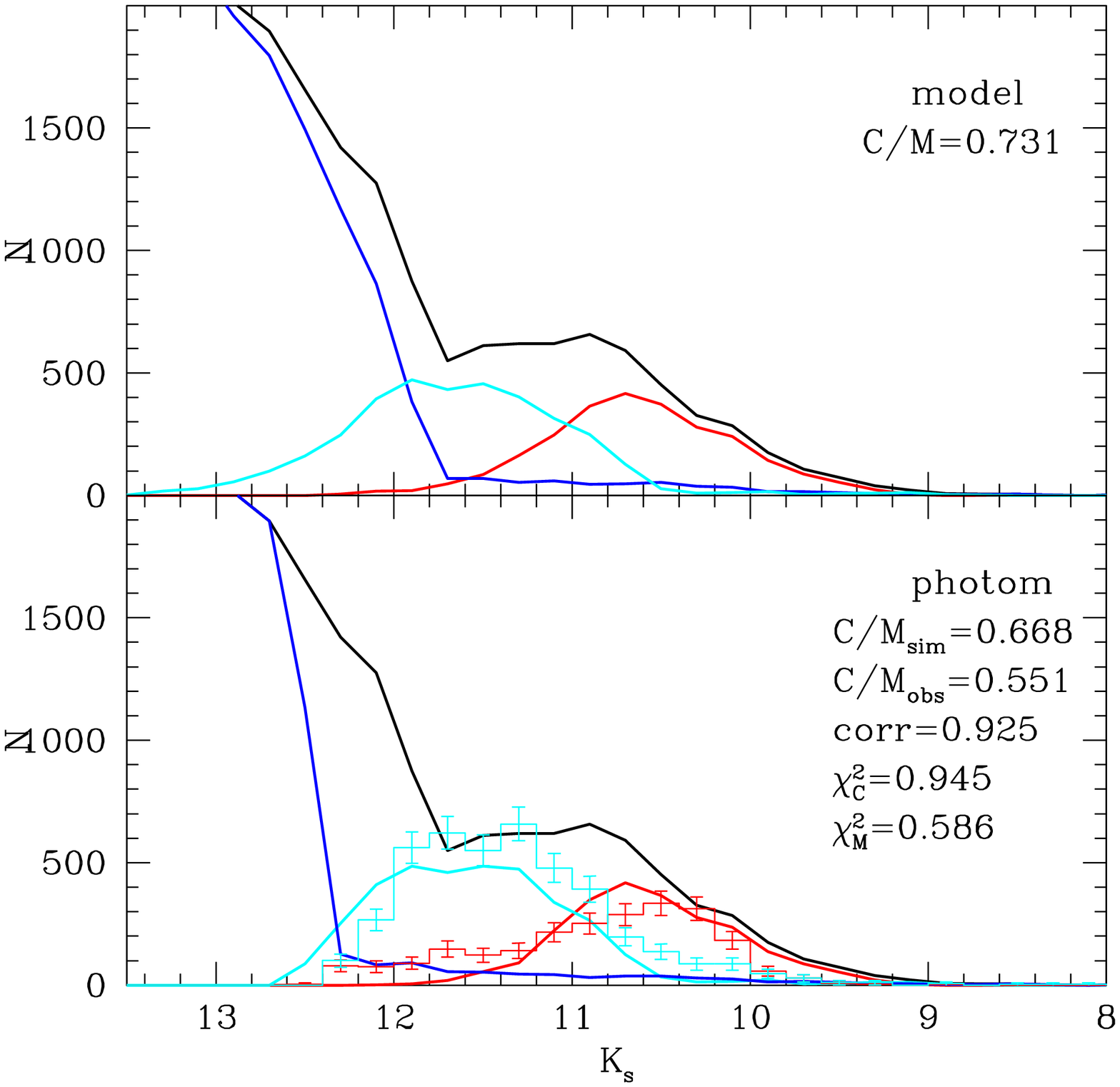}}
\end{minipage}
\caption{LFs for a realistic LMC model ({\bf left})
and for a model that fits rather well both the
distribution of C and M stars ({\bf right}) of sector E-NE of ring $0$. 
The continuous lines show model results: cyan =
O-rich TP-AGB, red = C-rich TP-AGB, blue = other stars, black =
total. Upper panel: using the modelling criteria to classify
TP-AGB stars; bottom panel: using the photometric criteria. In the
latter case, LFs are compared to LMC data (histograms).
The number counts in the simulations have been normalized to
provide the same C-star number as in the data. In each panel of
the bottom row the labels inside each figure give  the observed
and photometric C/M ratios as well as the correlation coefficient
between the theoretical and observed distribution of C stars. The
last two numbers indicate the probability that the observed
distribution is represented by the given model distribution;  this
is the same probability of obtaining the $\chi^2$ derived from the
comparison between the two distributions.} \label{fig_LFs}
\end{figure*}

We have defined clear criteria for extracting C- and M-type TP-AGB
stars from both data and theoretical simulations. Let us now
compare the model and data for the LMC in more detail. 
First of all we compare the observed and theoretical LFs for an 
inner section. We have chosen the ring 0 and sector E-NE 
(cf. Sect.~\ref{sec_surface}) because for this region 
the number of stars is large and crowding is not a
problem. The left panel of Fig.~\ref{fig_LFs} shows the LF 
results compared to our ``realistic''
LMC model discussed in Sect.~\ref{sec_realistic} using
the SFR and AMR as published. The number of simulated stars has
been normalized to give the same number of C-type stars as in the
observations.

In Fig.~\ref{fig_LFs} (left) the
photometric criteria for selecting C and M AGB stars, shown in the
bottom panel, provide LFs in good
agreement with the model distribution in the upper panel.
Differences between the two criteria appear only at the faintest
magnitude bins, as expected. Looking at the bottom panels, we can
notice the good description of the LF shapes for both M and C-type
stars. But the number of M-type AGB stars predicted by the model 
is just about half of the observed ones. 
Thus, this particular model strongly overpredicts the C/M
ratio of AGB stars.

We simulated the Galaxy foreground with the TRILEGAL code -- as
described in Girardi et al. (2004) -- and verified that the
contamination of the CMD in the M-star region by foreground
objects is very modest, and has no effect on the observed C/M
ratio. Thus, we conclude that our ``realistic'' LMC 
model has a problem. One possibility is that the TP-AGB
models by Marigo et al. in some way overestimate the number of C
stars; however this seems quite unlikely since their lifetimes in the
C-star phase agree with those derived from C-star counts in LMC
clusters (Girardi \& Marigo 2002). 
Alternatively, the fraction of M-type AGB stars may be underestimated,
a possibility that is presently under investigation.
Another possibility is that the SFR we use -- as derived
from other authors using completely independent data, and
regarding different regions of the LMC -- is not
appropriate for the region we are looking at.

Since we cannot, at this stage, clarify the origin of the problem,
we proceed in the following way: we fix the 
AMR of the LMC, and change the family of SFR given
by Eq.~\ref{eq_sfr}. The parameter $\alpha$ is let vary and
then for each simulation we compare the derived and observed LF,
separately for C and M stars. The comparison is based on a
$\chi^2$ test. For example in Fig. \ref{fig_LFs} $\chi^2_C$ and
$\chi^2_M$ indicate the probability that the observed distribution
is represented by the theoretical distribution (or the probability
of getting the $\chi^2$ value obtained comparing both
distributions). It is clear that the particular model shown in the 
left panel of Fig. \ref{fig_LFs} represents rather well the distribution of C
stars at the $96.1$\% level while the distribution of M stars is
far from the one predicted. For comparison, in the right panel we plot
the results for the SFR that best fits the LFs for both C- and M-type 
stars ($\alpha=1000$); in this case, the model well represents the
distribution of C stars at the $94.5$\% level and the one of M stars at the 
$58.6$\% level. 

\begin{figure*}
\epsfxsize=0.195\hsize \epsfbox{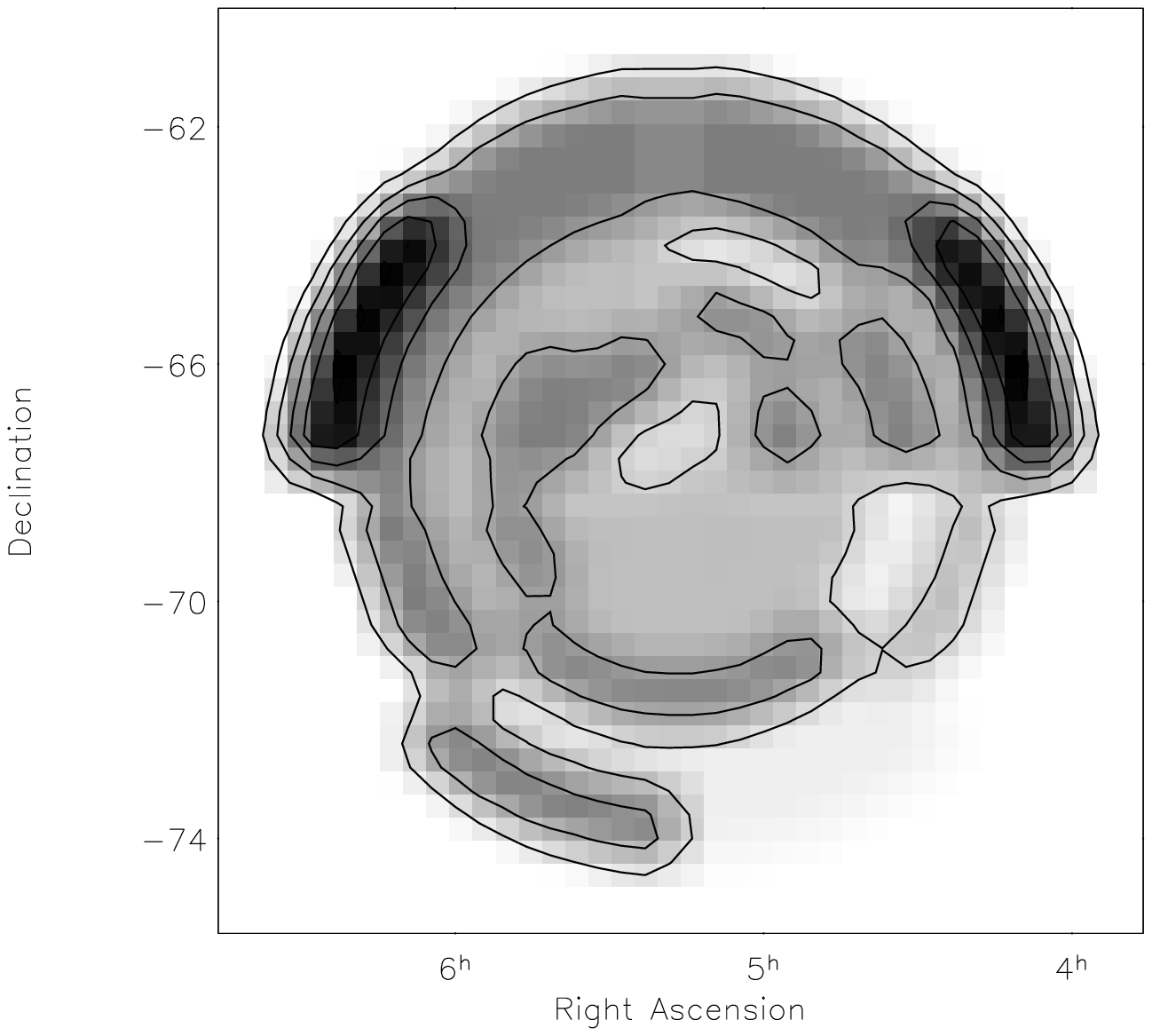}
\epsfxsize=0.195\hsize \epsfbox{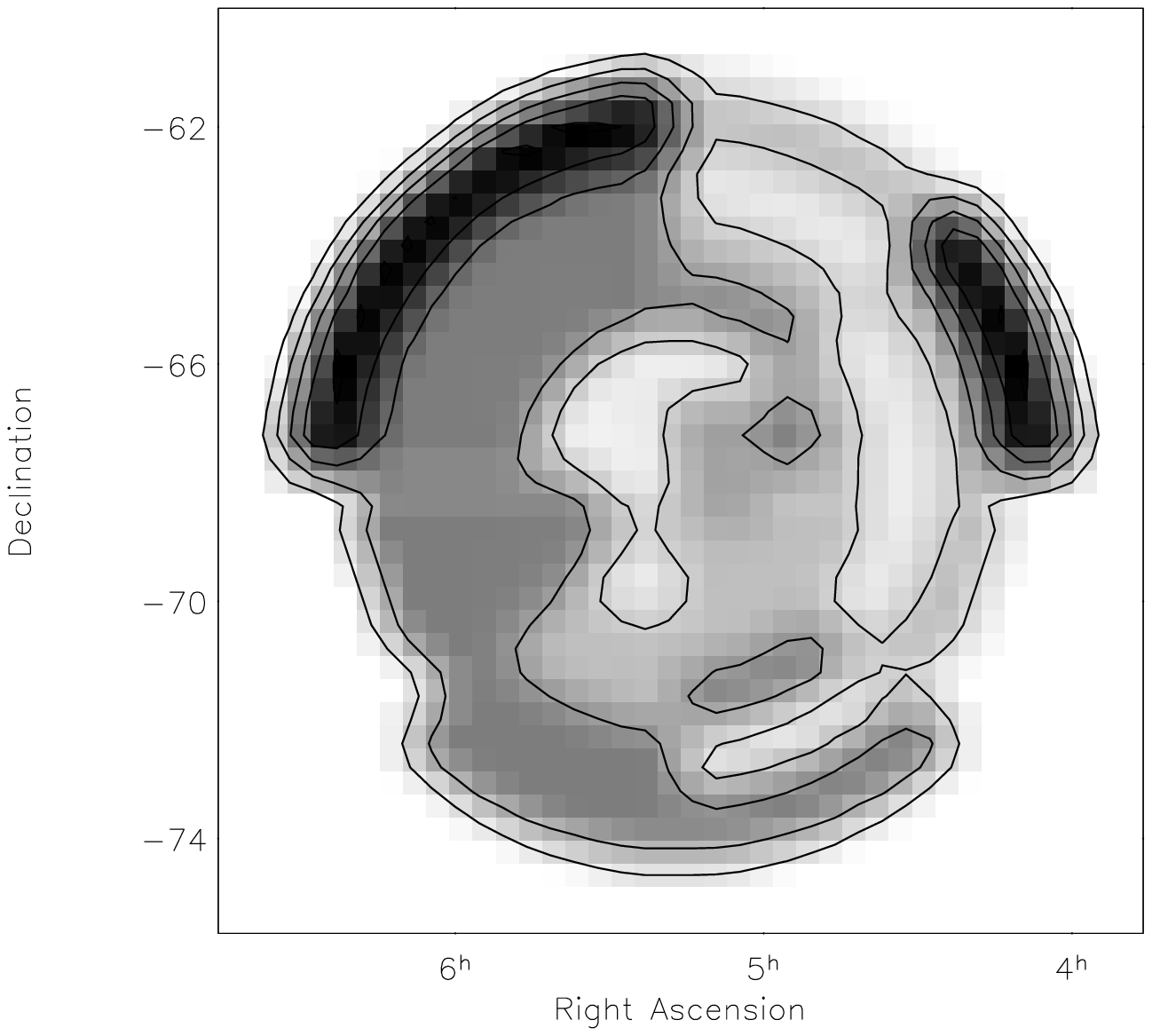}
\epsfxsize=0.195\hsize \epsfbox{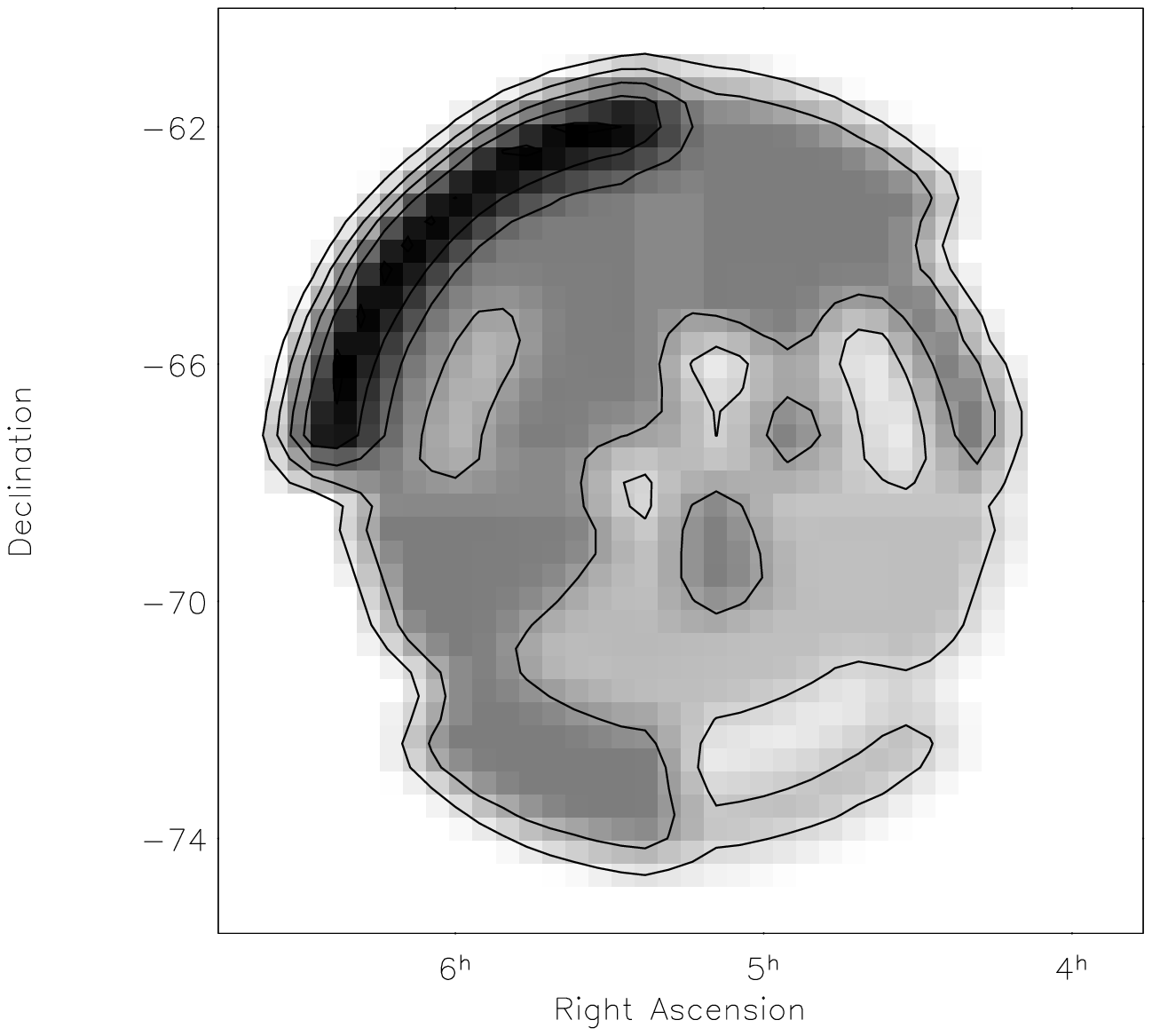}
\epsfxsize=0.195\hsize \epsfbox{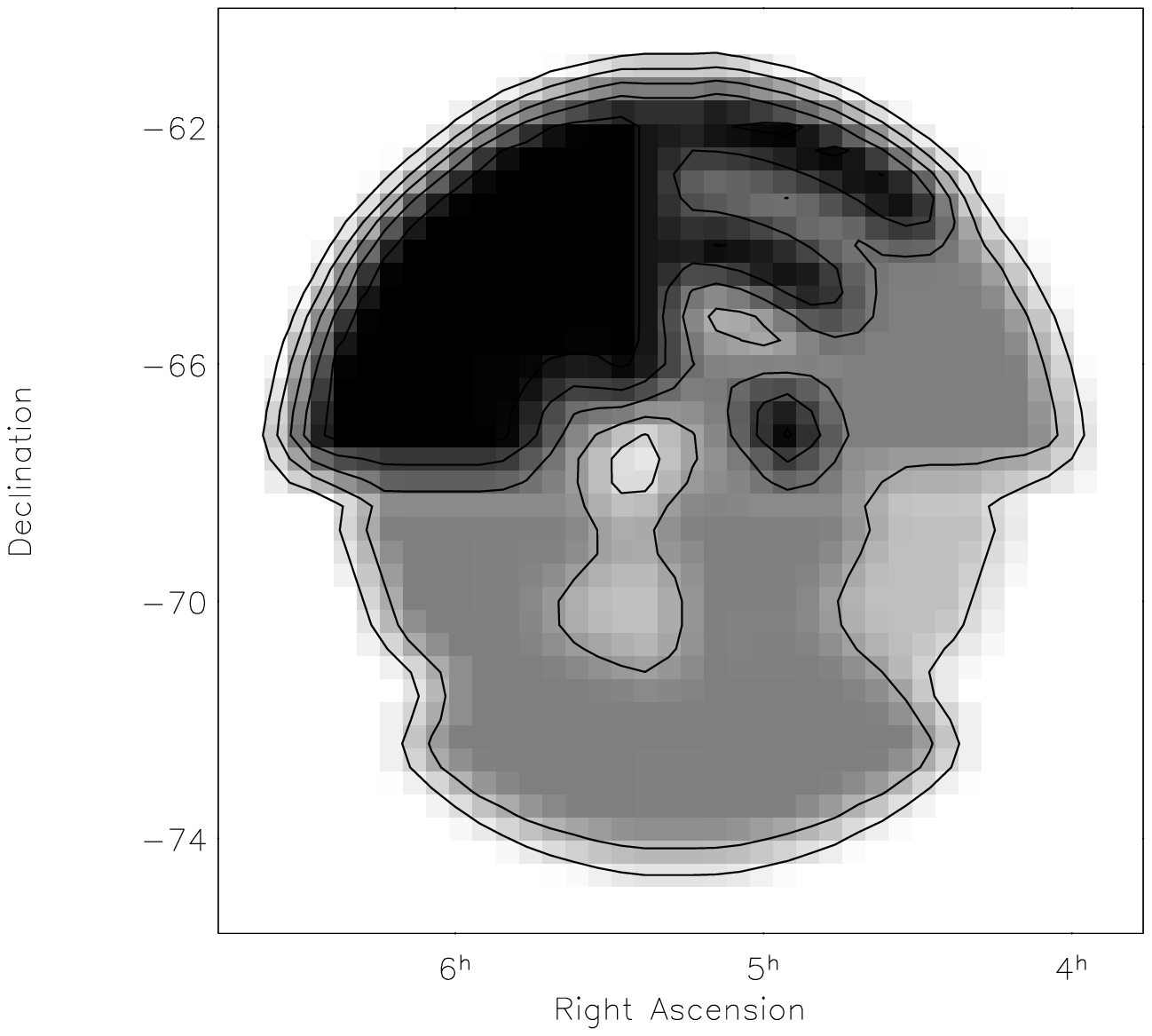}
\epsfxsize=0.195\hsize \epsfbox{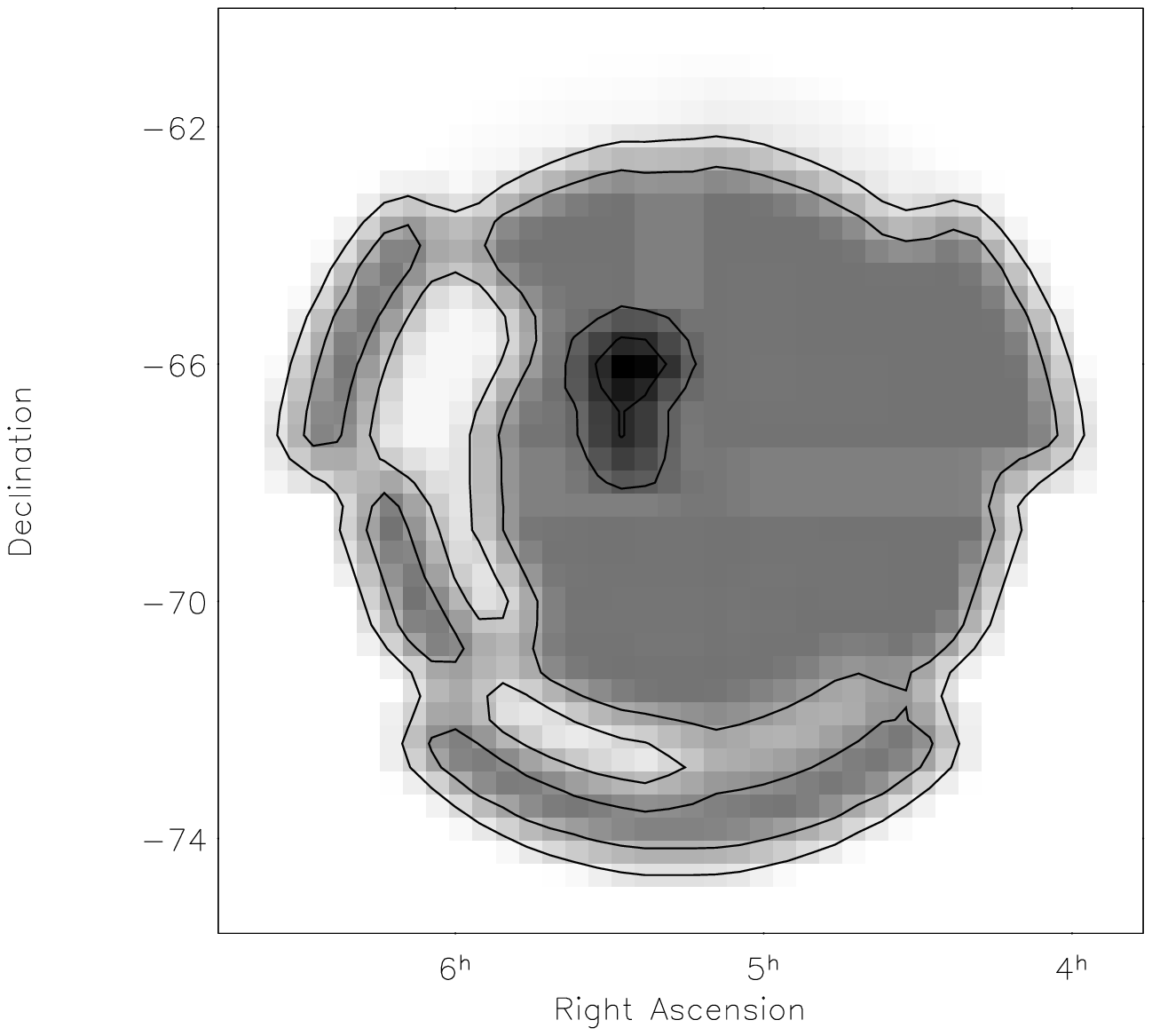}

\epsfxsize=0.195\hsize \epsfbox{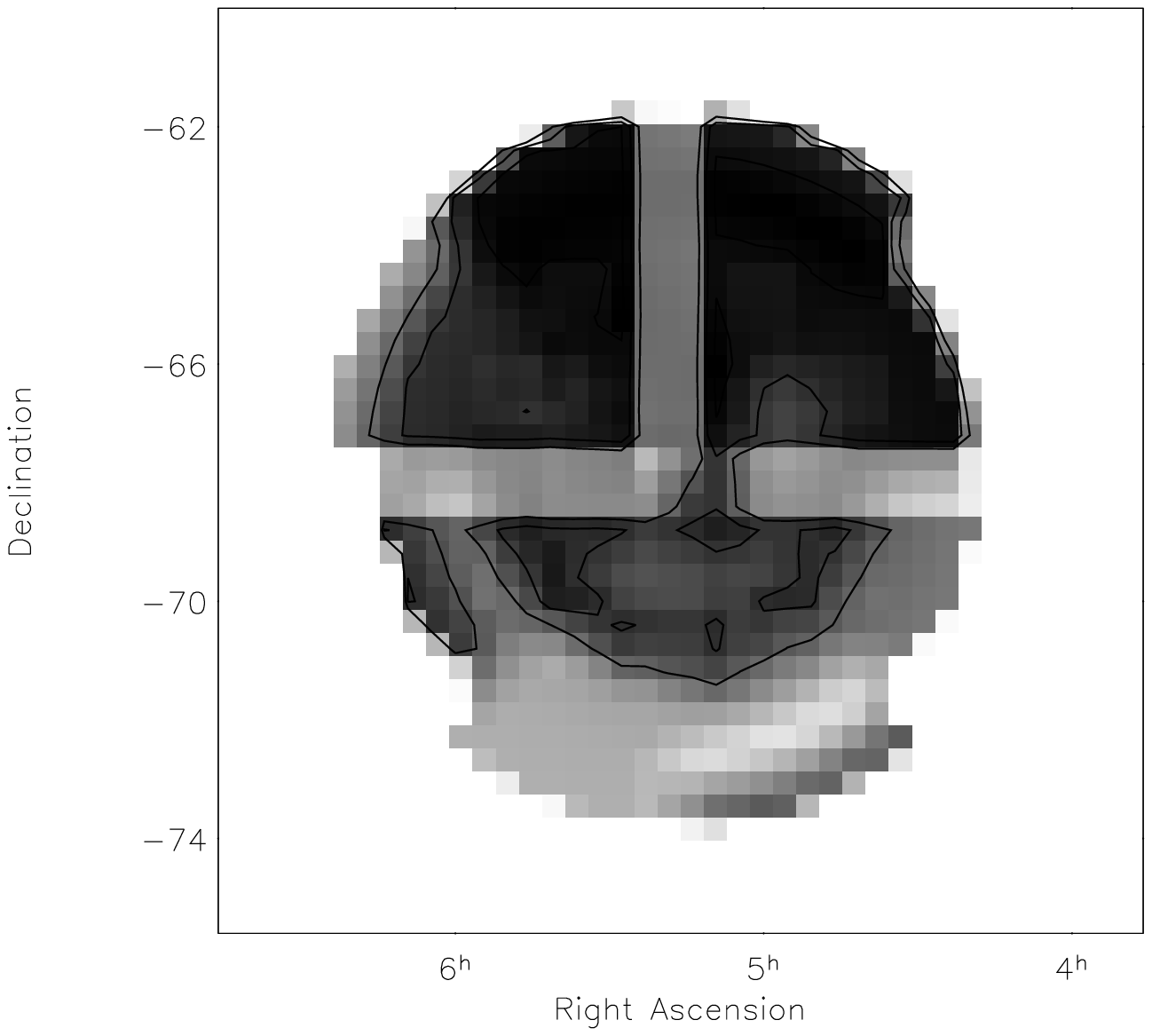}
\epsfxsize=0.195\hsize \epsfbox{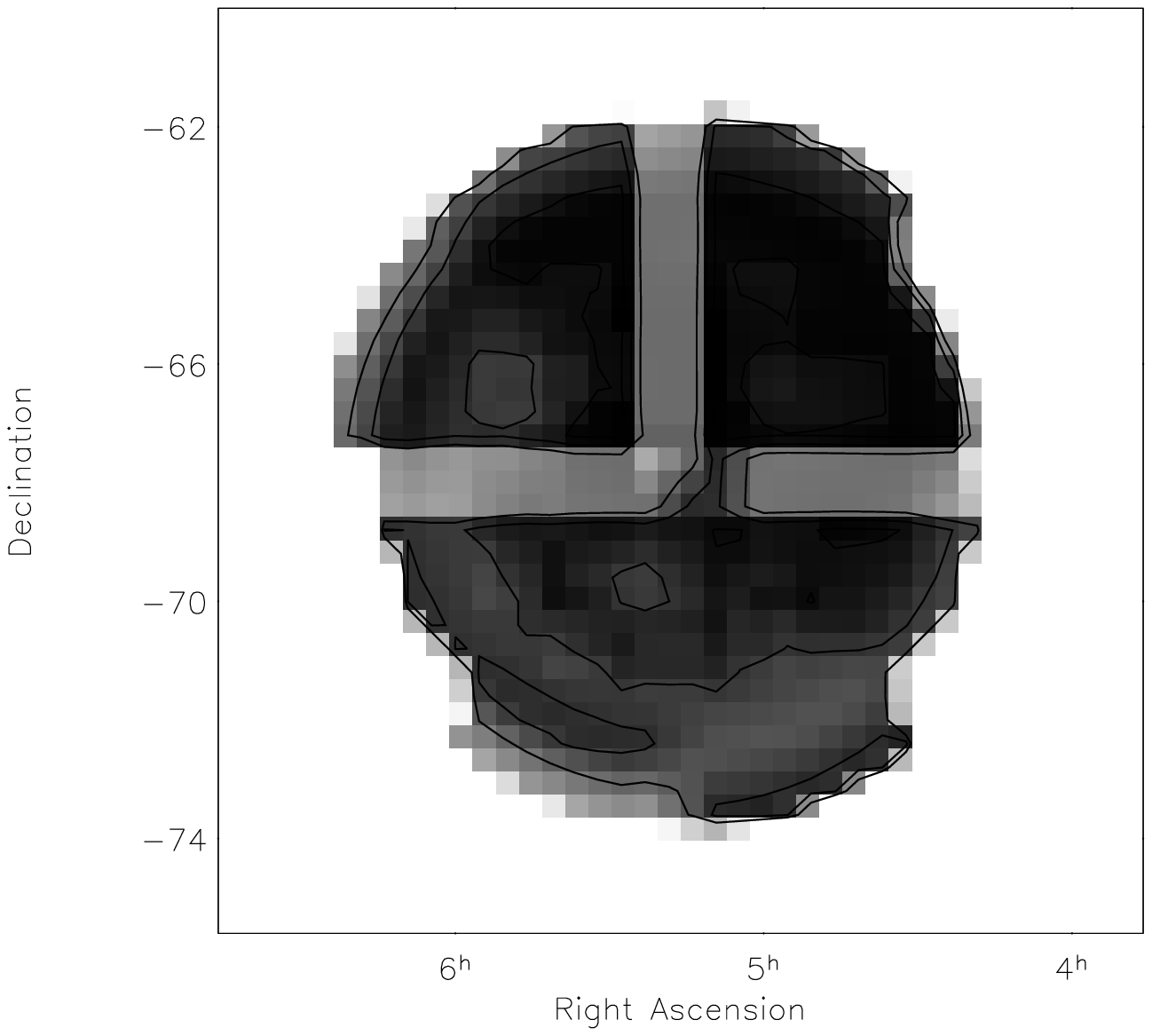}
\epsfxsize=0.195\hsize \epsfbox{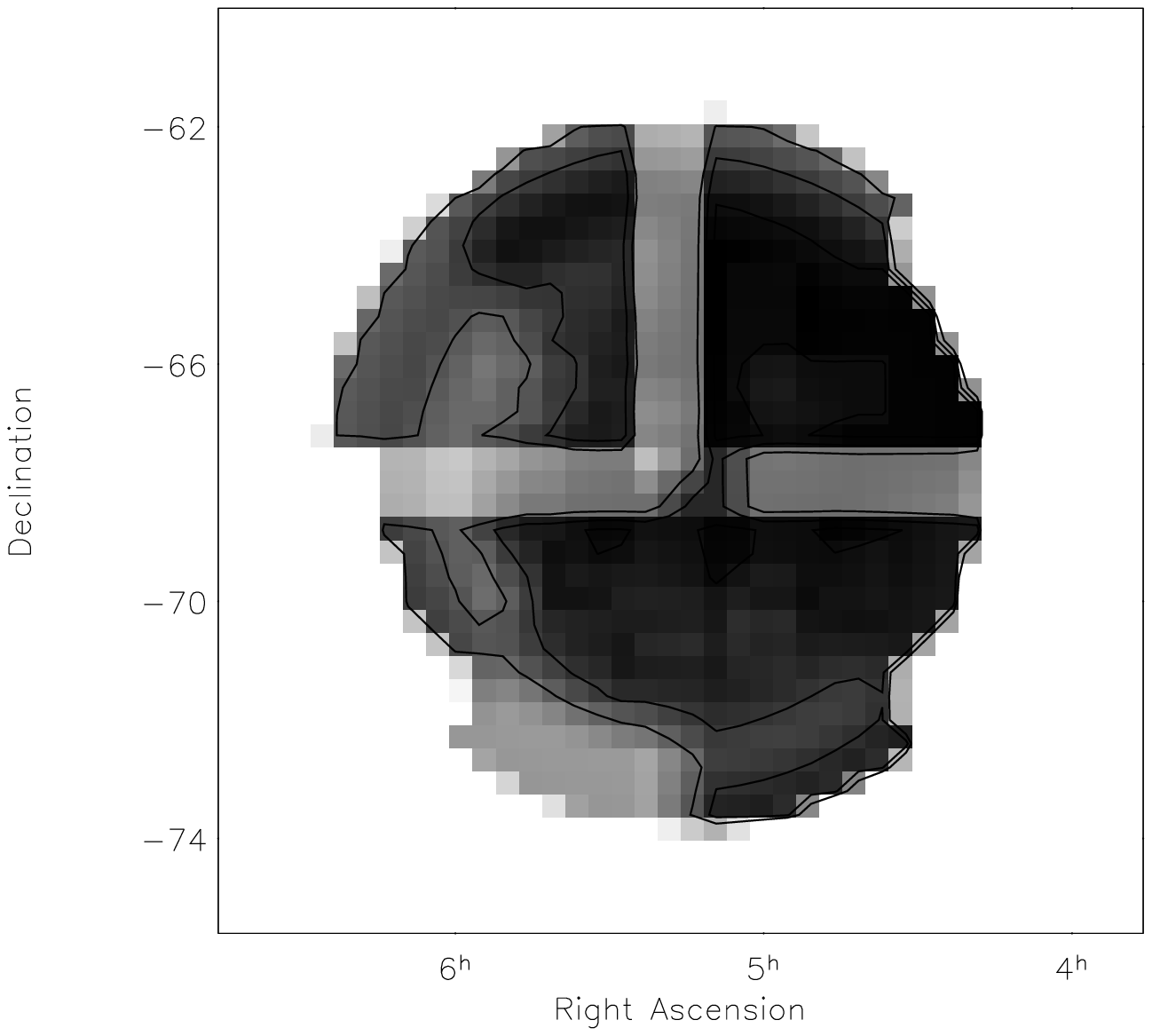}
\epsfxsize=0.195\hsize \epsfbox{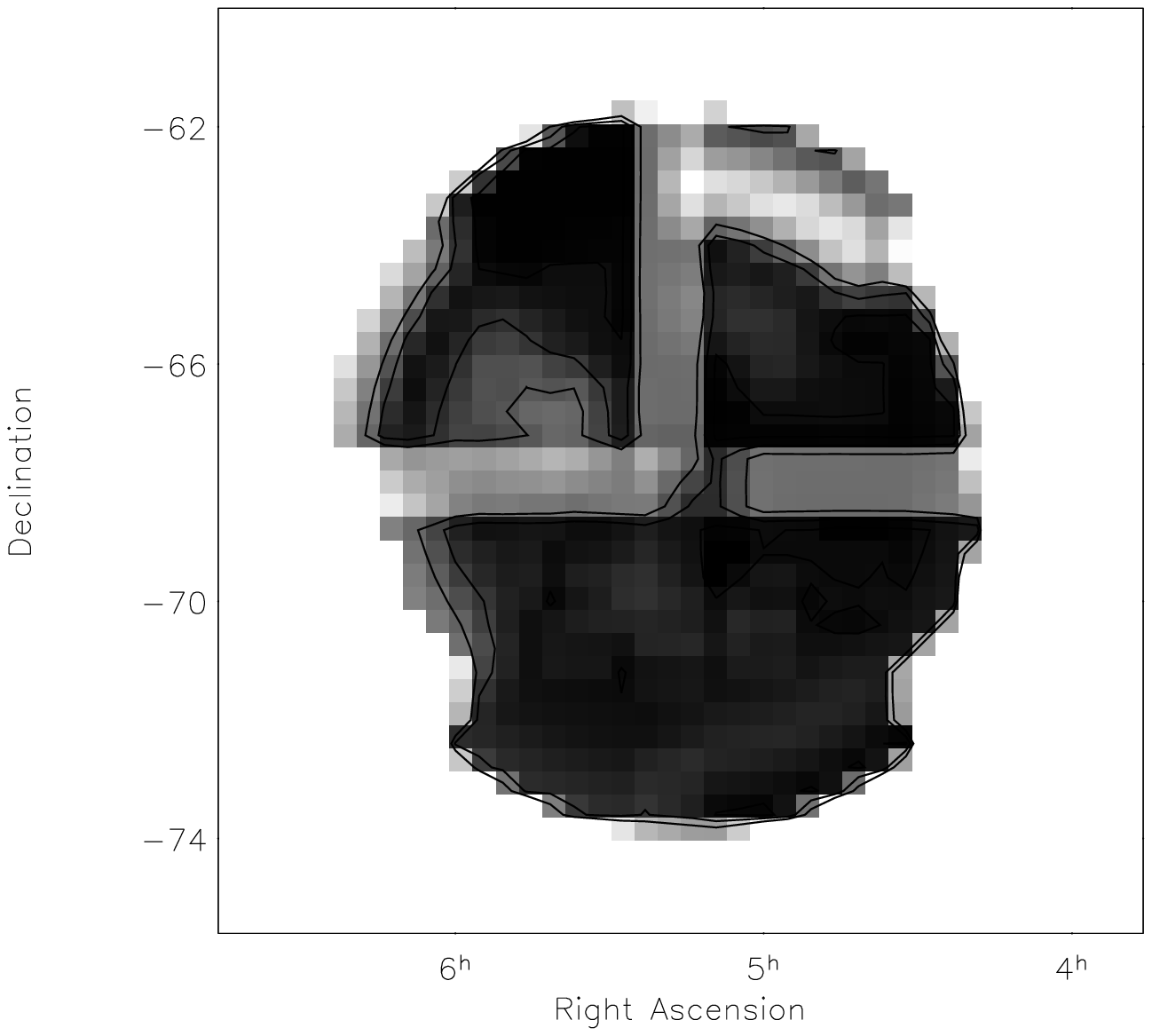}
\epsfxsize=0.195\hsize \epsfbox{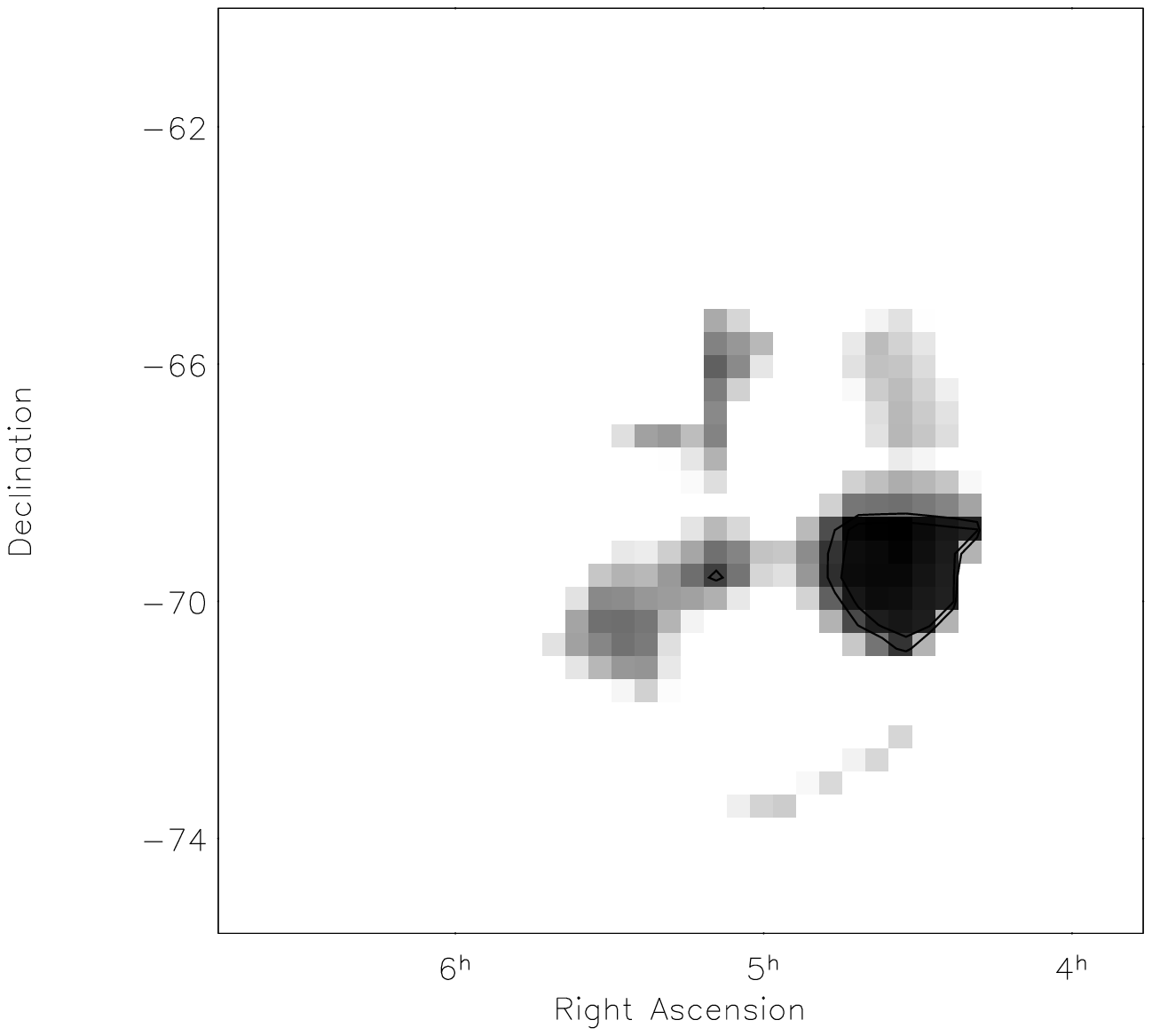}

\epsfxsize=0.195\hsize \epsfbox{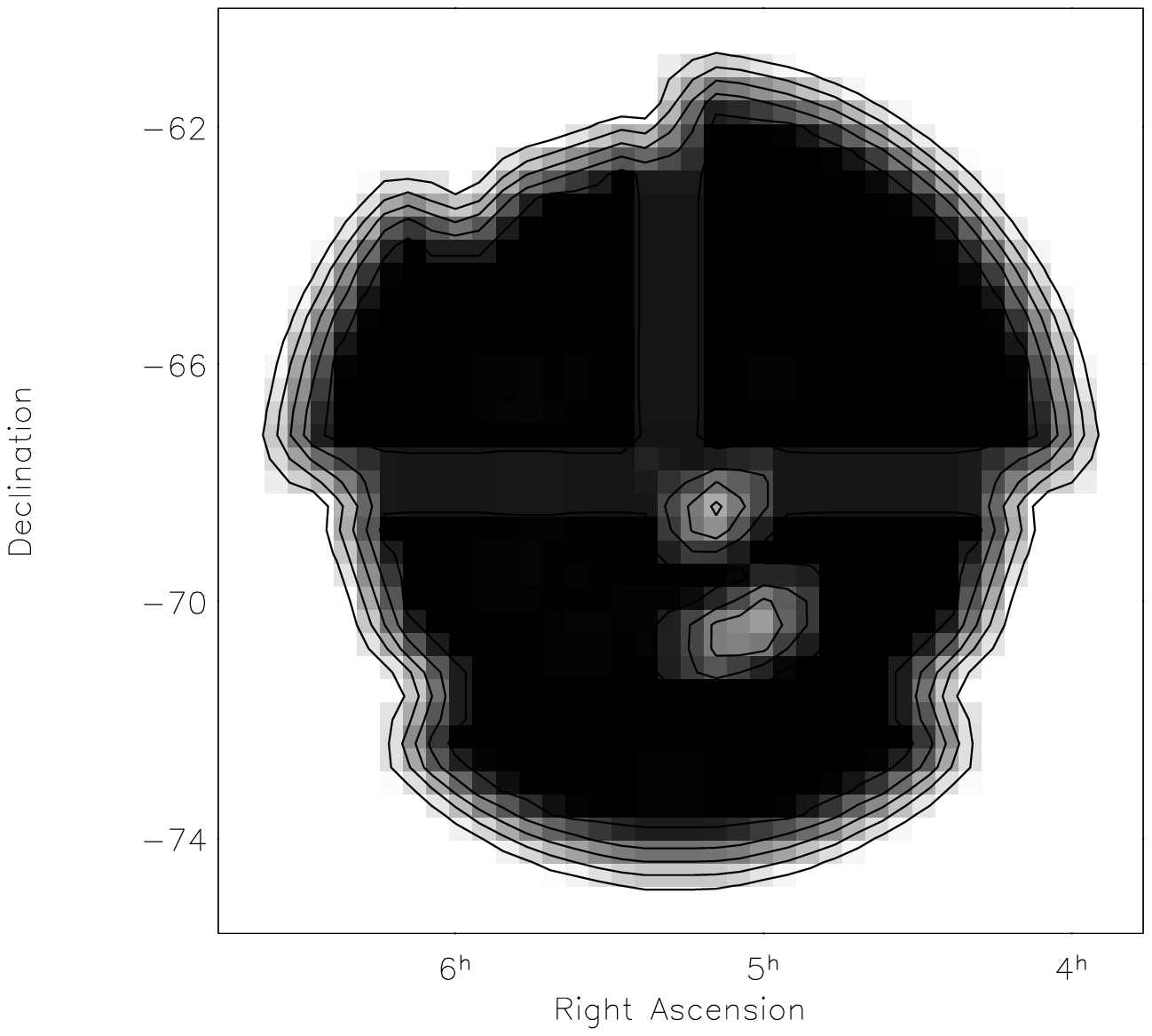}
\epsfxsize=0.195\hsize \epsfbox{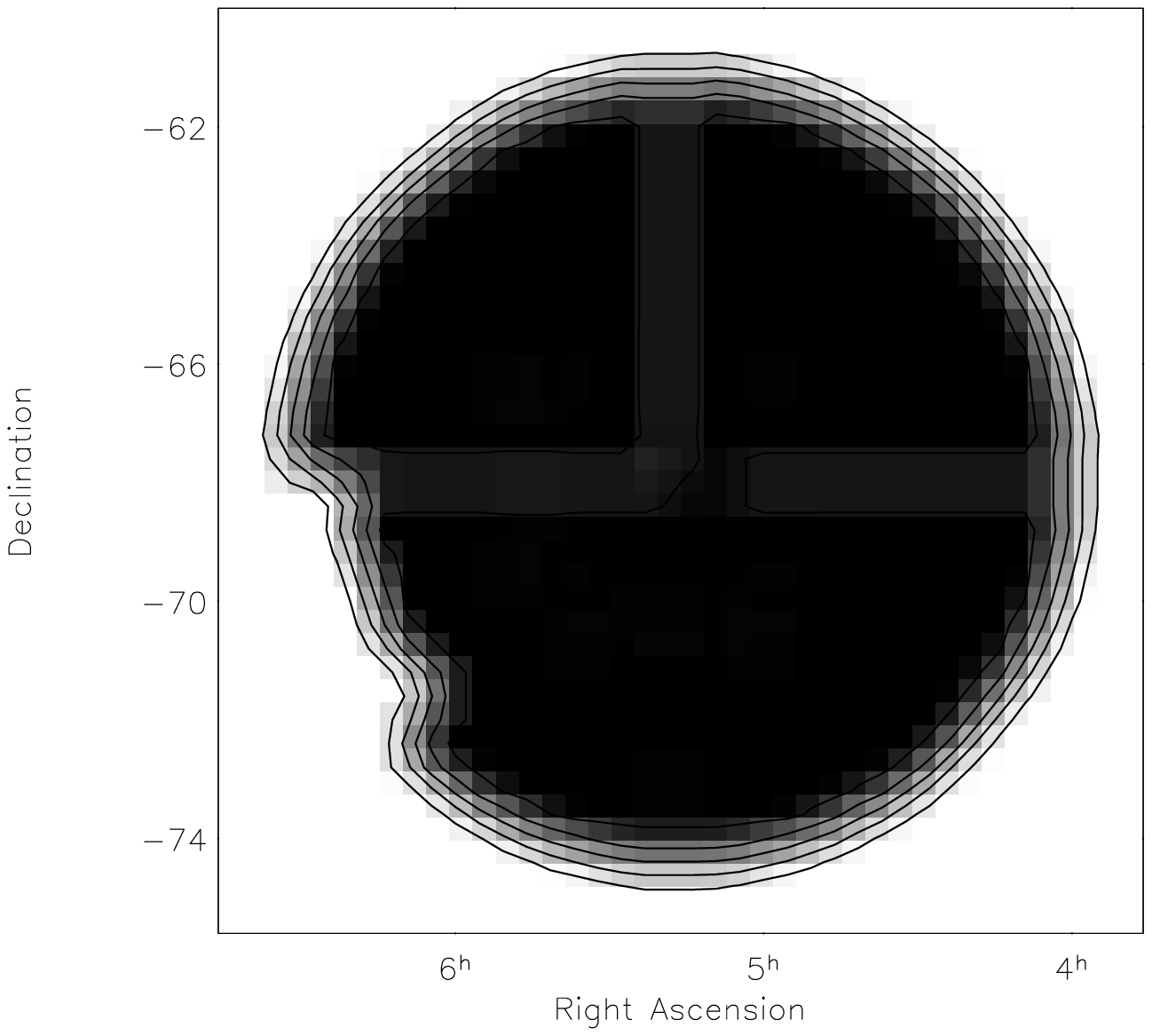}
\epsfxsize=0.195\hsize \epsfbox{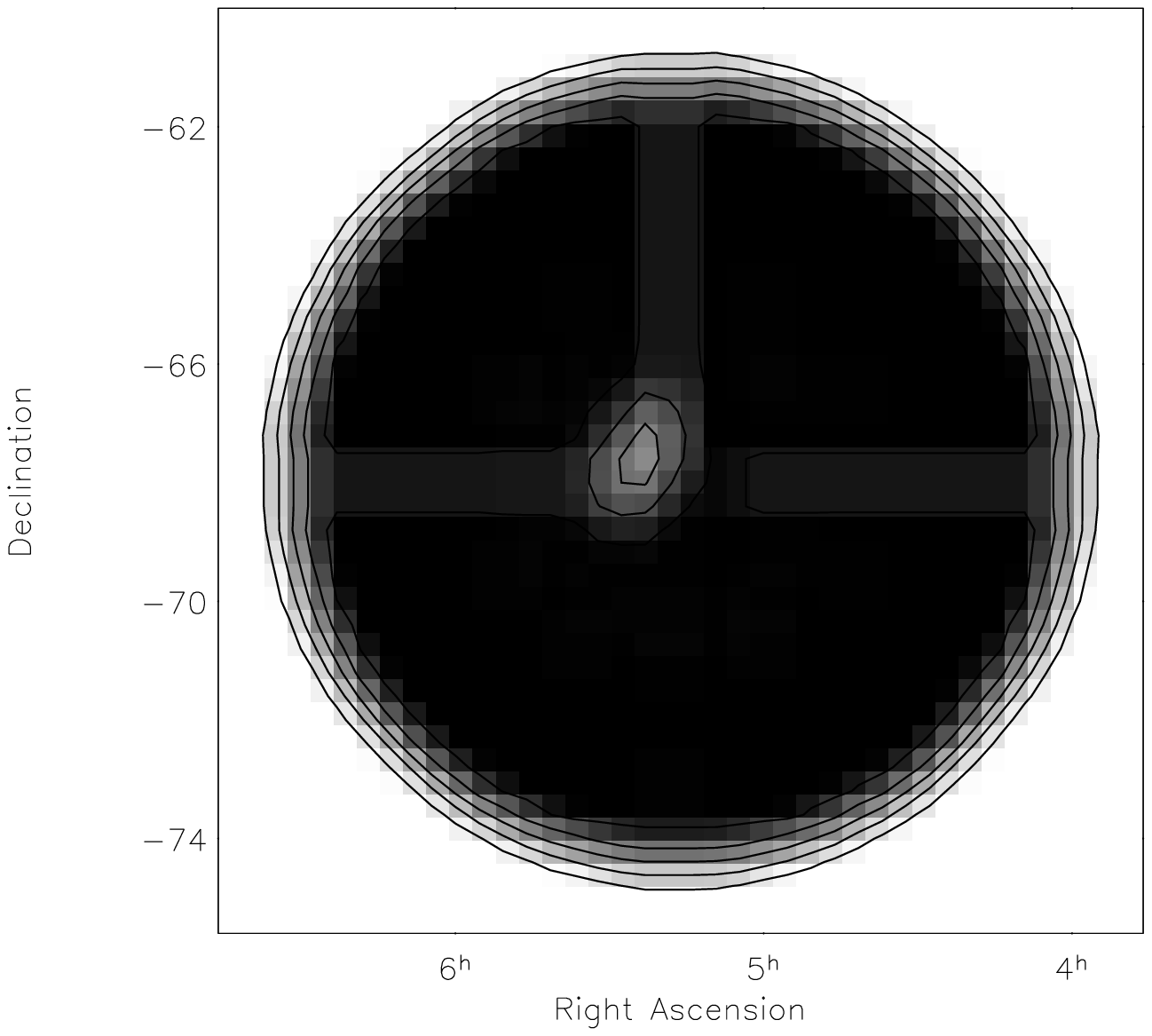}
\epsfxsize=0.195\hsize \epsfbox{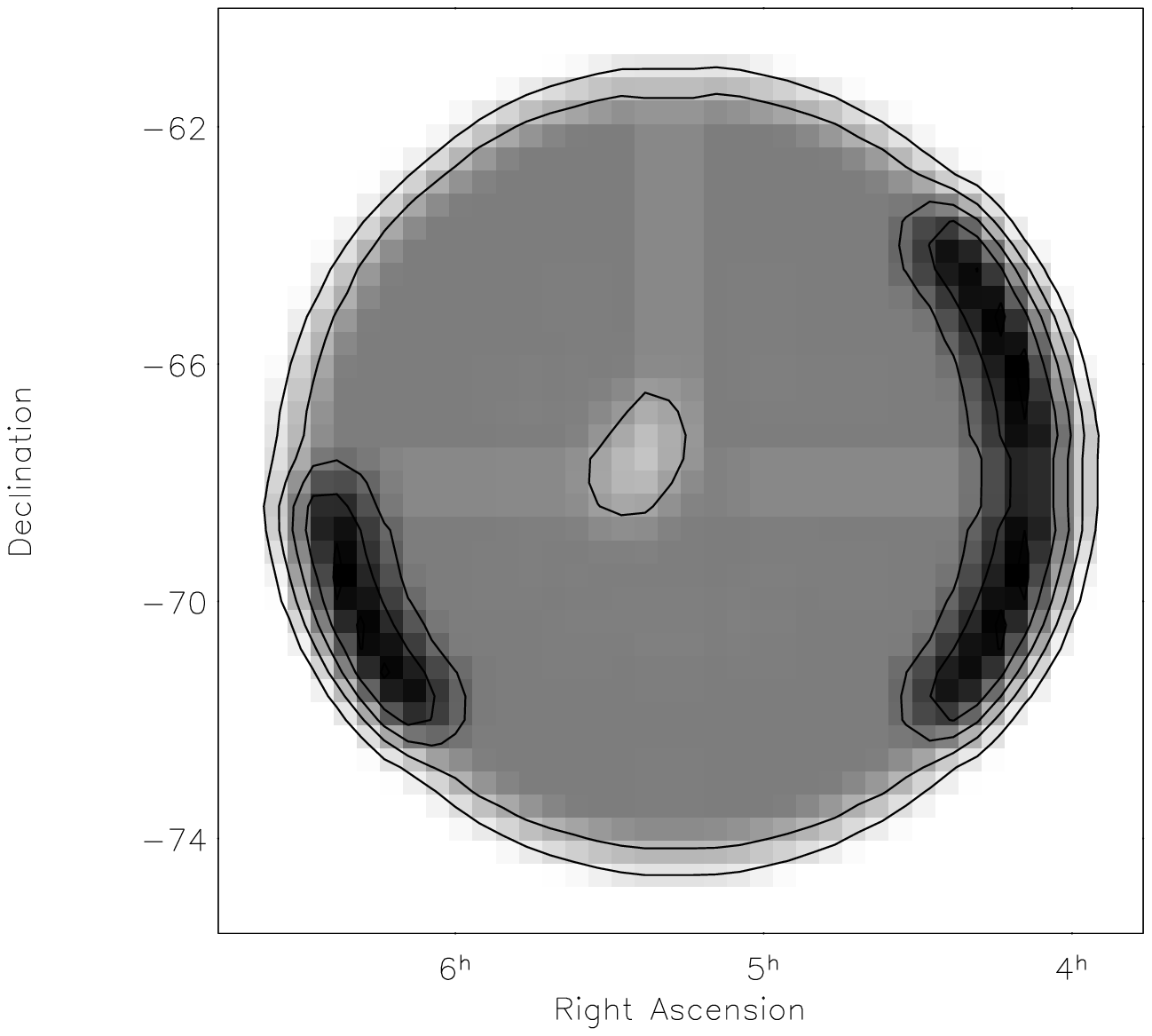}
\epsfxsize=0.195\hsize \epsfbox{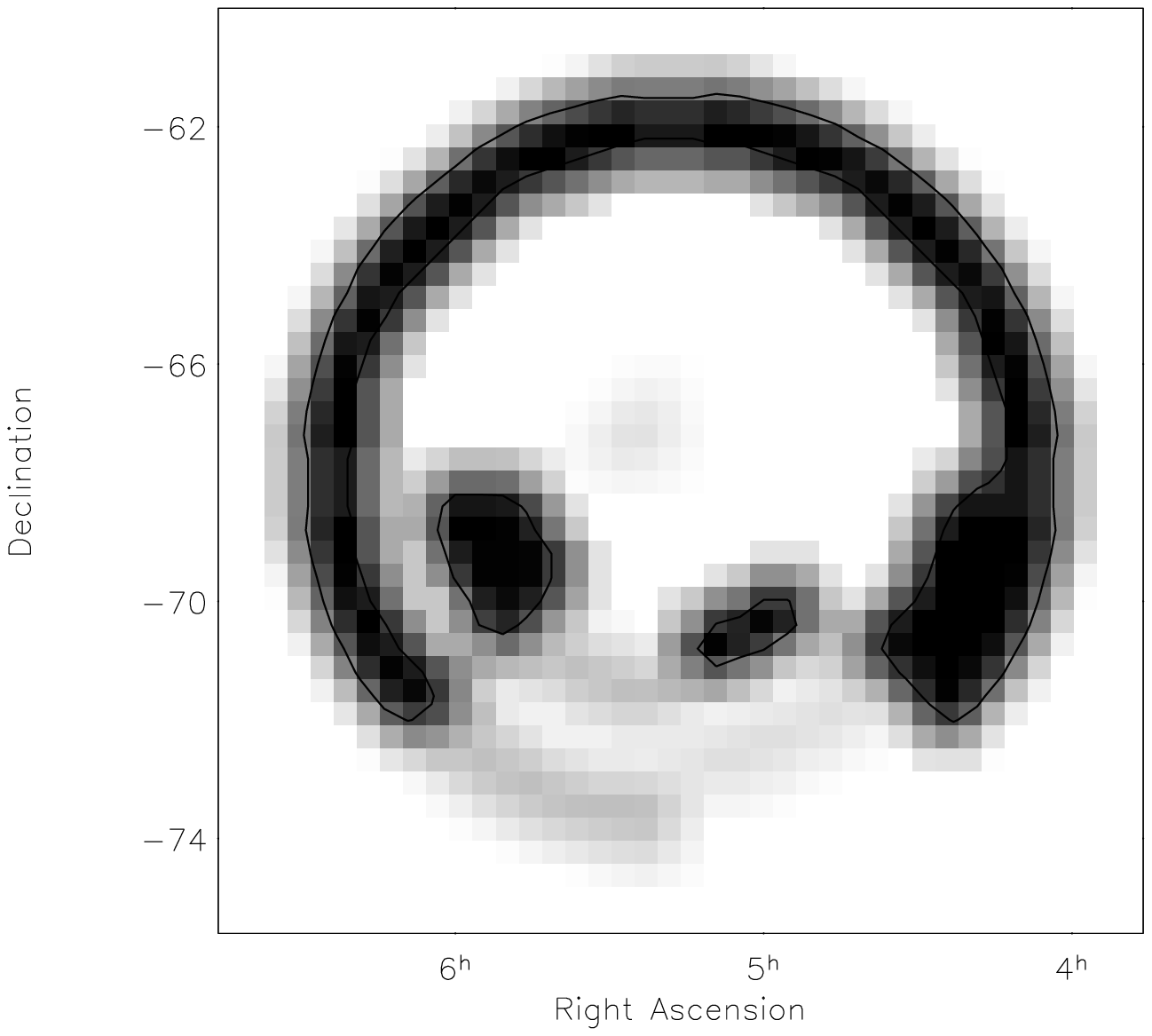}

\epsfxsize=0.195\hsize \epsfbox{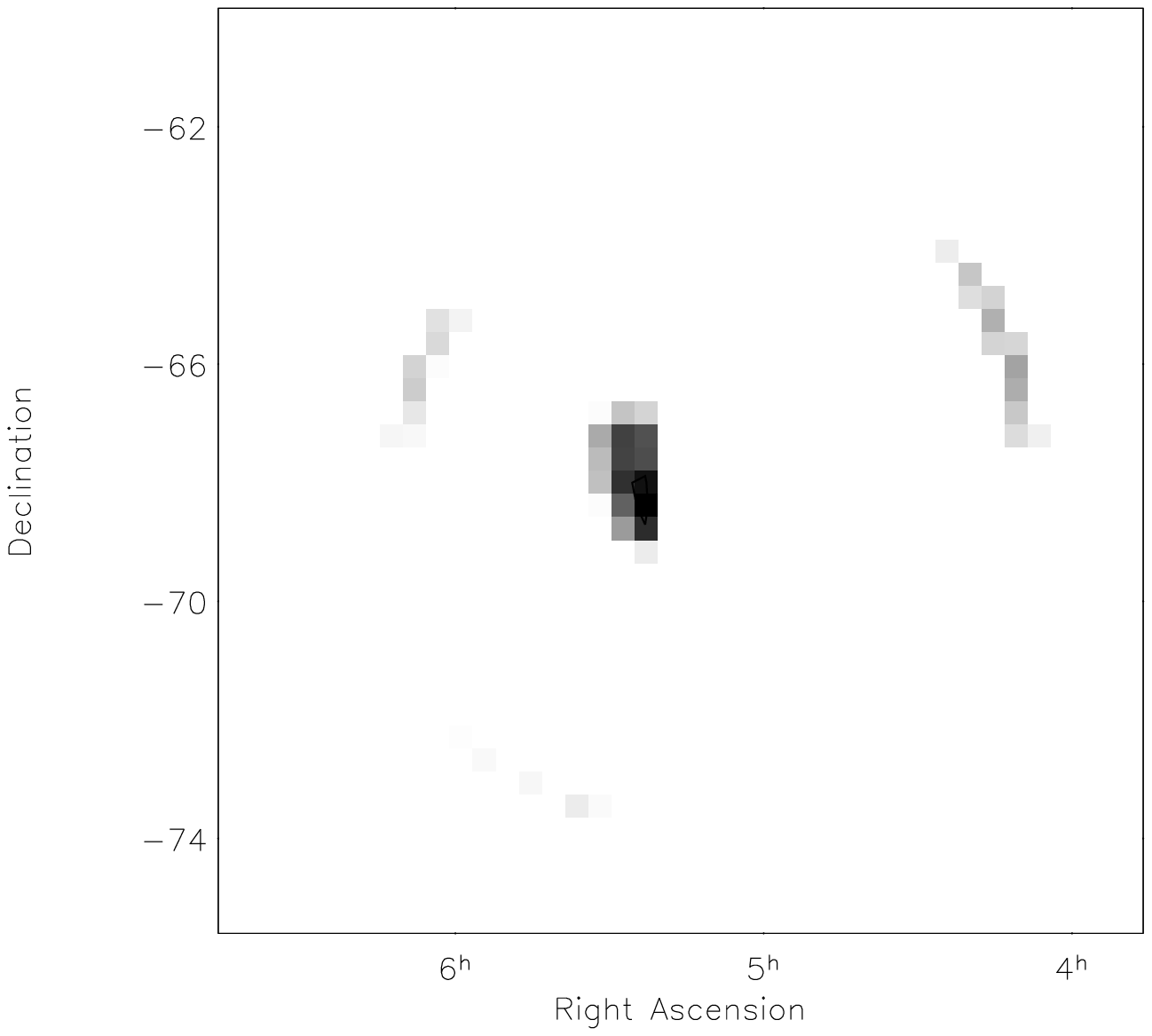}
\epsfxsize=0.195\hsize \epsfbox{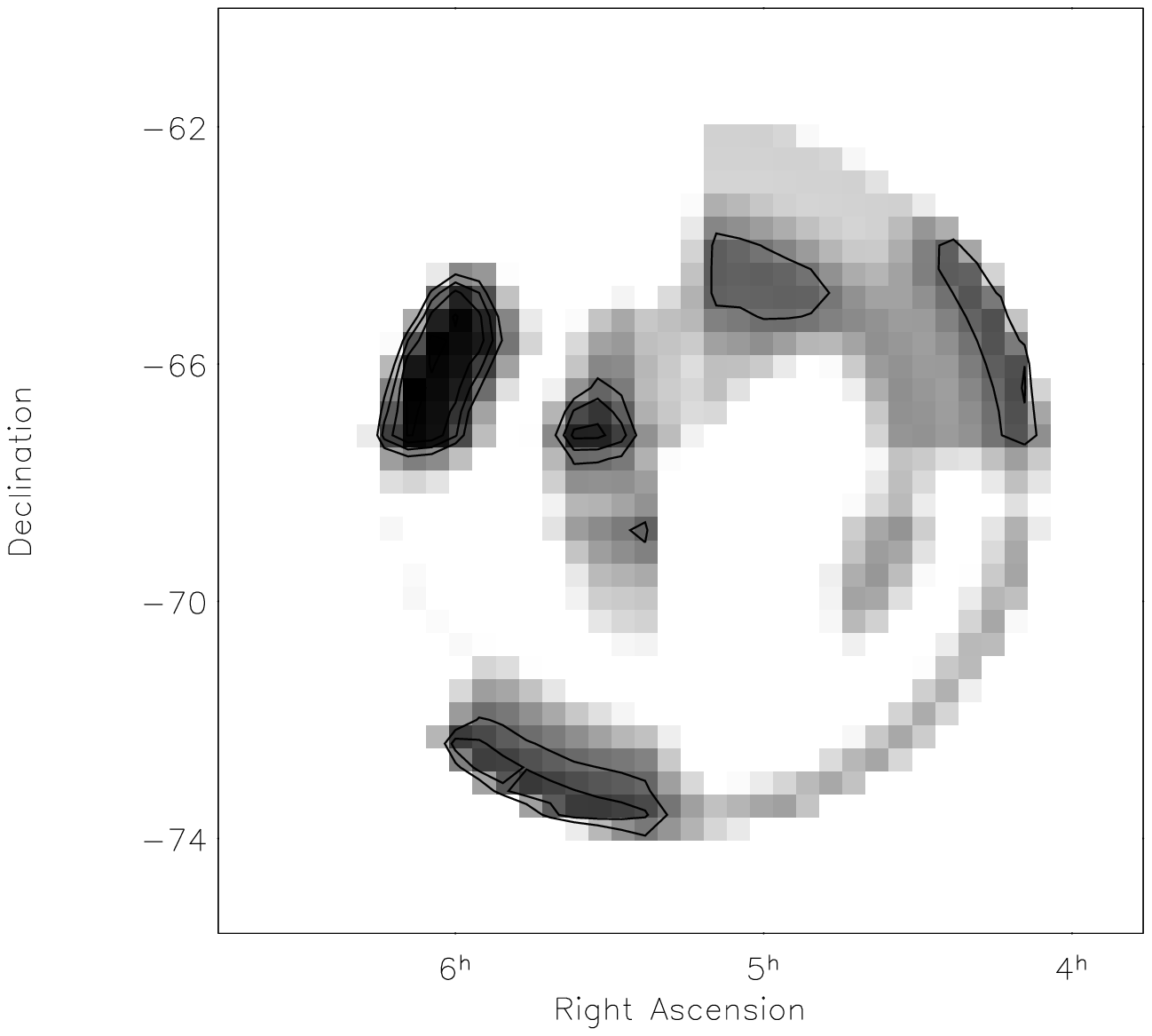}
\epsfxsize=0.195\hsize \epsfbox{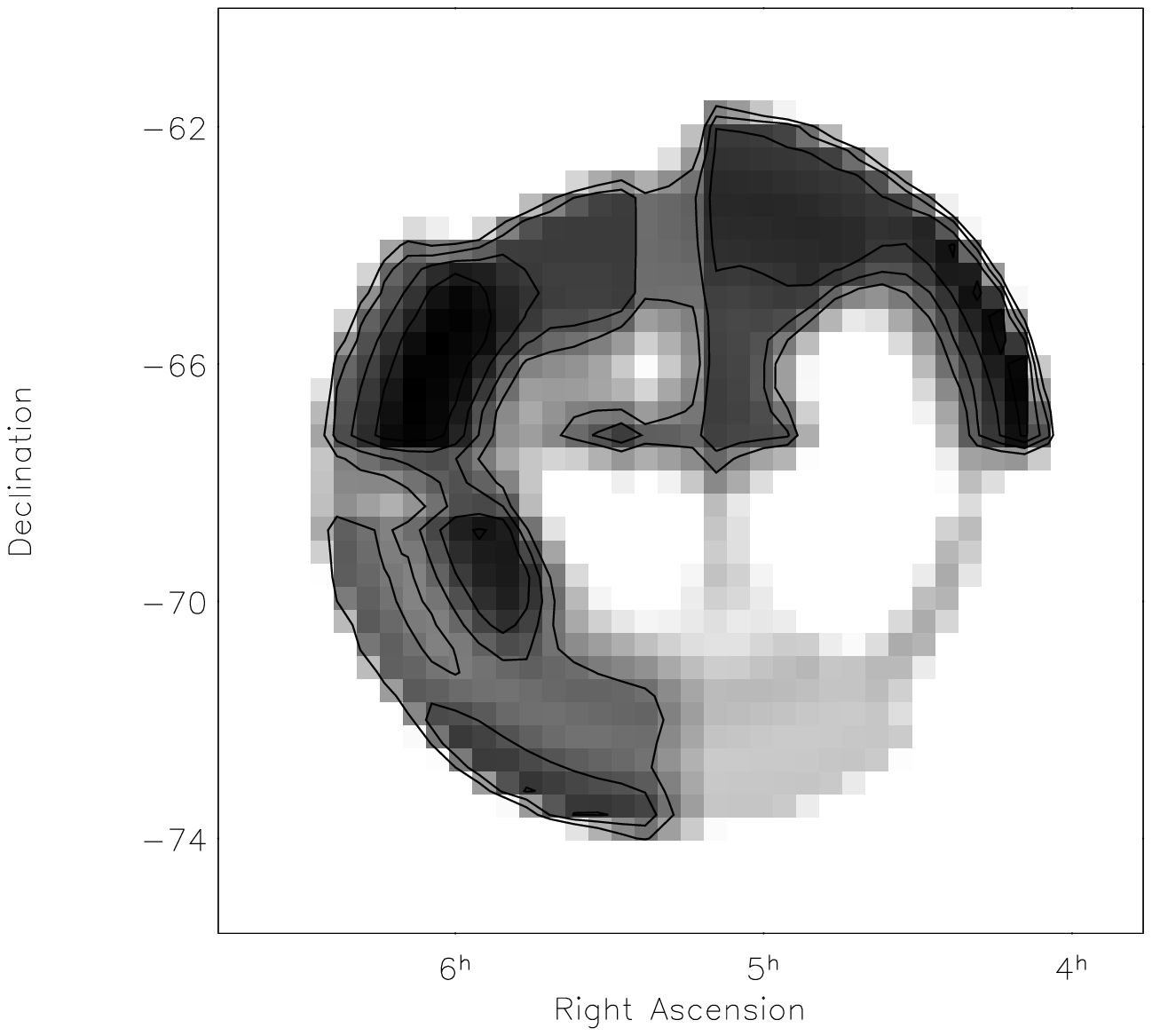}
\epsfxsize=0.195\hsize \epsfbox{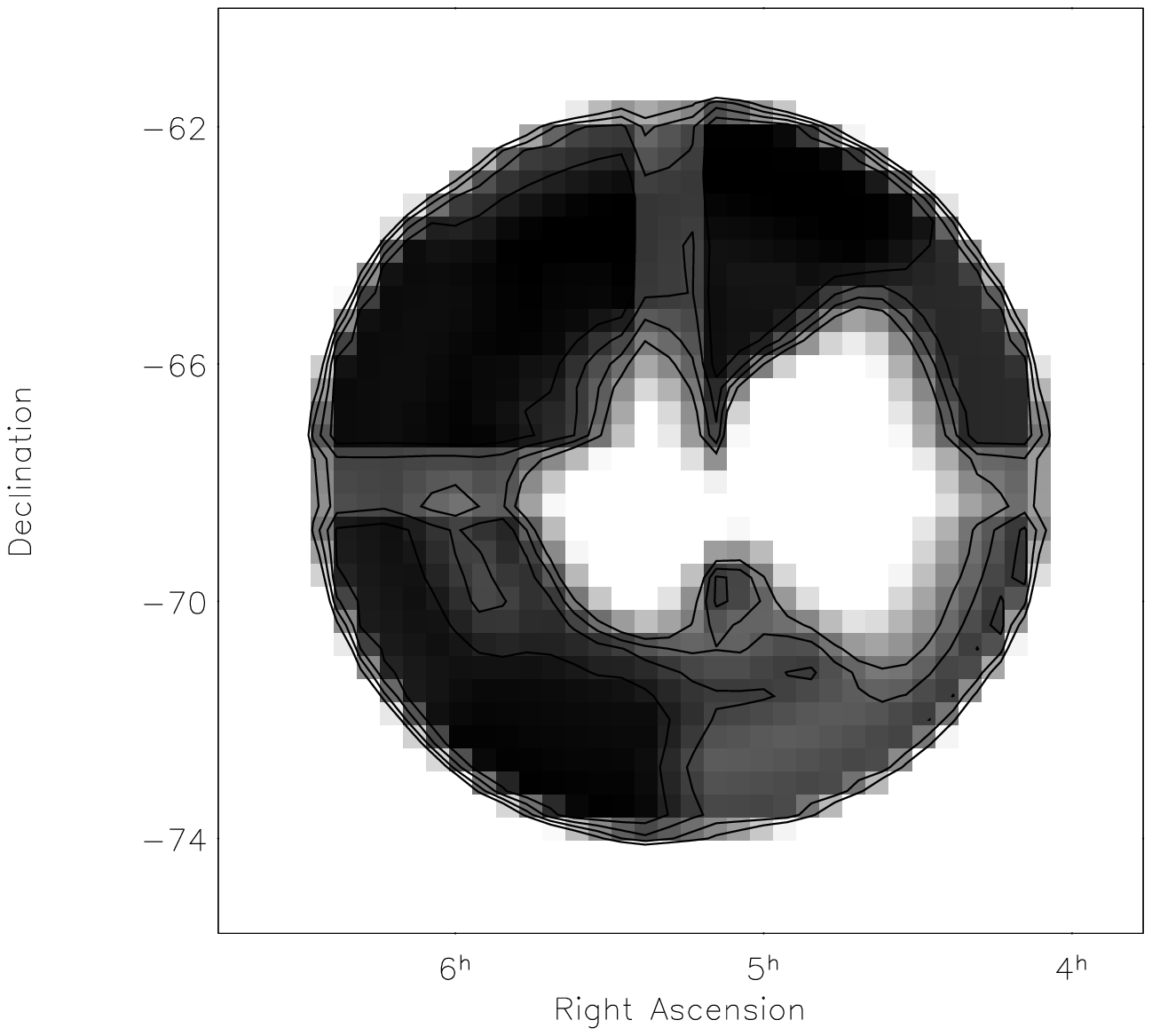}
\epsfxsize=0.195\hsize \epsfbox{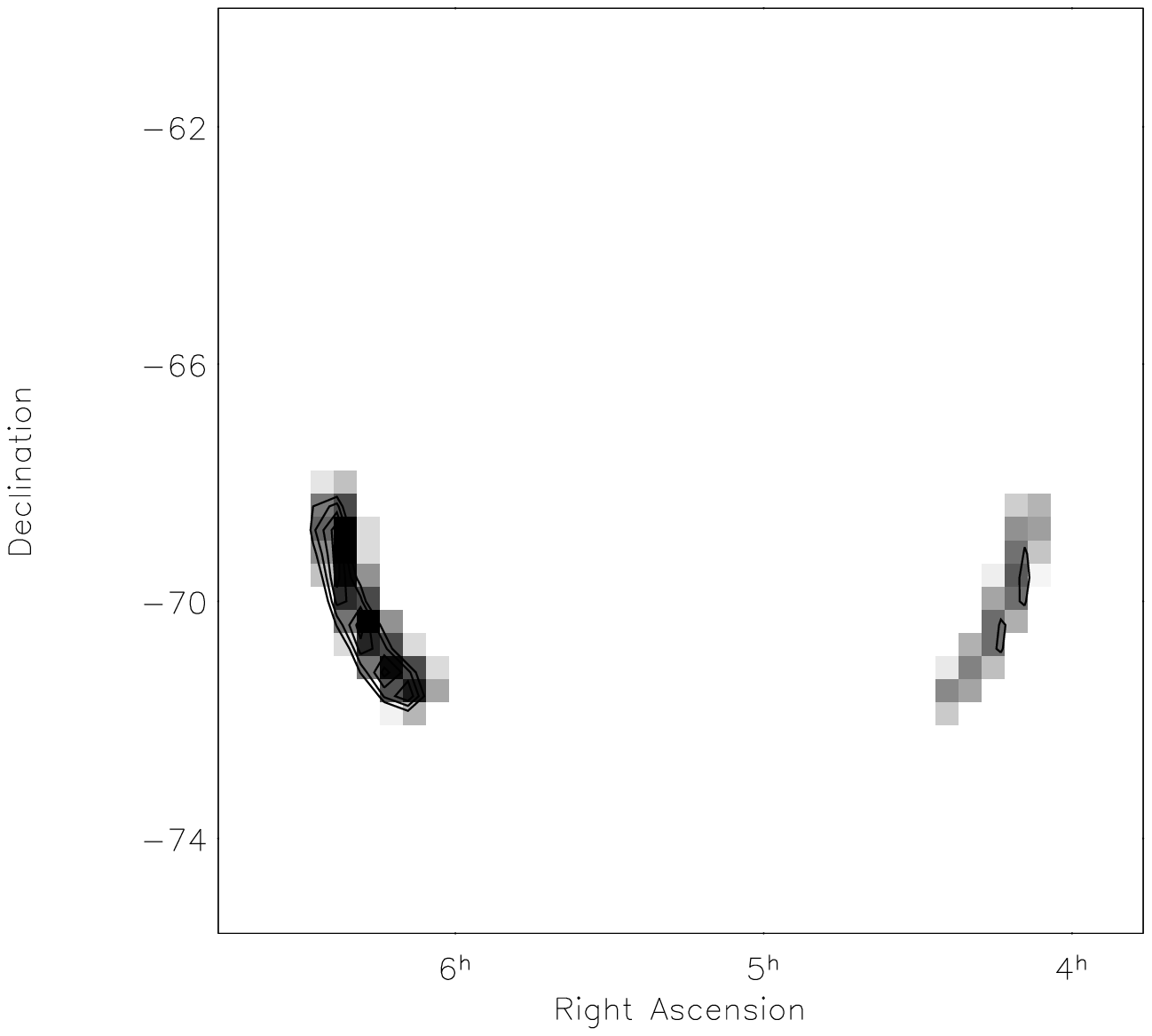}
\caption{Distribution of the metallicity that corresponds to the
maximum probability for a given SFR for C stars ({\bf first row}) and M stars
({\bf third row}). From left to right the SFR corresponds to a
population with a mean age of $2$, $3.9$, $6.3$, $8.7$ and $10.6$ Gyr. 
Metallicity contours are at Z $= 0.003-0.015$ with a step of $0.003$.
Darker regions correspond to higher values. The
probability distributions associated with each SFR are shown in the 
{\bf second row} for C stars (the grey scale shows only values above
$0.8$ and contours are at $0.93$, $0.96$, and $0.99$) and in the 
{\bf fourth row} for M stars (the grey scale shows only values above
$0.6$ and contours are at $0.8$, $0.85$, $0.9$, $0.95$).}
\label{lumall}
\end{figure*}

For each sector in each ring we have calculated the probability as
above for each combination of SFR and metallicity $Z$.  
The $Z$ that gives the best fit is then interpreted
as the best model for the stellar population in that particular
sector.  We adopt a constant AMR, i.e. the same $Z$ for all ages, 
which is an acceptable approximation since the metallicities that count
most in defining the model LFs are those at young ages 
(say less than $3$ Gyr) for which little galactic chemical evolution 
is expected. Note also that we select C and M stars
with the help of lines in the diagrams, lines that depend on the
chosen metallicity $Z$. 

Since our best fitting model results from a series of approximations
-- both in the TP-AGB model, and in the assumption of exponential
SFRs -- the best fit we get for a single LMC region may not be
 the best possible solution. Therefore, we should not 
overinterpret the results in terms of the derived metallicities
and mean ages. Much more important than the absolute values of these
parameter, are their {\em variations} across the LMC.
Our next goal is to map these variations.

In Fig. \ref{lumall} maps of metallicity 
across the LMC area are shown separately for C and M stars for each SFR. 
Contour values directly correspond to metallicity expressed as values
of $Z$. The distribution of the maximum probability is also shown  
and indicates how significant a metallicity map is. 

In order to build these maps, 
because we dispose of numbers (probability, metallicity, SFR) for
a given sector (this is necessary in order to obtain a statistical
significant LF), we first created a grid of $25921$ points in the
plane ($-8:8$, $-8:8$) equally spaced with a step of $0.1$. Second
we assigned to each point the quantity in which we are interested (for
example the most probable metallicity) accordingly to which sector
a point belongs. Third we re-binned the distribution of values
in bins equal to $0.4$ (this corresponds to a resolution of $0.16$
deg$^2$) and applied a boxcar smoothing of width$=2$. Finally, we
constructed greyscale maps where darker regions correspond to higher
numbers. 

\subsection{The most probable metallicity and SFR distributions}
\begin{figure*}
\epsfxsize=0.32\hsize \epsfbox{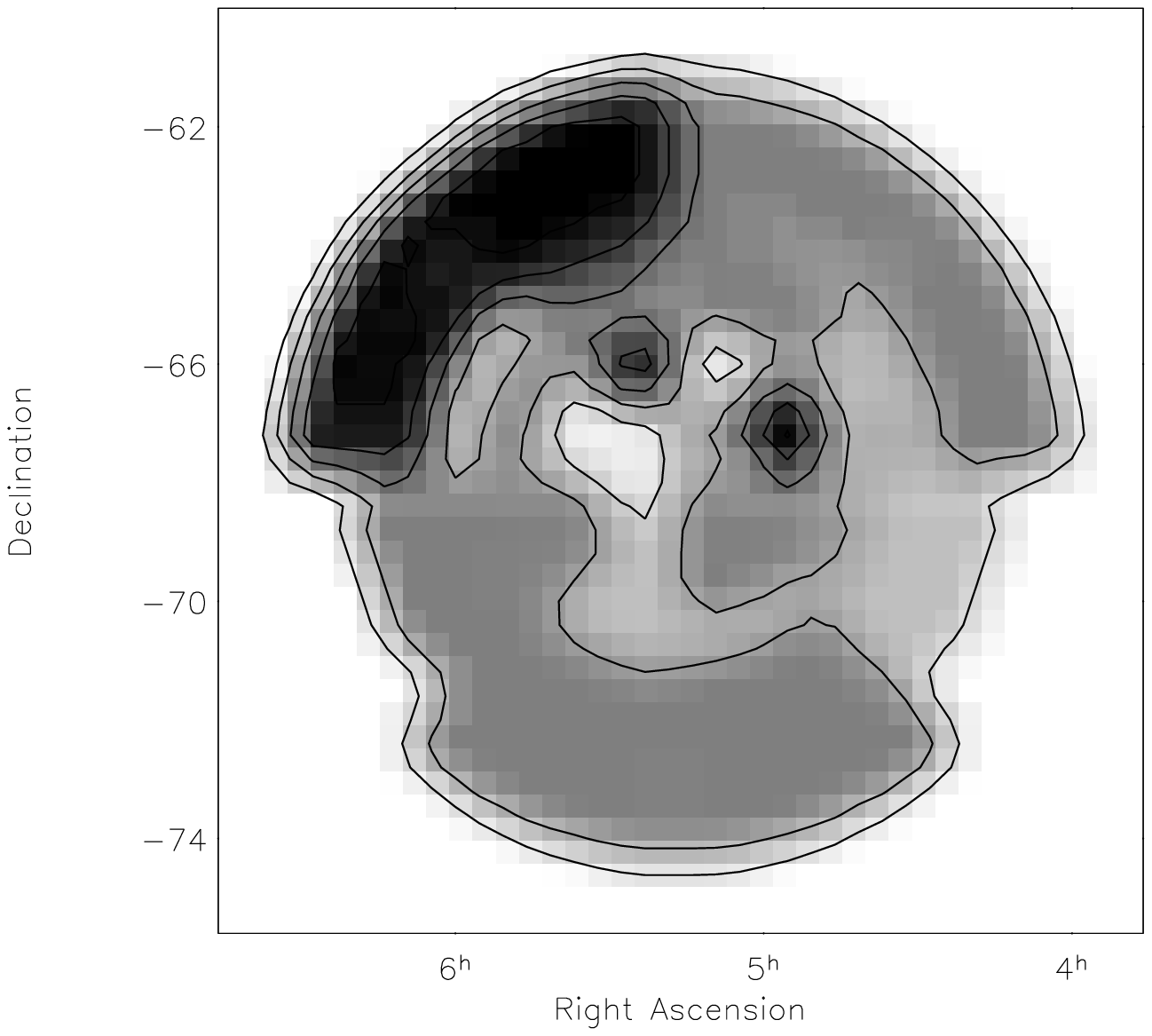}
\epsfxsize=0.32\hsize \epsfbox{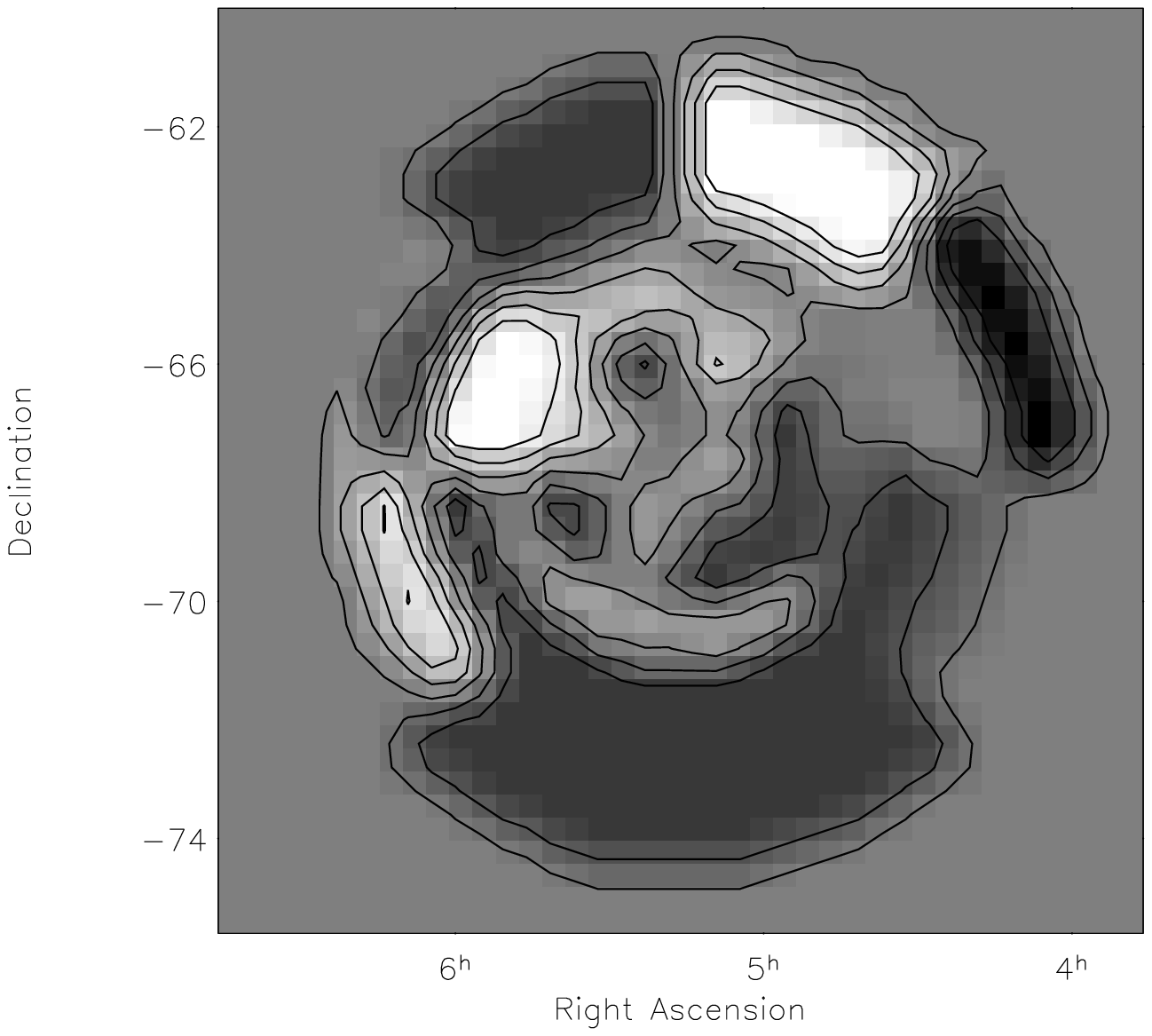}
\epsfxsize=0.32\hsize \epsfbox{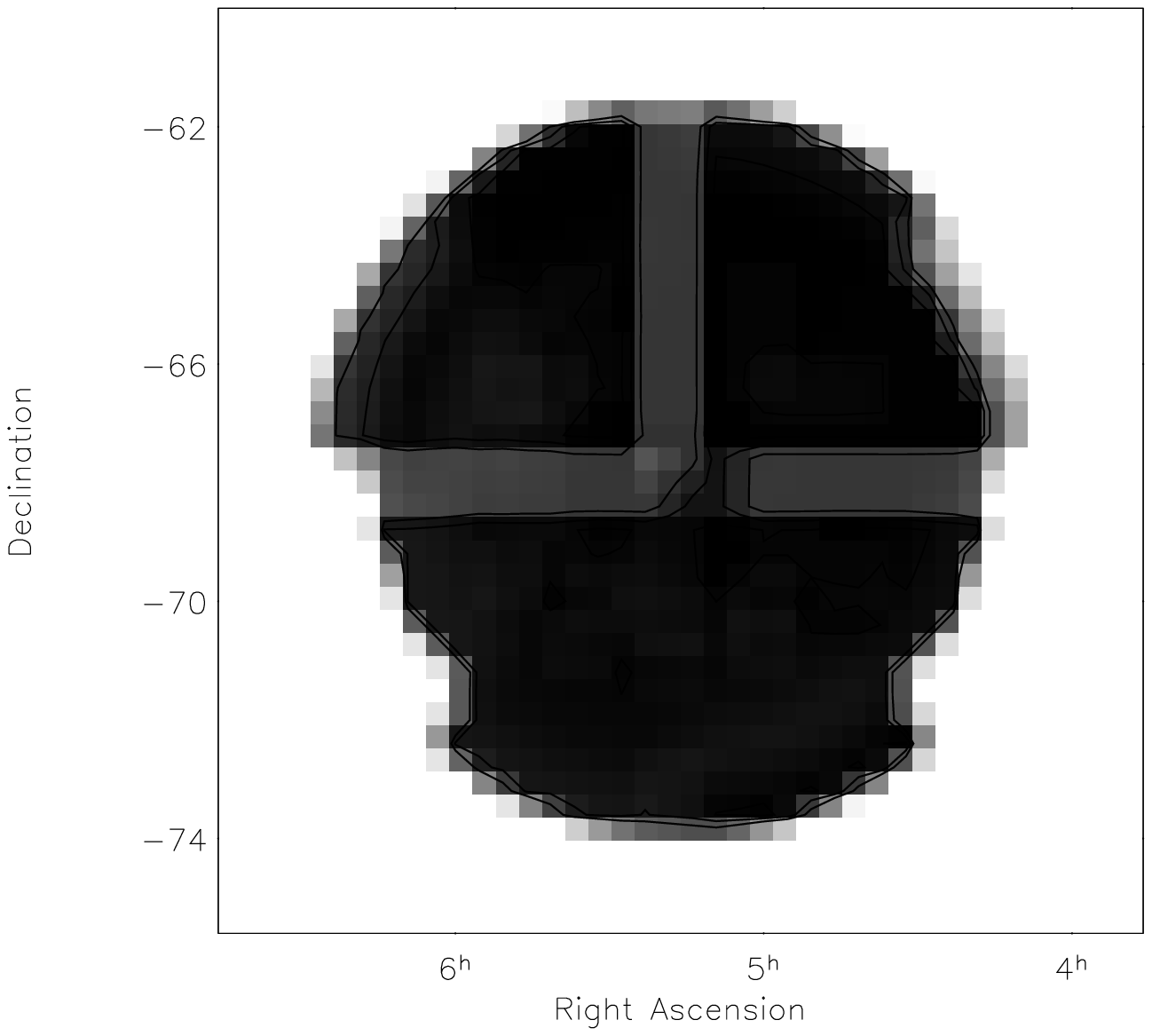}

\epsfxsize=0.32\hsize \epsfbox{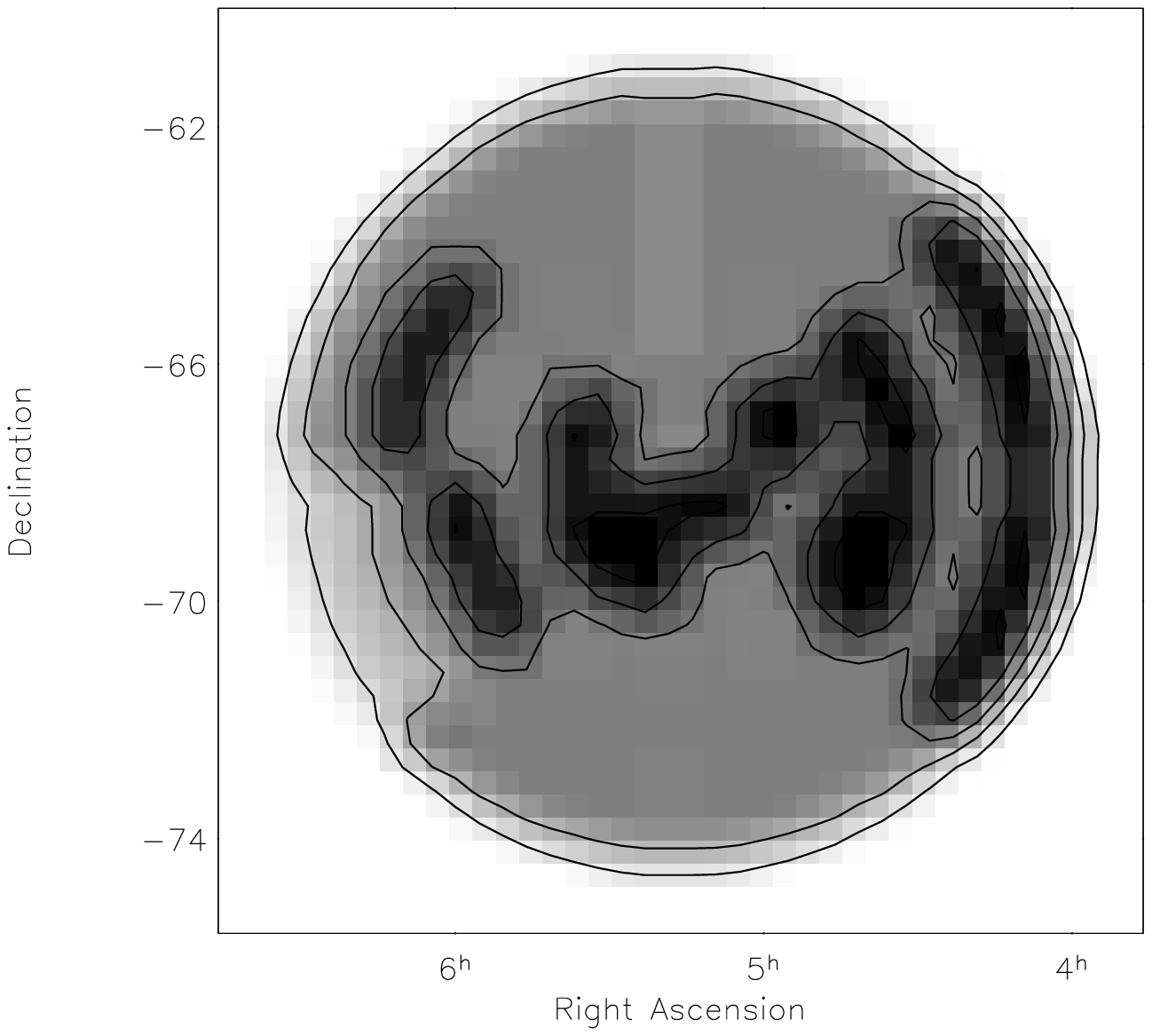}
\epsfxsize=0.32\hsize \epsfbox{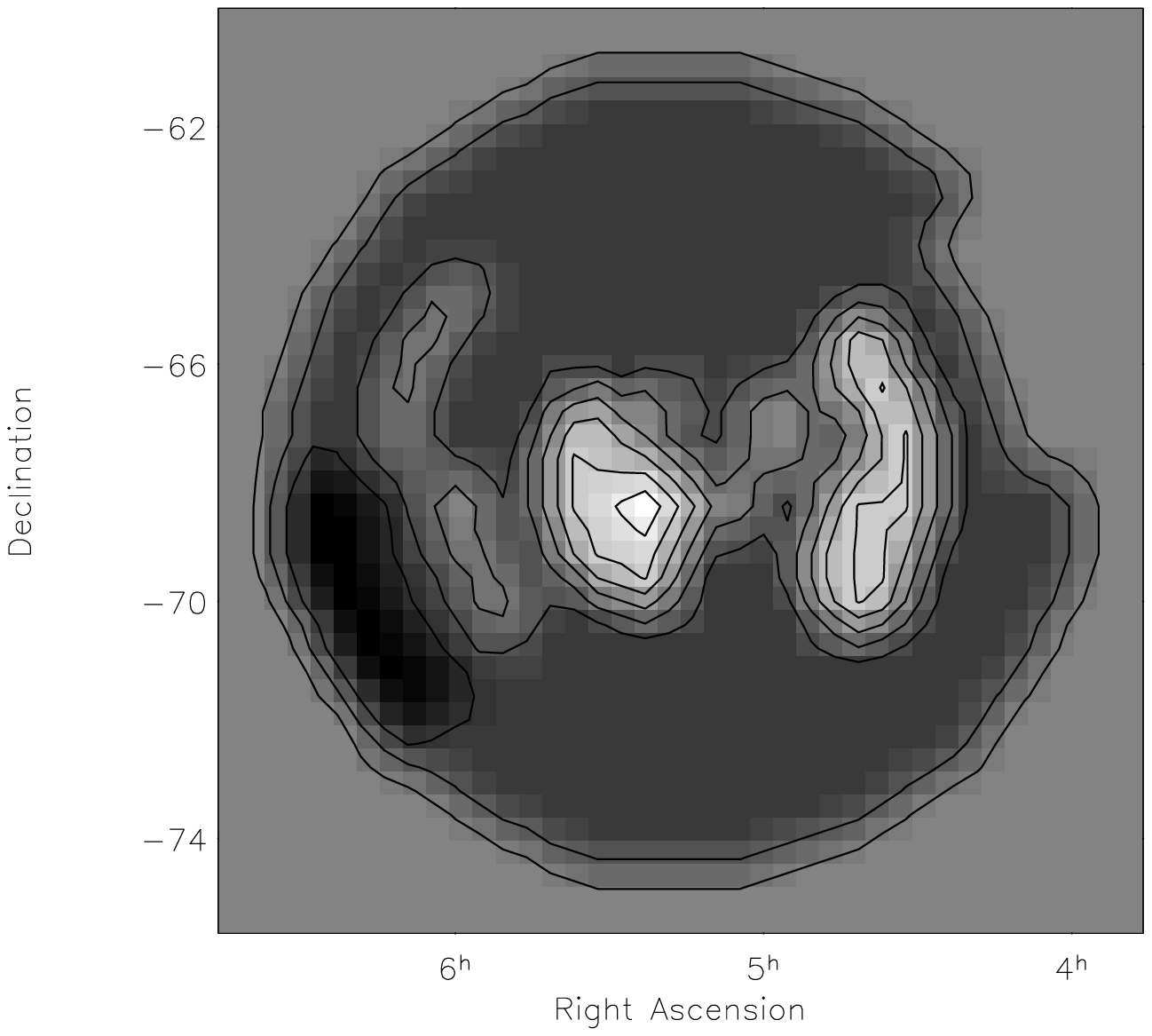}
\epsfxsize=0.32\hsize \epsfbox{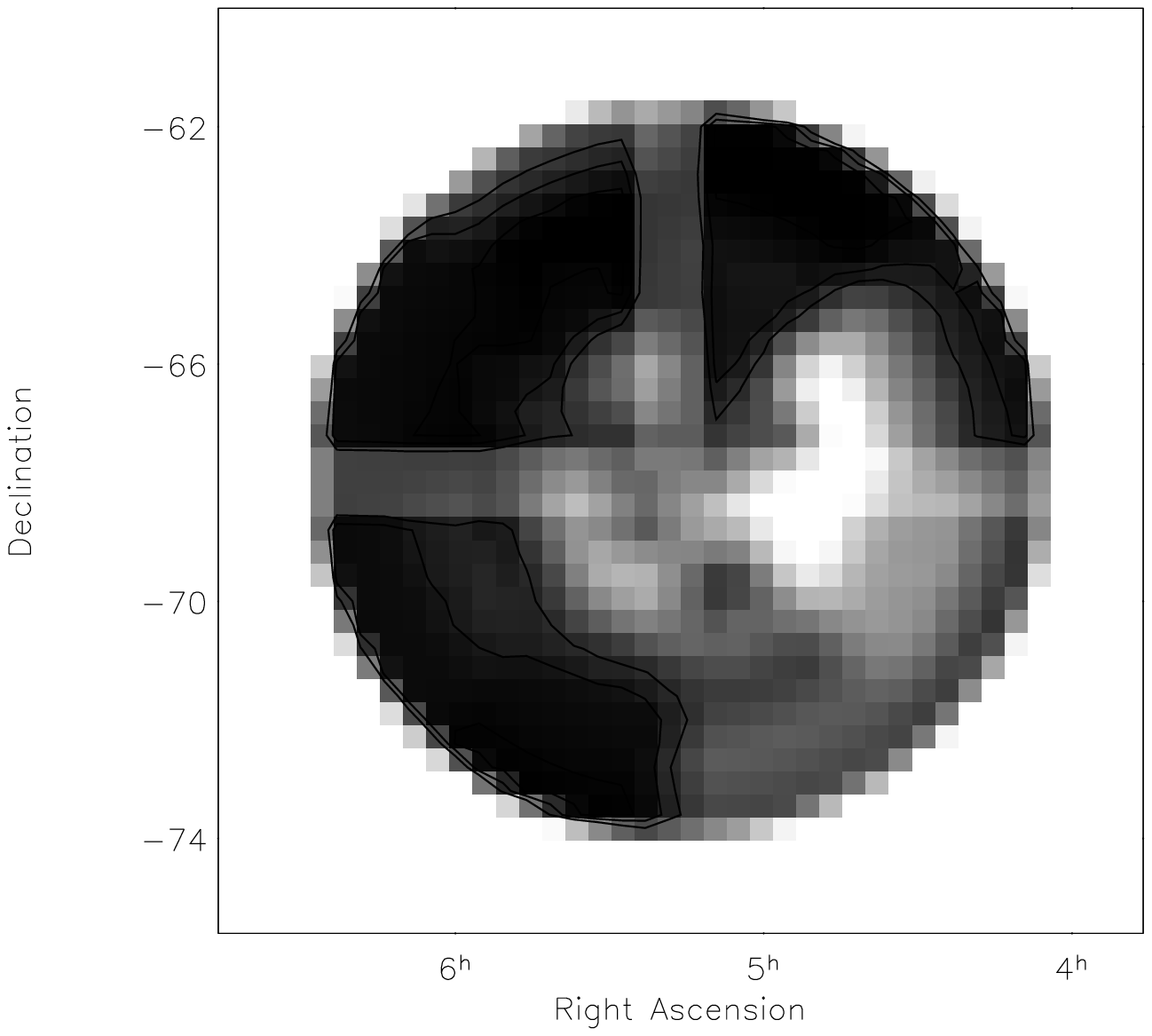} 
\caption{{\bf Top row}: distribution of the most probable metallicity
({\bf left} -- contours are at A$=0.003-0.015$ with a step of $0.003$),
SFR ({\bf middle} -- contours are from $3$ to $9$ Gyr with a step of $1$) and the
corresponding probability ({\bf right} -- 
the grey scale shows values above $0.8$ and
contours are at $0.97$, $0.98$ and $0.99$) for C stars. 
{\bf Bottom row}: the same distributions, with the same contour
values, for M stars. Darker regions correspond to higher numbers.}
\label{lumza}
\end{figure*}

Combining the information in each of the maps in Fig. \ref{lumall}
we have constructed maps that for C and M stars separately 
show the best combination of metallicity and SFR. In practice we show
the metallicity that corresponds to the maximum
probability among the different SFRs (i.e. for the C stars in the top
row of Fig. \ref{lumall} the most probable value in each bin).
The result is shown in Fig. \ref{lumza}
where the map of best metallicity, SFR and corresponding probability
are shown, again separately for C (top row) and M stars (bottom row). 
Note that to each SFR we have
assigned the mean age of all stars that formed accordingly to that
particular SFR because this quantity, contrary to
the mean age of AGB stars only, does not depend on metallicity
(Table \ref{sfrage}). Therefore contour values correspond to the 
mean age of the
local population. 

\begin{table}
\caption{Mean age of all stars formed accordingly to a specific
rate of star formation.}
  \[
    \begin{array}{cc}
      \hline
      \noalign{\smallskip}
      \alpha & \langle\mathrm{age (Gyr)}\rangle \\
      \noalign{\smallskip}
      \hline
      \noalign{\smallskip}
        -2    &  2.0 \\
        -5    &  3.9 \\
      +\infty &  6.3 \\
         5    &  8.7 \\
         2    & 10.6 \\
      \noalign{\smallskip}
      \hline
    \end{array}
  \]
\label{sfrage}
\end{table}

\begin{figure}
\hspace{-0.6cm}
\epsfxsize=0.6\hsize \epsfbox{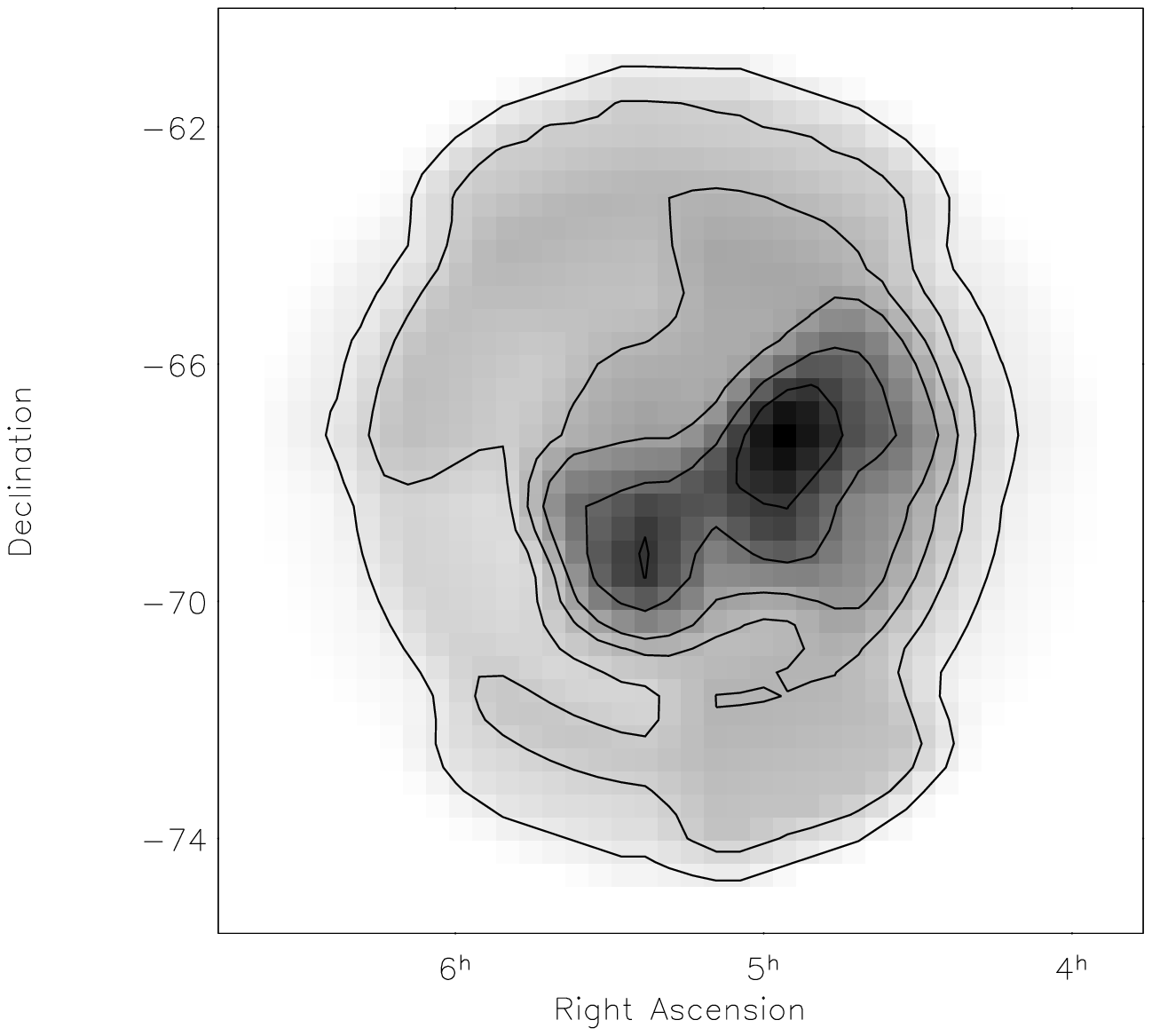}
\hspace{-1.4cm}
\epsfxsize=0.6\hsize \epsfbox{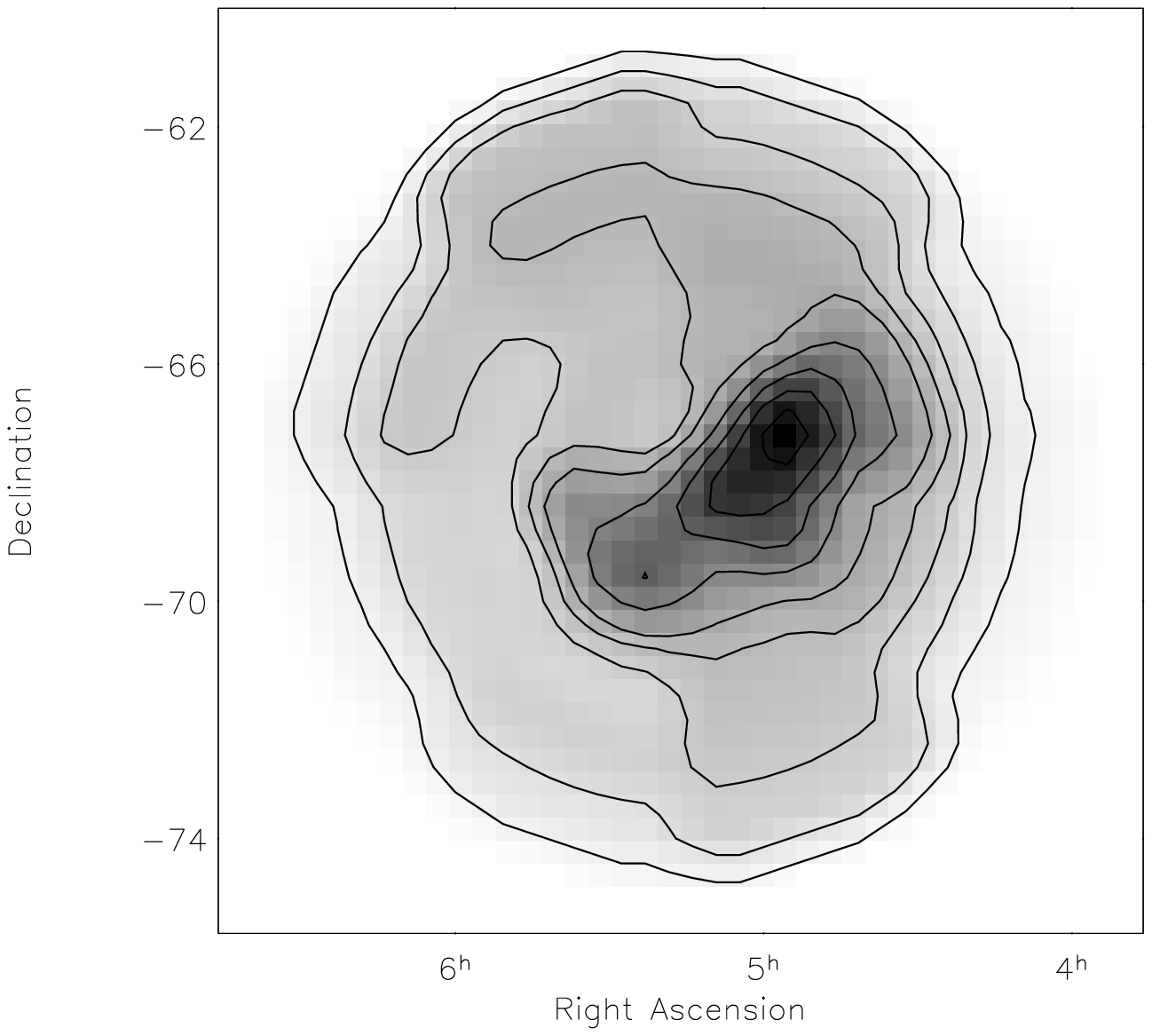}

\hspace{-0.6cm}
\vspace{-0.1cm}
\epsfxsize=0.6\hsize \epsfbox{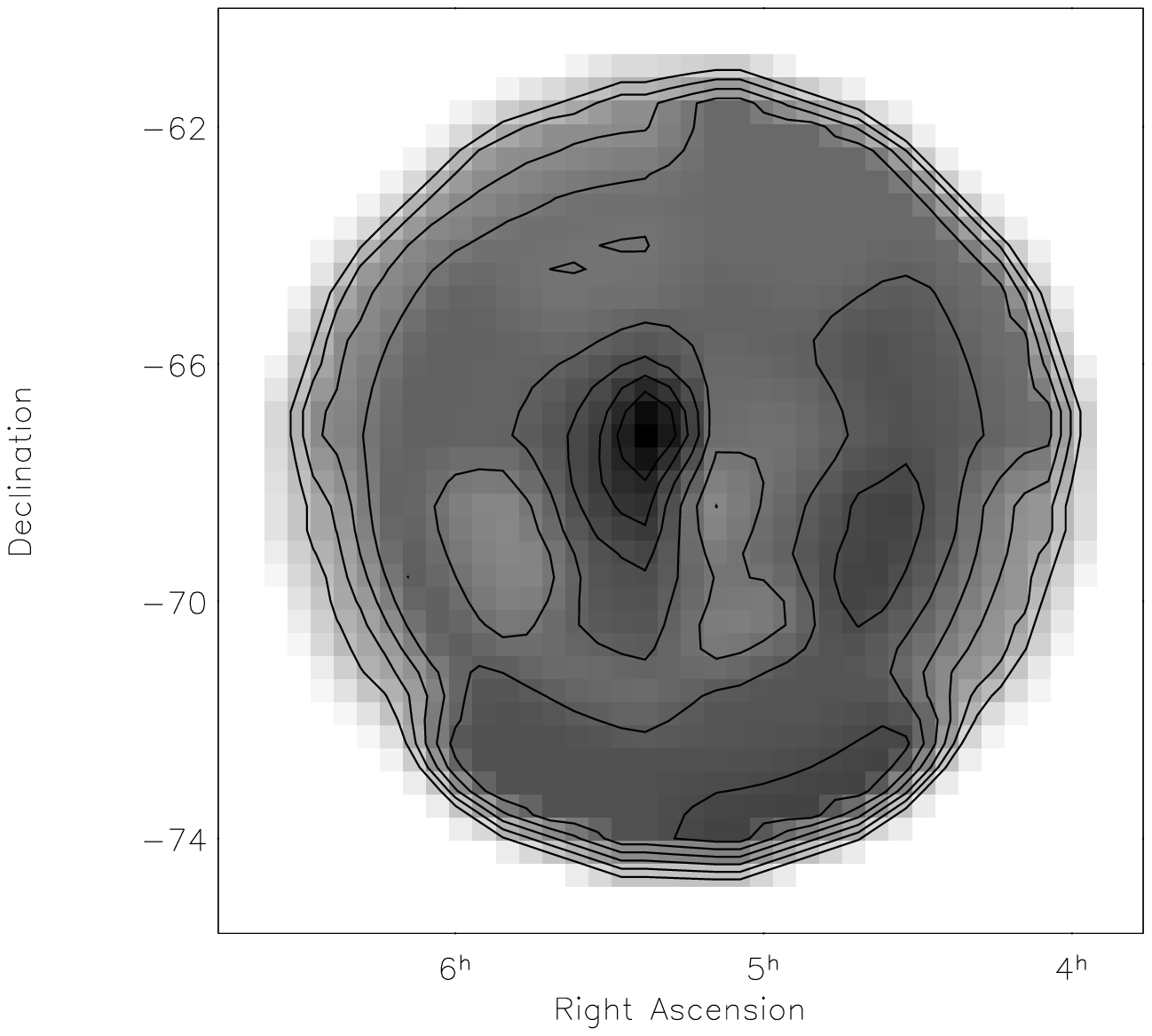}
\hspace{-1.4cm}
\epsfxsize=0.6\hsize \epsfbox{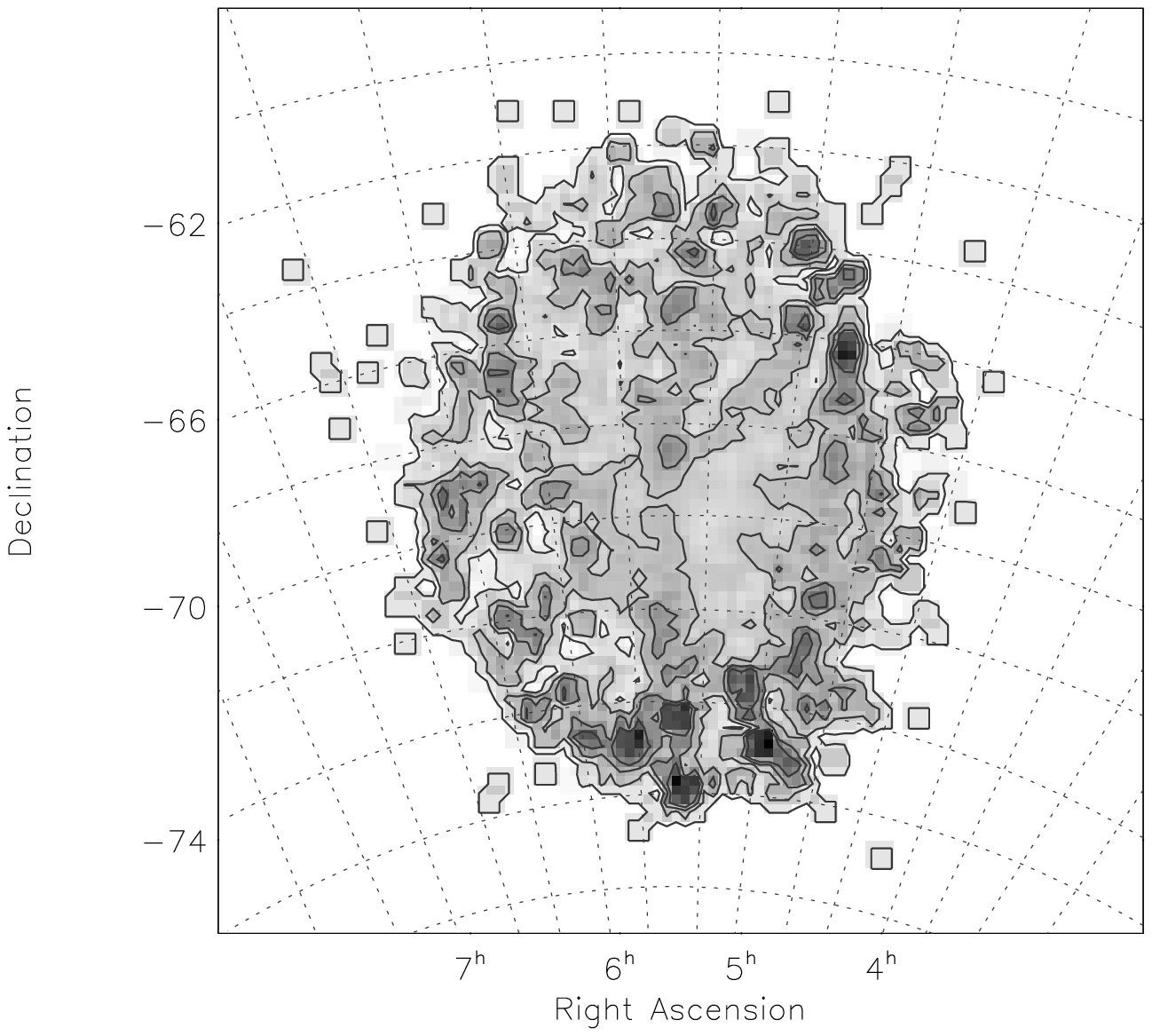} 

\caption{{\bf Top row}: Distribution of the number of C (contours are
at $50$, $100$, $150$ and then at $200-500$ with a step of $100$)
and M (contours are at $100-500$ with a
step of $100$ and then at $700-1300$ with a step of $200$) stars. {\bf
Bottom row}: distribution of the C/M ratio binned as in this work
($0.16$ deg$^2$/bin -- contours are at $0.2-0.55$ 
with a step of $0.05$) and as in paper I ($0.04$ deg$^2$/bin -- zoom
into original figure). Note that here because
we need a significant number of stars to fit the LF we do not
reach a resolution as high as in paper I while mantaining the
same global features (see text).}
\label{num}
\end{figure}

Figure \ref{num} shows the distribution of the number of C and
M stars and of their C/M ratio across the LMC with the same binning as
in the other maps. A high resolution version of the distribution of the
C/M ratio as obtained in paper I is also shown. The
effect of binning on smoothing galaxy features is clear, but major
trends are conserved. For example a comparison of both C/M ratio
distributions shows a dark region (i.e. a region with a high C/M
ratio) to the S of the LMC extending to the NW, another
approximately circular region is present NW of the galaxy
centre. The first dark patch is recognizable also in the
high-resolution map as well as the patch close to the centre. The
latter is emphasized in the low-resolution map because of the
larger bin size although it remains surrounded by a region of much
lower C/M ratio values. High resolution maps of all AGB stars (C
plus M stars) can be found in Cioni et al. (\cite{cmor}).

The distribution of C stars is overall best described by $Z=0.006$
except in the NE region (towards the Milky Way plane) as well as in two
approximately circular inner regions where the metallicity appears to
be higher, and in the SW, towards the SMC, encompassing the bar region
where the metallicity is perhaps lower. The stellar population has a
rather patchy distribution of age: it is on average $8$ Gyr old in the
S and NE with inner areas with stars at least $3$ Gyr younger. 
The probability distribution obtained from fitting the LF of
C-rich AGB stars is overall above $80$\% with most values above
$97$\%; this provides an excellent constraint to the distribution of
metallicity and look-back-time. Note that the bar of the galaxy does
not emerge from these maps, perhaps indicating that its formation was
not coeval. 
 
Similar maps for M stars indicate an overall metallicity, in agreement
with that derived from C stars, of $Z=0.006$ except along the bar of
the galaxy and continuing to the outer W region, where the metallicity
could be higher.
The distribution of the mean age of the stellar population is rather
uniform and indicates a mean age of $8$ Gyr except for three well
defined regions: an older population to the E and a younger population
close to the center and; on the W end of the bar region. However the
probability distribution provides a weaker constraint especially in
this region.  

\subsection{Projection effect}

The orientation of the LMC disk has been determined by van der Marel
\& Cioni (\cite{maci}) analysing the brightness variation of features in the
colour-magnitude diagram as a function of position within the galaxy.
In particular the mode of the distribution of AGB stars,
independently of their spectral type, of C-rich stars only and of the
tip of the RGB. All these tracers produced a sinusoidal variation
with an amplitude of about $0.25$ mag as a
function of angular position in the galaxy. A best fit geometrical
model was used to infer a value for the inclination and position angle
of the line of nodes.

The detected amplitude is of the same order of size of the bins in the maps
presented above, therefore it may affect the shape of the distribution
of either C-rich or O-rich AGB stars. For example a shift of $0.1$ mag
implies that half of the sources in a given bin will move to the
adjacent bin. In order to investigate the strength of this effect we
have applied a correction to each sectors within rings $2$ and $5$
following the best fit model used by van der Marel \& Cioni
(\cite{maci}). The correction is the same for each ring and is listed
in Table \ref{ortab}. Although van der Marel \& Cioni
(\cite{maci}) also detected a radial variation (i.e. as a
function of ring) it is not very well constrained. Note also that no
correction was applied to the inner most rings ($0$ and
$1$). This region, strongly influenced by the effect of the bar,
requires further study.

\begin{table}
\caption{Magnitude shift due to the LMC orientation. The shift is the
same for rings number $2$, $3$, $4$ and $5$.}
  \[
    \begin{array}{lrclr}
      \hline
      \noalign{\smallskip}
      \mathrm{Sector} & \mathrm{Shift} & | & \mathrm{Sector} & \mathrm{Shift}\\
      \noalign{\smallskip}
      \hline
      \noalign{\smallskip}
        \mathrm{E-NE} & -0.12 & | & \mathrm{W-SW} &  0.12 \\
        \mathrm{NE-N} & -0.10 & | &\mathrm{SW-S} &  0.10 \\
        \mathrm{N-NW} & -0.02 & | & \mathrm{S-SE} &  0.02 \\
        \mathrm{NW-W} &  0.07 & | & \mathrm{SE-E} & -0.09 \\
      \noalign{\smallskip}
      \hline
    \end{array}
  \]
\label{ortab}
\end{table}

\begin{figure}
\resizebox{\hsize}{!}{\includegraphics{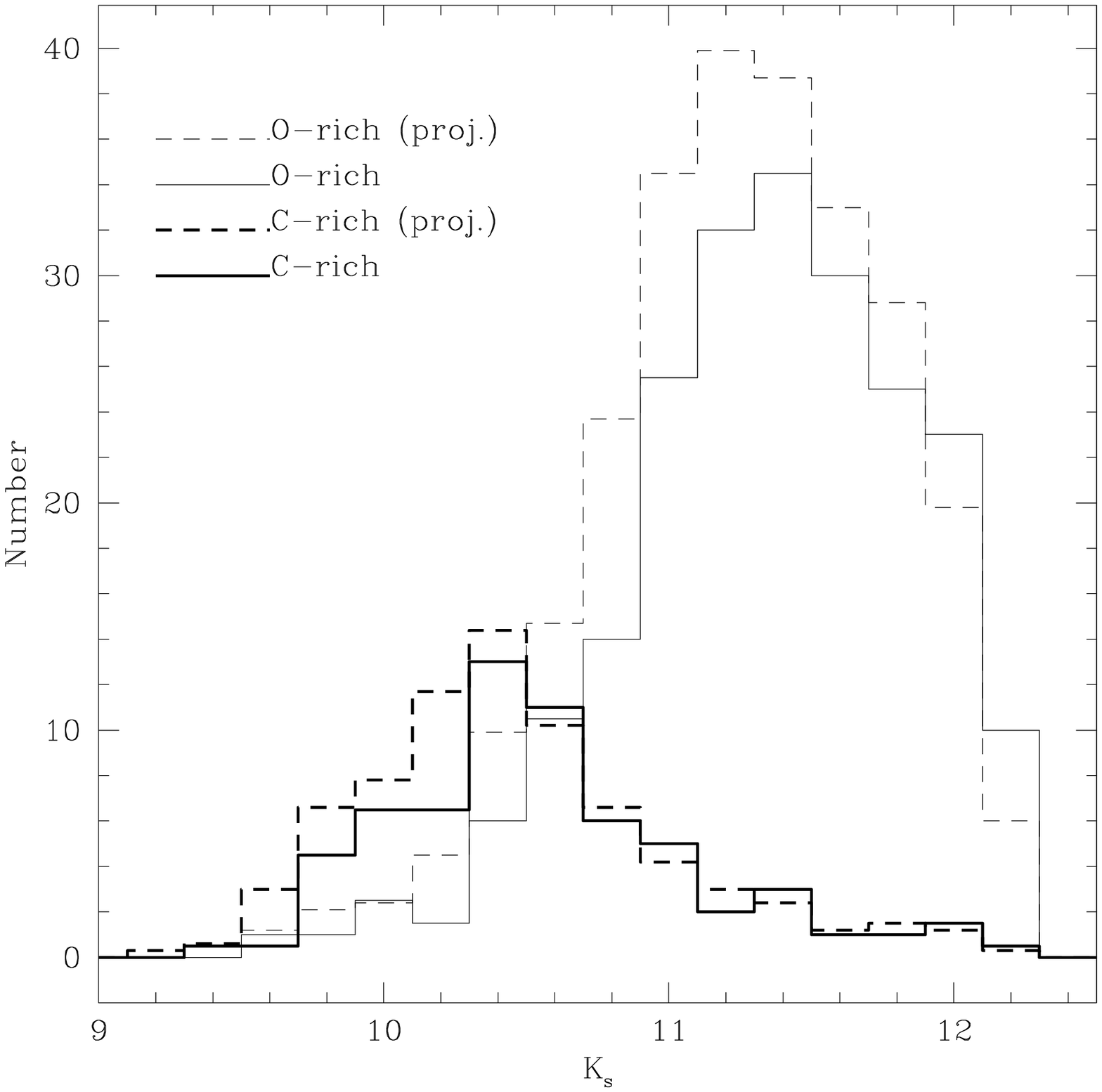}}
\caption{Magnitude distribution of C-rich and O-rich AGB stars in
sector E-NE of ring $2$ with (dashed lines) and without (continuous line)
correcting for the orientantion of the LMC.}
\label{orfig}
\end{figure}

Figure \ref{orfig} shows as an example the distribution of C-rich
and O-rich AGB stars with and without correction. In this case (sector
E-NE of ring $2$) a shift of $-0.12$ mag is applied and the
histogram moves to brighter magnitudes as well as changing in shape.
Note the distribution of C-rich stars around $K_\mathrm{s}=10.2$.

\begin{figure*}
\epsfxsize=0.32\hsize \epsfbox{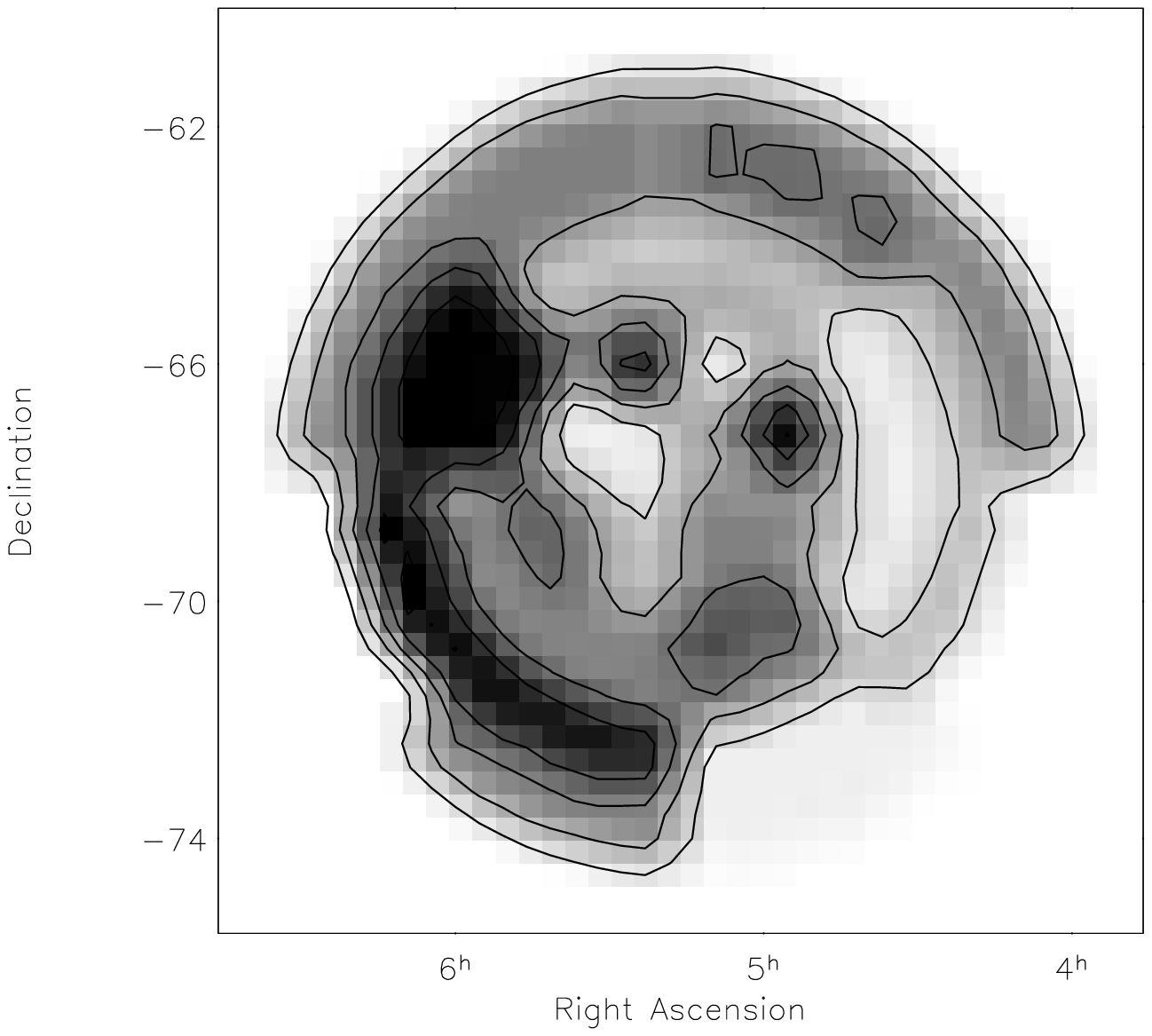}
\epsfxsize=0.32\hsize \epsfbox{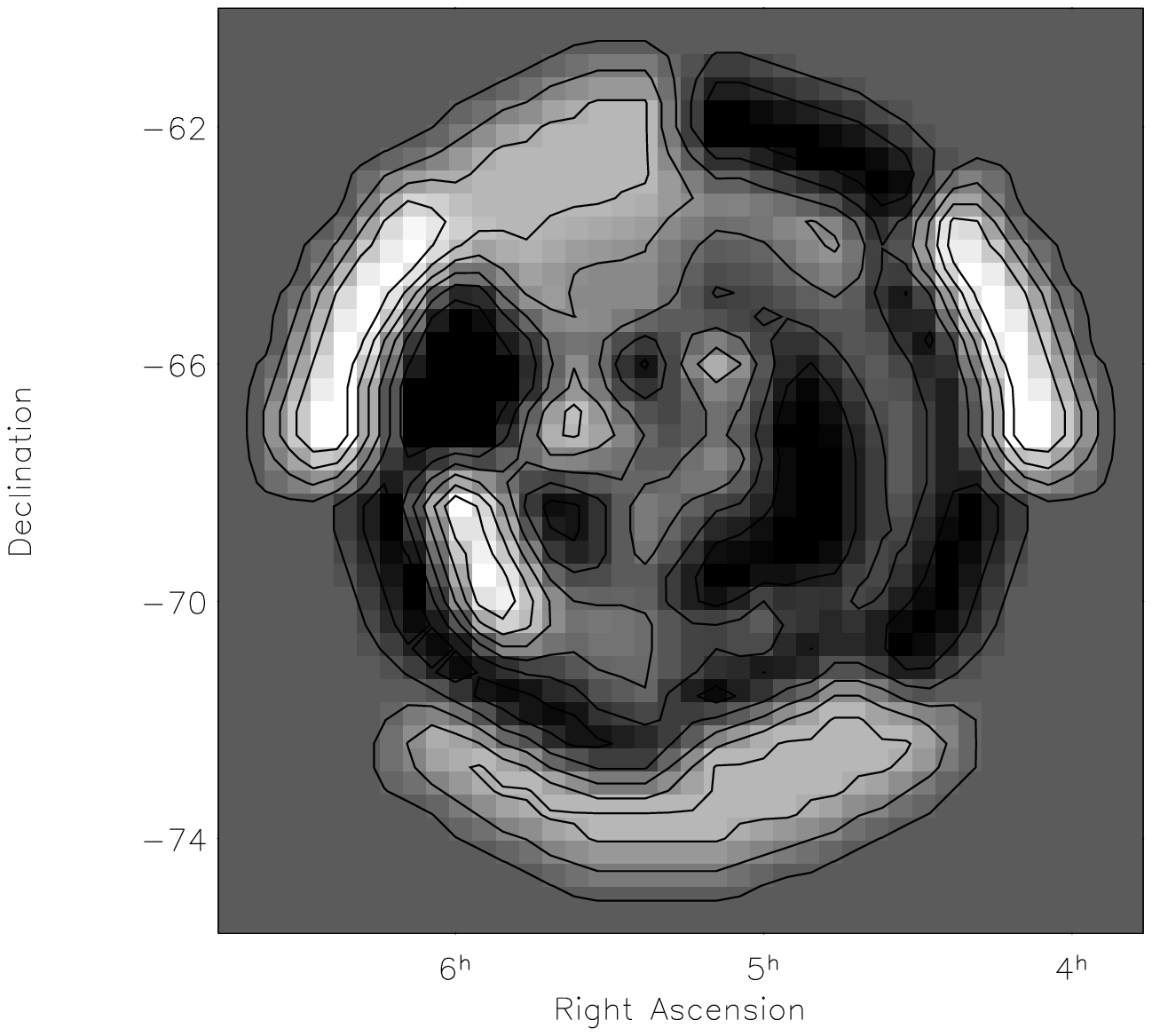}
\epsfxsize=0.32\hsize \epsfbox{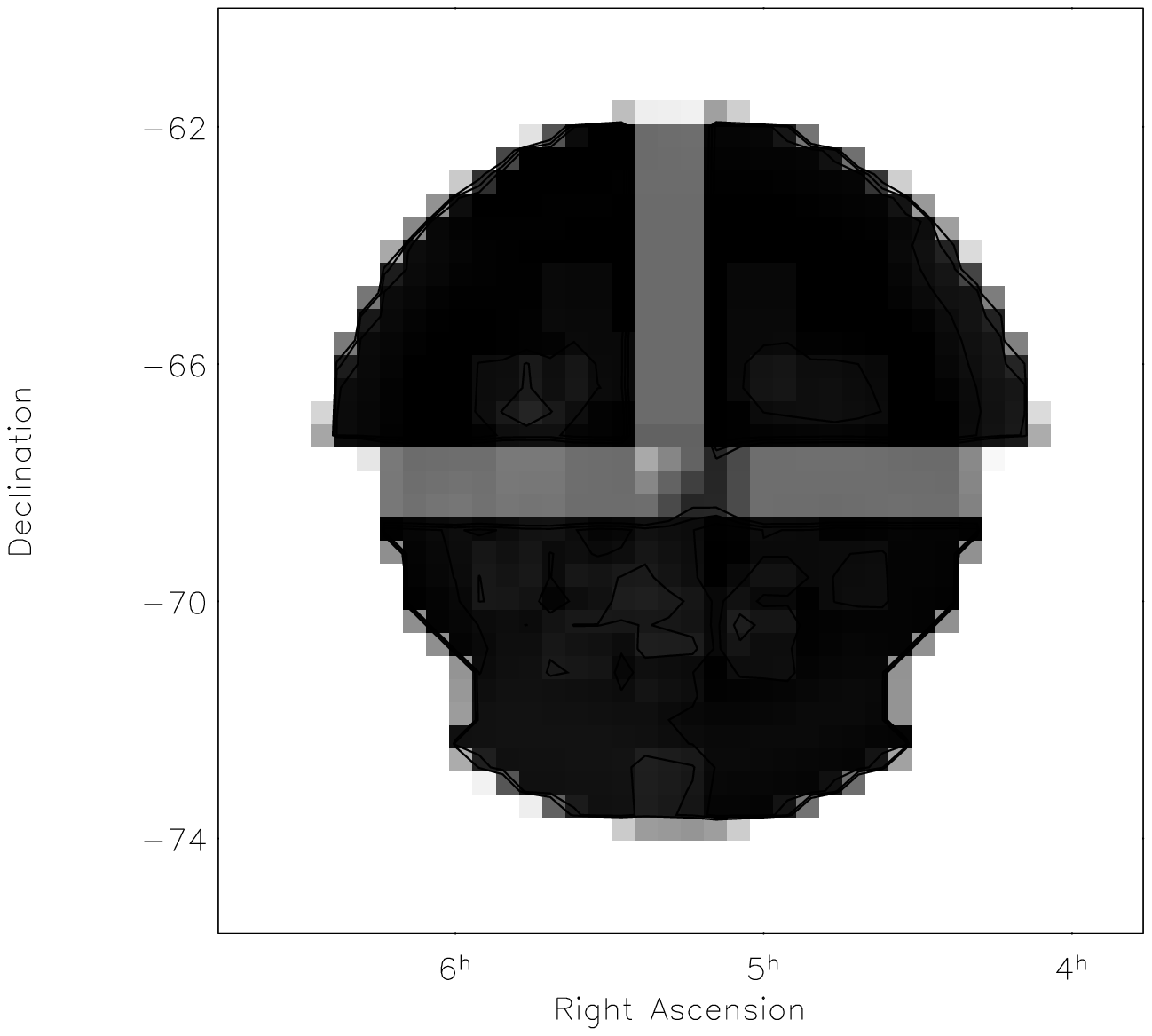}

\epsfxsize=0.32\hsize \epsfbox{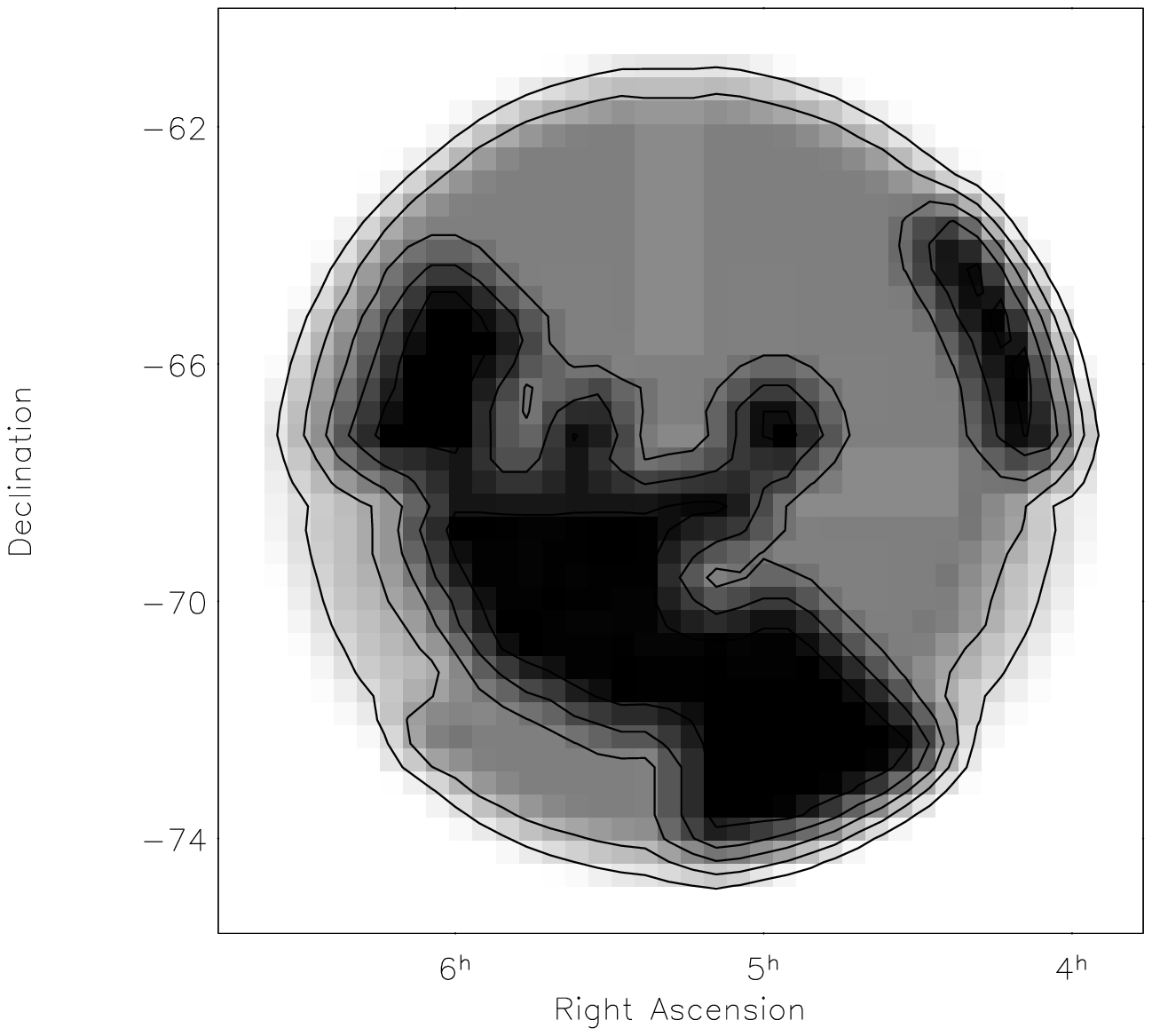}
\epsfxsize=0.32\hsize \epsfbox{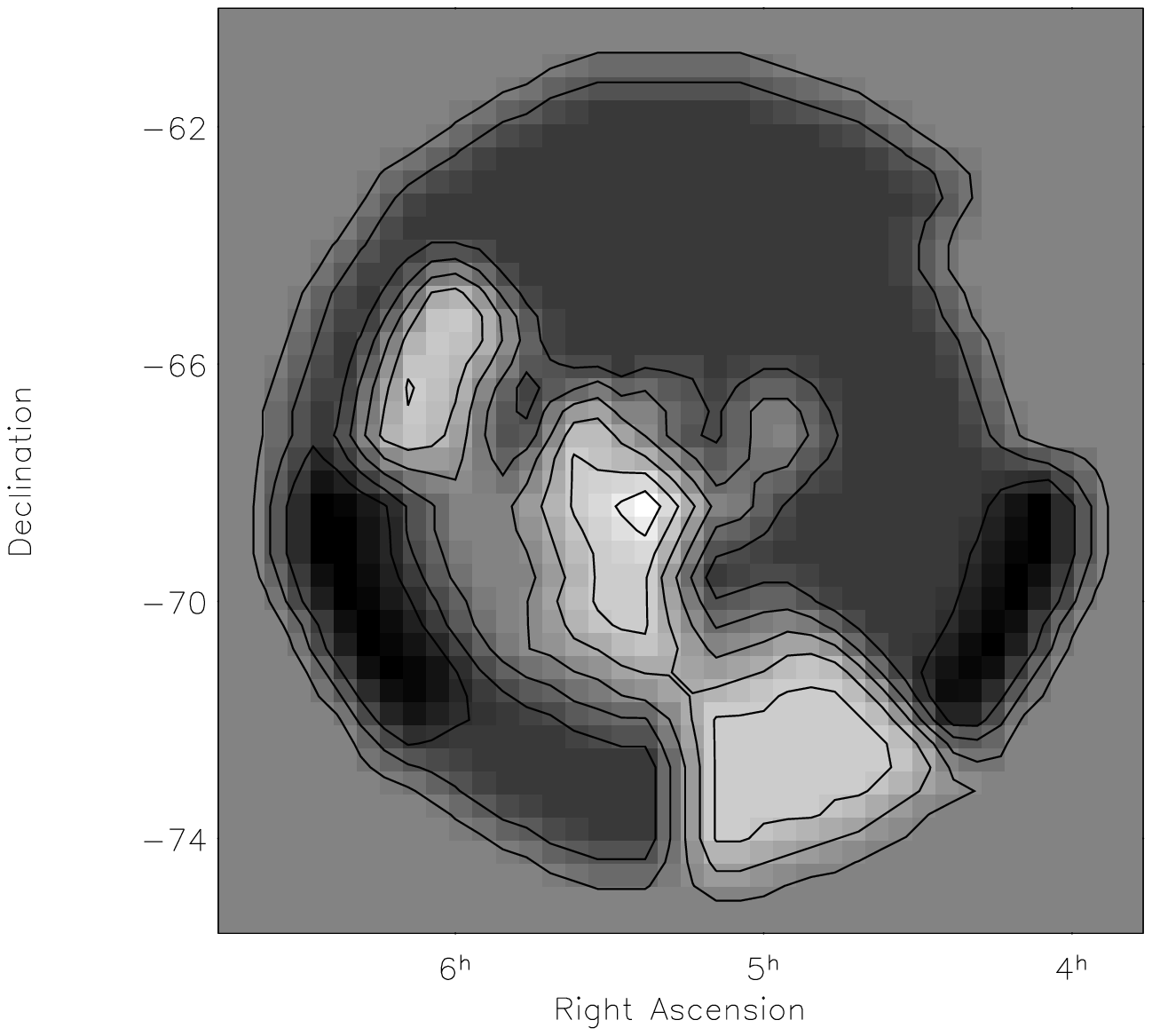}
\epsfxsize=0.32\hsize \epsfbox{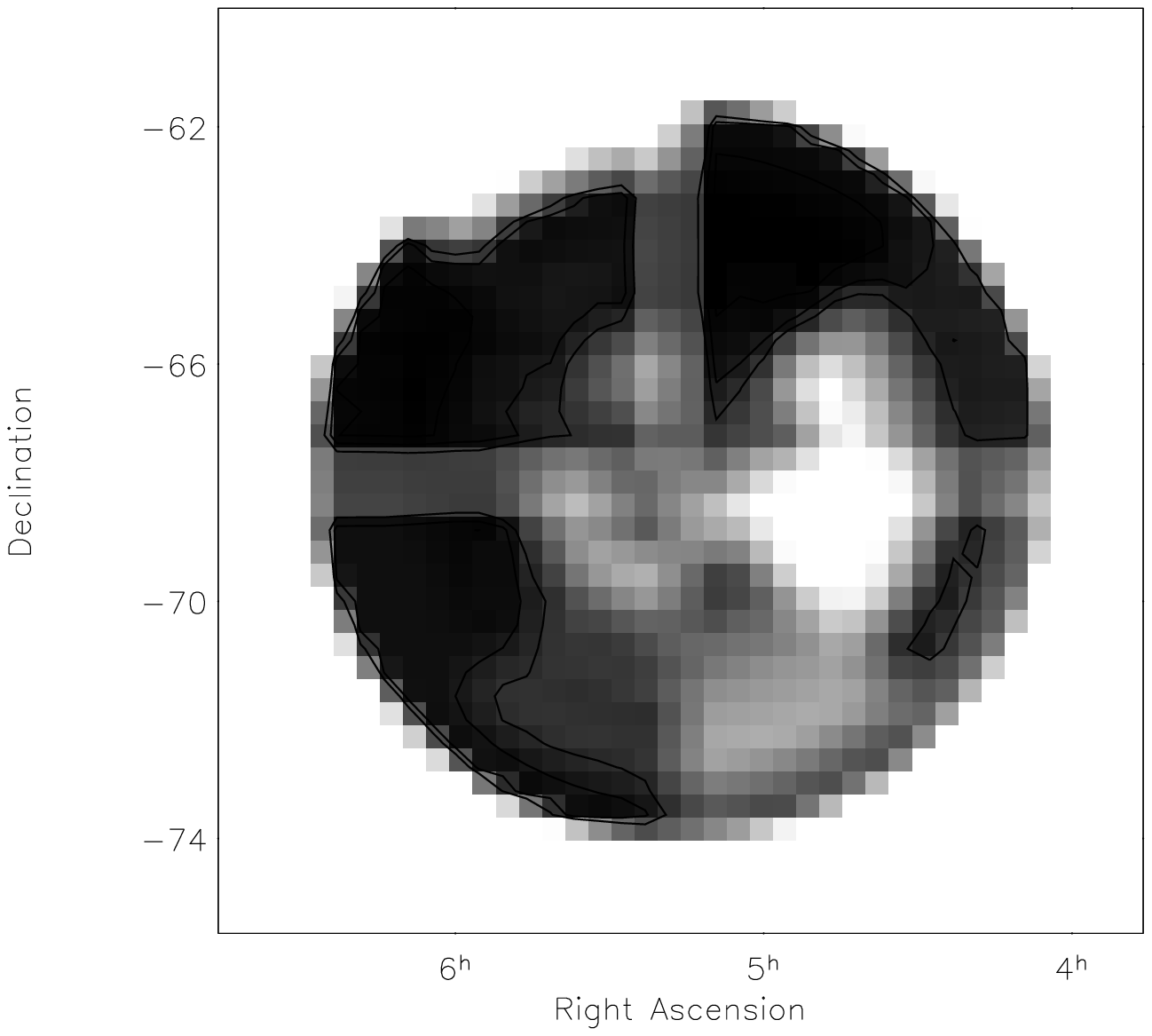} 
\caption{The same as Fig. \ref{lumza} but after correcting each
sector in rings number $2-5$ for the LMC orientation as in Table
\ref{ortab}. Contour values are the same as in Fig. \ref{lumza}.}
\label{lumzan}
\end{figure*}

Figure \ref{lumzan} shows the same maps as in Fig. \ref{lumza} but
after correcting for the orientation of the LMC. The fit of the
observed distribution of both C and M stars occur at almost equal
probabilities while there are considerable variations in the
distribution of both metallicity and mean-age across the area of the
galaxy. In particular the outermost metal rich E-NE region suggested by
C stars has moved slightly S. The metallicity remains low towards
the SMC. The distribution of the mean-age of the stellar population,
although still appearing rather patchy, indicates an inner ring/region
more metal rich than the outer ring/region. M stars support the same
overall metallicity $Z=0.006$ but the region of high metallicity that
previously coincided with the bar of the galaxy is now stretching from
E to SW encompassing the E and central bar region but not the W end
of it. The same structure corresponds to a younger population (about
$3$ Gyr on average) compared to the population of the overall disk,
although in the SE a region with an older population is still present.

\section{Discussion}
\label{dis}
\subsection{How well does the C/M ratio alone trace metallicity?}
In paper I the distribution of the C/M ratio has been interpreted
purely as a tracer of metallicity concluding that the LMC shows the 
typical metallicity pattern, as
in other galaxies, of decreasing metallicity towards its outskirts. 
In this paper we have shown that the C/M ratio depends
also on the SFR and we have created maps of metallicity, separately
for C and M stars, that account for this effect. After this
analysis are the conclusions of paper I still valid?

Magnitude variations of the order of those applied in the
previous section have a minor effect on the distribution of the C/M
ratio. At the brightest magnitudes there are overall very
few sources while at the faintest magnitudes the 
contamination with RGB stars and the larger photometric errors
involved play a larger role in assigning a source to a given bin than a
shift due to the viewing angles of the galaxy. Only in regions with 
overall low number statistics is the C/M ratio influenced
by the orientation of the galaxy. A precise correction would also
include a radial component that in the case of van der Marel \& Cioni
(\cite{maci}) was poorly constrained although present. This means that only
the tangential component of the orientation has been corrected for. 
Considering also that for the SMC (paper III) a correction is not
known at present, and we would like to analyse both galaxies
consistently, in the following discussion we compare the distribution
of the C/M ratio (Fig. \ref{num}) with the maps presented in
Fig. \ref{lumza}.    

The distribution of the C/M ratio in the LMC (Fig. \ref{num}) shows a
higher ratio to the S. This indeed corresponds to regions of
low metallicity in both the metallicity map obtained from C and M
stars. We can safely conclude that in this region the C/M ratio alone
is a good tracer of metallicity. Extending to the SW high C/M values
indicate a low metallicity in agreement with the metallicity derived
from C-stars but O-rich stars support a higher metallicity in this
region, although the probability is low ($<60$\%). On the periphery of
the LMC, still in the SW, the C/M ratio decreases supporting a higher
metallicity in agreement with that traced by M stars. In this region 
there are not enough C stars to produce a reasonable fit of their luminosity
function, the same is true in the SE. Note that the distribution of the number
of C and M stars shows that M stars extend in number much further
out compared to C stars.
The well defined region of high metallicity (NE) obtained from C
stars is also high in metals from M stars although the latter spreads
across a much larger area. In this region the C/M ratio has
intermediate values but clearly decreases to the periphery it is thus
plausible to conclude that towards the NE of the LMC the metallicity
is high. In the central part of the galaxy the C/M ratio indicates a
region of low metallicity. The C star metallicity distribution
supports this finding although there is 
 a dichotomy in the sense that the northern most part of this
small region could be metal rich while the southernmost part of it
metal poor. 
Even though the binning plays a major role in exactly constraining 
the location of this inner region, it may 
correspond to Shapley constellation
III where star formation is currently taking place. This explains the
young population inferred from M stars as well as an underlying older
population inferred from C stars. Therefore in this region the C/M ratio is
strongly influenced by age. The C/M ratio is not a good tracer of metallicity
if the chemical composition of a system varies on time scales shorter than
about 1 Gyr (Mouhcine \& Lan\c{c}on, \cite{moulan}).

Although at first sight the metallicity maps obtained from C and M
stars and the distribution of the C/M ratio do not look much alike
they have similar trends. This is the strength of the C/M
ratio alone while for a more accurate evaluation of metallicity
a knowledge of the age of the population is essential. 
A comparison between the C/M ratio (Fig. \ref{num}) and the 
metallicity distributions shown in Fig. \ref{lumzan} allow us
to reach similar conclusions in those regions where the probability
(i.e. the level of significance) is high, in particular for C stars. 

Finally the spread in
metallicity within both Magellanic Clouds derived in paper I
amounts to $0.75$ dex, this is in very good agreement with $0.7$
dex derived from C stars and M stars (i.e. the difference between
the low and high contour converted into [M/H] assuming
Z$_\odot=0.019$).

\subsection{The mean age distribution}
%Information about the star formation history of the LMC has been
%obtained from the study of many relatively small regions located
%in the outer and inner disk as well as along the bar (Sect.
%\ref{intro}). There are regions of the galaxy that can be
%described by a relatively uniform SFR across several Gyr. The dominant
%stellar population is of intermediate-age ($>2.5$ Gyr) and extends
%to the remote periphery. A burst of star formation has occurred
%between $1$ and $3$ Gyr, although some authors attribute to this
%event the formation of the bar, there is evidence that stars as
%old as $4-8$ Gyr exist in the bar as well as in the disk. Only a
%detailed kinematic study of the stellar population will reveal its
%distinct components. This as well as more recent bursts of star
%formation are probably due to a close passage with the Mikly Way
%and the SMC. In the outer disk searches for stars similar to those
%in the halo of the Milky Way have shown that metal poor old giants
%are lacking. AGB stars formed essentially during two major epochs:
%around $10^8$ yr ago for the most massive and a few Gyr ago for
%lower masses. Globular clusters are older than about $11$ Gyr or
%younger than about $4$ Gyr.

Interpreting the magnitude distribution of C and M stars we derive that most of
the LMC population is on average about $5-6$ Gyr old (Fig. \ref{lumza}),
supporting a population that is predominatly of intermediate age.
There are, however, well defined regions with older or younger stars
especially in the northen hemisphere of the galaxy compared to the
southerm hemisphere that appears more homogeneous, unless a correction
for the LMC viewing angles is applied. There is a clear indication
that the bar of the galaxy contains a composite stellar population
(i.e. it did not form at the same time). C stars indicate that the
periphery of the galaxy is not systematically older than the inner
region, contrary to M stars, although the indication provided by M
stars is less significant (i.e. they correspond to a lower $\chi^2$ value). 
If the distribution of stars is corrected for the projection of the
galaxy there appear to be younger C stars in its periphery. Note that this is
further out than the observations by Costa \& Frogel (\cite{cost}) who
claimed a similar result.

\section{Summary and conclusions}
In this paper we compare the observed $K_\mathrm{s}$ magnitude
distribution of C and M stars within sectors entirely covering 
the LMC, of a suitable size to provide a statistically 
significant sample, with theoretical distributions. These have
been obtained using the population synthesis code TRILEGAL (Girardi et al.
\cite{gi05}), employing suitable stellar evolutionary tracks for stars of
different mass and in particular using Marigo et al. (\cite{ma03}) models for
thermally pulsing AGB stars. This allows us to well describe the behaviour of
both C and M type stars as a function of $K_\mathrm{s}$ magnitude.
The quality of the comparison has been quantified using the $\chi^2$ test.

Surface maps indicating the most probable metallicity and mean age
distribution have been obtained varying one or
both parameters at the same time. Results have been
compared with the information available from previous studies, although
they covered only a limited area of the LMC.
With respect to the conclusions reached in Paper I, the C/M ratio is
influenced by the age of the population where it is calculated, however it
remains a good tracer of metallicity only if the population is on average
older than a few Gyr. Note that here we refer to the mean age of the
stellar 
population in a given spatial location and not to the precise age at which
that particular population began forming stars. According to the LMC
structure parameters derived by van der Marel \& Cioni (\cite{maci})
we have created new surface maps that account for the viewing angles
of the LMC. Non-negligible differences were obtained in the
distribution of both metallicity and mean age of the underlying
stellar population.

Our best fit models indicate that stars in the LMC are on average 
$5-6$ Gyr old, and that the metallicity of most stars is consistent with
$Z=0.006$. However, since TP-AGB models are not perfect, and since 
we are testing just a family (the exponential ones) among many possible 
SFR histories, these average values of age and metallicity may be
affected by systematic errors, and are likely to change as we adopt 
improved TP-AGB models, and test other possible SFR functions.
Instead, we consider to be more significant the detected variations of 
these mean age and metallicity values across the LMC disk.
In fact, we find distinct regions in the LMC with an older or a younger
mean age (i.e. Shapley III). There is also a clear indication that the
metallicity is higher towards the plane of the Milky Way and
lower towards the SMC. 
This perhaps traces the dynamical history of the galaxy (i.e. the
interaction with its companions) for which a final
understanding will be provided from a detailed kinematic study of its
stellar population.

\label{fin}

\begin{acknowledgements}

L.G. is indebted to M. Salaris and M. Groenewegen for the frequent
sharing of information on several aspects of simulating CMDs for
Local Group galaxies.
\end{acknowledgements}

\end{document}